\documentclass[aps,pra,onecolumn,amssymb,amsfonts,floatfix,superscriptaddress,longbibliography,notitlepage,nofootinbib ]{revtex4-1}
\usepackage[utf8]{inputenc}
\pdfoutput=1
\usepackage{color}
\usepackage{float}
\usepackage{xcolor}
\usepackage{mathtools,amsmath,amsthm,amsfonts,amssymb}
\usepackage{commath}
\usepackage{physics}
\usepackage{bm}
\usepackage[pdftex]{graphicx}
\usepackage[colorlinks=false]{hyperref}
\usepackage{enumitem}

\begin{document}

\title{Chaos and Thermalization in the Spin-Boson Dicke Model}

\author{David Villase\~nor}
\affiliation{Instituto de Ciencias Nucleares, Universidad Nacional Aut\'onoma de M\'exico, Apdo. Postal 70-543, Mexico City 04510, Mexico}
\affiliation{Instituto de Investigaciones en Matem\'aticas Aplicadas y en Sistemas, Universidad Nacional Aut\'onoma de M\'exico, Mexico City 04510, Mexico}
\author{Sa\'ul Pilatowsky-Cameo}
\affiliation{Center for Theoretical Physics, Massachusetts Institute of Technology, Cambridge, MA 02139, USA}
\author{Miguel A. Bastarrachea-Magnani}
\affiliation{Departamento de F\'isica, Universidad Aut\'onoma Metropolitana-Iztapalapa, Av. Ferrocarril San Rafael Atlixco 186, Mexico City 09340, Mexico}
\author{Sergio Lerma-Hern\'andez}
\affiliation{Facultad de F\'isica, Universidad Veracruzana, Circuito Aguirre Beltr\'an s/n,  Xalapa 91000, Mexico}
\author{Lea F. Santos} 
\affiliation{Department of Physics, University of Connecticut, Storrs, CT 06269, USA}
\author{Jorge G. Hirsch}
\affiliation{Instituto de Ciencias Nucleares, Universidad Nacional Aut\'onoma de M\'exico, Apdo. Postal 70-543, Mexico City 04510, Mexico}


\begin{abstract}
We present a detailed analysis of the connection between chaos and the onset of thermalization in the spin-boson Dicke model. This system has a well-defined classical limit with two degrees of freedom, and it presents both regular and chaotic regions. Our studies of the eigenstate expectation values and the distributions of the off-diagonal elements of the number of photons and the number of excited atoms validate the diagonal and off-diagonal eigenstate thermalization hypothesis (ETH) in the chaotic region, thus ensuring thermalization. The validity of the ETH reflects the chaotic structure of the eigenstates, which we corroborate using the von Neumann entanglement entropy and the Shannon entropy. Our results for the Shannon entropy also make evident the advantages of the so-called ``efficient basis'' over the widespread employed Fock basis when investigating the unbounded spectrum of the Dicke model. The efficient basis gives us access to a larger number of converged states than what can be reached with the Fock basis.
\end{abstract}

\maketitle

\section{Introduction}
\label{sec:Introduction}

The onset of thermalization in isolated quantum systems, which evolve unitarily and are described by pure states, was discussed in von Neumann's 1929 paper on the quantum ergodic theorem~\cite{vonNeumann1929,vonNeumann2010,Goldstein2010,Goldstein2010b}. In Pechukas's 1984 paper ``Remarks on Quantum Chaos''~\cite{Pechukas1984}, one finds a brief summary of von Neumann's work and the subsequent developments against it, particularly by Bocchieri and Loinger~\cite{Bocchieri1958}, which, Pechukas claims, actually make von Neumann's results sharper~\cite{Pechukas1984b}. In Pechukas's words,  
\begin{quote}
if one selects a state
“at random” from an energy shell and determines, as a function
of time, the probability that it lies in a “typical” subspace of the
shell, the time average of this probability is liable to be much closer
to the statistical expectation than is its instantaneous value, the
more so the less degenerate the spectrum of the Hamiltonian.
\end{quote}
The ideas in this quotation are connected with the notion of ``typicality''~\cite{Goldstein2010,Goldstein2010b}, that has been one of the directions in studies of thermalization. 

In retrospect, one can identify in the 1985 work by Jensen and Shakar~\cite{Jensen1985} the seeds for what later became known as the eigenstate thermalization hypothesis (ETH)~\cite{Dalessio2016,Deutsch2018}. There, the authors study the infinite-time average and the eigenstate expectation value of the magnetization of a finite spin-1/2 chain for different initial states and different energy eigenstates, and show that both exhibit very good agreement with the microcanonical average when the system is chaotic. They state that in the chaotic system, 
``the magnetization is a fairly smooth and monotonic function of energy'', and in this case,
\begin{quote}
even an initial
energy eigenstate will exhibit a constant value for the
observable which is very close to that predicted by the
statistical theory,
\end{quote}
while for the integrable Hamiltonian, the ``deviations
from equilibrium tend to be large''. This means that an observable $O$ should thermalize when its expectation value $\langle E_k |\hat{O} |E_k \rangle$ is a smooth function of energy, since in this situation, its infinite-time average, $\overline{O}$, will be very close to the microcanonical average, $O_{\text{mic}}$, that is~\cite{Rigol2008}
\begin{equation}
\overline{O} = \sum_{k}|c_{k}|^{2}O_{k,k} \simeq O_{\text{mic}}  = \frac{1}{W_{E,\Delta E}}\sum_{k}O_{k,k}  \ \ ,  
\label{Eq:ETH}
\end{equation}
where $|E_k \rangle$ are the eigenstates of the system's Hamiltonian, $O_{k,k}=\langle E_k|\hat{O}|E_k\rangle$ are the diagonal elements of the operator $\hat{O}$,  $c_k = \langle E_k|\Psi(0)\rangle $ are the coefficients of the initial state $|\Psi(0)\rangle = \sum_k c_k |E_k \rangle$, and $W_{E,\Delta E}$ is the number of eigenstates $|E_{k}\rangle$ contained in the energy window $E_{k}\in[E-\Delta E,E+\Delta E]$ with $|E-E_{k}|<\Delta E$. Under the assumption of the smooth behavior in energy, even a single eigenstate inside the microcanonical window should give  $O_{k,k}$ very close to the microcanonical average, therefore the term ``eigenstate thermalization hypothesis'' (ETH) coined by Srednicki in his 1994 paper~\cite{Srednicki1994}. Notions of quantum chaos~\cite{Jensen1985}, random matrices~\cite{Deutsch1991}, and the Berry's conjecture~\cite{Srednicki1994} have been invoked to justify the validity of Eq.~\eqref{Eq:ETH}. Starting with Rigol {\em et al}'s 2008 paper~\cite{Rigol2008}, several numerical studies have confirmed Eq.~\eqref{Eq:ETH} for chaotic systems. Since then, various studies further elaborated the framework of the ETH to take into account the convergence of $\overline{O}$ and $O_{\text{mic}}$ as the system size increases~\cite{Beugeling2014,Dalessio2016}, the analysis of the off-diagonal elements $O_{k,k'} = \langle E_k|\hat{O}|E_{k'}\rangle$,  known as off-diagonal ETH~\cite{Dalessio2016,LeBlond2019}, the dependence on the energy of the initial state~\cite{Torres2013,He2013}, and the structure of the eigenstates in realistic many-body quantum systems~\cite{Santos2010PREa,Santos2010PREb,RigolSantos2010}.

The analysis in Ref.~\cite{Santos2010PREa} was inspired by the 1990's papers from previous members of the Novosibirsk school, including Casati~\cite{Benenti2001}, Flambaum~\cite{Flambaum1994,Flambaum1996b}, Izrailev~\cite{Flambaum1997,Borgonovi1998}, Shepelyansky~\cite{Jacquod1997}, and Zelevinsky~\cite{Horoi1995,Zelevinsky1996b}, who focused on the structure of the eigenstates to define quantum chaos and explain the onset of thermalization~\cite{Borgonovi2016}. In realistic interacting many-body quantum systems, the energy eigenstates can be rather complicated, but they are not random vectors, as the eigenstates of Gaussian random matrices or the random superpositions of plane waves with random phases and Gaussian random amplitude stated by the Berry’s conjecture~\cite{Berry1977}. For such random vectors,  Eq.~\eqref{Eq:ETH} is trivially satisfied, but they are associated with unphysical models. In the above mentioned 1990's papers, chaotic eigenstates are those that, when written in the mean field basis, exhibit coefficients that are random variables following a Gaussian distribution around the envelope defined by the energy shell. They emerge away from the edges of the spectrum of quantum systems with many strongly interacting particles and ensure the validity of the Fermi-Dirac~\cite{Flambaum1997} and Bose-Einstein~\cite{Borgonovi2017} distributions. These ideas were employed in Ref.~\cite{Santos2010PREa} to corroborate the validity of the ETH for few-body observables when the eigenstates are chaotic.

The present article is dedicated to Professor Giulio Casati on the occasion of his 80th birthday in 2022 and to Professor Felix Izrailev's 80th birthday in 2021. Not only their past, but also their recent works~\cite{Wang2020b,Balachandran2021,Wang2022a,Wang2022b,Santos2012PRL,Borgonovi2017,Borgonovi2019a,Borgonovi2019b} continue to have an enormous impact in the developments of quantum chaos and its connections with thermalization, quantum statistical mechanics, and the quantum-classical correspondence.

Here, we investigate the validity of the diagonal and off-diagonal ETH, in connection with the structure of the eigenstates, for the spin-boson Dicke model~\cite{Dicke1954,Garraway2011,Kirton2019}. This system has a well-defined classical limit and two-degrees of freedom, so it does not quite fall within the requirement for thermalization of a large number of coupled degrees of freedom~\cite{Deutsch1991}. 

Depending on the parameters and energy region, the Dicke model may be regular or chaotic. Since its introduction in the 1950's to explain superradiance in spontaneous radiation processes~\cite{Dicke1954}, it has been employed in a variety of theoretical studies, that include quantum phase transitions~\cite{Emary2003,Emary2003PRL,Brandes2013},classical and quantum chaos~\cite{Furuya1998,Lobez2016,Chavez2016,Sinha2020,Valencia2022}, non-equilibrium quantum dynamics~\cite{Altland2012NJP,Kloc2018,Lerma2018,Lerma2019,Kirton2019,Villasenor2020}, the evolution of out-of-time-ordered correlators (OTOCs)~\cite{Chavez2019,Lewis-Swan2019,Pilatowsky2020}, quantum scarring~\cite{Deaguiar1992,Furuya1992,Bakemeier2013,Pilatowsky2021NatCommun,Pilatowsky2021}, and quantum localization measures in phase space~\cite{Wang2020,Pilatowsky2021NatCommun}. Experimentally, the model  can be realized with superconducting circuits~\cite{Jaako2016}, cavity assisted Raman transitions~\cite{Baden2014,Zhang2018}, trapped ions~\cite{Cohn2018,Safavi2018}, and other systems~\cite{Chelpanova2021}. In the particular context of the relationship between chaos and thermalization, the Dicke model was studied in Ref.~\cite{Kirkova2022} with emphasis on the behavior of the fidelity OTOC and the agreement between the long-time average of the collective spin observable and the microcanonical ensemble. The chaos-thermalization connection was also explored for the kicked Dicke model in Ref.~\cite{Ray2016}. 

We analyze the diagonal and off-diagonal ETH for the number of photons and the number of excited atoms in the regular and chaotic regions of the Dicke model, and compare the results with the structure of the eigenstates analyzed with the entanglement entropy and the Shannon entropy. The latter is a delocalization measure that depends on the basis in which the eigenstates are written. The Hilbert space of the Dicke model is infinite, because its number of bosons is unbounded. Our results show that the values of the Shannon entropy grows rapidly with the eigenvalues when the eigenstates are written in the Fock basis, but are more restricted when the ``efficient basis'' is used, which makes evident the advantages of the latter when one wants to study large systems and high energies.

The paper is organized as follows. We introduce the Dicke model in Sec.~\ref{sec:Dicke_Model} and  briefly review the onset of classical and quantum chaos in Sec.~\ref{sec:Chaos_Dicke_Model}. In Sec.~\ref{sec:ETH}, we study the diagonal and off-diagonal ETH and  analyze the eigenstates in Sec.~\ref{sec:Entropies}. Our conclusions are summarized in Sec.~\ref{sec:Conclusions}.

\section{Dicke Model}
\label{sec:Dicke_Model}

The Dicke model~\cite{Dicke1954} describes a system of $\mathcal{N}$ two-level atoms  coupled with a single mode of a quantized radiation field. Setting $\hbar=1$, the Hamiltonian is given by
\begin{align}
    \label{eqn:dicke_hamiltonian}
    \hat{H}_{\text{D}} & = \hat{H}_{\text{F}} + \hat{H}_{\text{A}} + \hat{H}_{\text{I}}, \\ 
    \label{eqn:dicke_hamiltonian_hf}
    \hat{H}_{\text{F}} & = \omega\hat{a}^{\dagger}\hat{a}, \\ 
    \label{eqn:dicke_hamiltonian_ha}
    \hat{H}_{\text{A}} & = \omega_{0}\hat{J}_{z}, \\ 
    \label{eqn:dicke_hamiltonian_hi}
    \hat{H}_{\text{I}} & = \frac{\gamma}{\sqrt{\mathcal{N}}}(\hat{a}^{\dagger}+\hat{a})(\hat{J}_{+}+\hat{J}_{-}),
\end{align}
where $\hat{H}_{\text{F}}$ defines the field's energy, $\hat{H}_{\text{A}}$ the energy of the two-level atoms, and $\hat{H}_{\text{I}}$ the atom-field interaction energy. In the equations above, $\hat{a}^{\dagger}$ ($\hat{a}$) is the bosonic creation (annihilation) operator of the single field mode and $\hat{J}_{+}$ ($\hat{J}_{-}$) is the raising (lowering) collective pseudo-spin operator, where $\hat{J}_{\pm}=\hat{J}_{x}\pm i\hat{J}_{y}$ and  $\hat{J}_{x,y,z}=(1/2)\sum_{k=1}^{\mathcal{N}}\hat{\sigma}_{x,y,z}^{k}$ for the Pauli matrices $\hat{\sigma}^k_{x,y,z}$ acting on the $k$'th two-level atom. Since the squared total pseudo-spin operator $\hat{\textbf{J}}^{2}=\hat{J}_{x}^{2}+\hat{J}_{y}^{2}+\hat{J}_{z}^{2}$ commutes with the Hamiltonian, $[\hat{H}_{D},\hat{\textbf{J}}^{2}]=0$, the eigenvalues of $\hat{\textbf{J}}^2$, given by $j(j+1)$, label the invariant subspaces in the Hilbert space. We work  within the totally symmetric subspace, which includes the collective ground state and is defined by the maximum pseudo-spin value $j=\mathcal{N}/2$. 

The Dicke Hamiltonian $\hat{H}_{\text{D}}$ commutes with the parity operator 
\begin{equation}
\hat{\Pi}=\text{exp}(i\pi\hat{\Lambda}) ,
\end{equation}
where 
\begin{equation}
\hat{\Lambda}=\hat{a}^{\dagger}\hat{a}+\hat{J}_{z}+j\hat{1}=\hat{n}+\hat{n}_{\text{ex}} ,
\label{eqn:lambda_operator}
\end{equation}
the number of photons is $\hat{n} = \hat{a}^{\dagger}\hat{a}$, and the number of excited atoms is $\hat{n}_{\text{ex}}=\hat{J}_{z}+j\hat{1}$. Because of this symmetry, the eigenstates have one of two different parities, $\hat{\Pi}|E_{k}\rangle=\pm|E_{k}\rangle$.

The three parameters of the Dicke Hamiltonian,  $\omega$, $\omega_{0}$, and $\gamma$, determine the energy scales of the system and the onset of chaos. The radiation frequency of the single-mode electromagnetic field (the boson) is given by $\omega$, the energy splitting of each two-level atom is $\omega_{0}$, and $\gamma$ is the coupling strength modulating the atom-field interaction. In the thermodynamic limit, $\mathcal{N}\rightarrow\infty$, the light-matter coupling divides the parameter space in the normal phase, when the strength $\gamma$ in smaller than the critical value $\gamma_{\text{c}}=\sqrt{\omega\omega_{0}}/2$, and the superradiant phase, when $\gamma > \gamma_{\text{c}}$~\cite{Hepp1973a,Hepp1973b,Wang1973,Emary2003}. The Dicke model also presents regular and chaotic regions depending on the parameters and excitation energies~\cite{Chavez2016}. We fix the coupling strength in the superradiant phase, $\gamma=2\gamma_{\text{c}}=1$, and the resonant frequencies at $\omega=\omega_{0}=1$, so that we ensure that both the classical dynamics and the quantum eigenspectrum are chaotic at excitation energies $\epsilon \geq -0.8$~\cite{Chavez2016}, where $\epsilon=E/j$ denotes the energy scaled to the system size $j$. 

The Hilbert space of the Dicke model is infinite. In the  App.~\ref{app:Diagonalization_Bases}, we provide details on how to diagonalize the system's Hamiltonian in an efficient basis.  We consider three system sizes, $j=30,60,100$, whose associated Hilbert space dimensions  in the efficient basis are given by $d_{\text{D}}^{\text{EB}}=8\,601,30\,371,80\,601$, ensuring the convergence of the  set of eigenstates from the ground-state energy $\epsilon_{\text{GS}}=-2.125$ until a truncated value $\epsilon_{\text{T}}$.

\section{Chaos in the Dicke Model}
\label{sec:Chaos_Dicke_Model}

\begin{figure}[H]
    \centering
    \includegraphics[width=\textwidth]{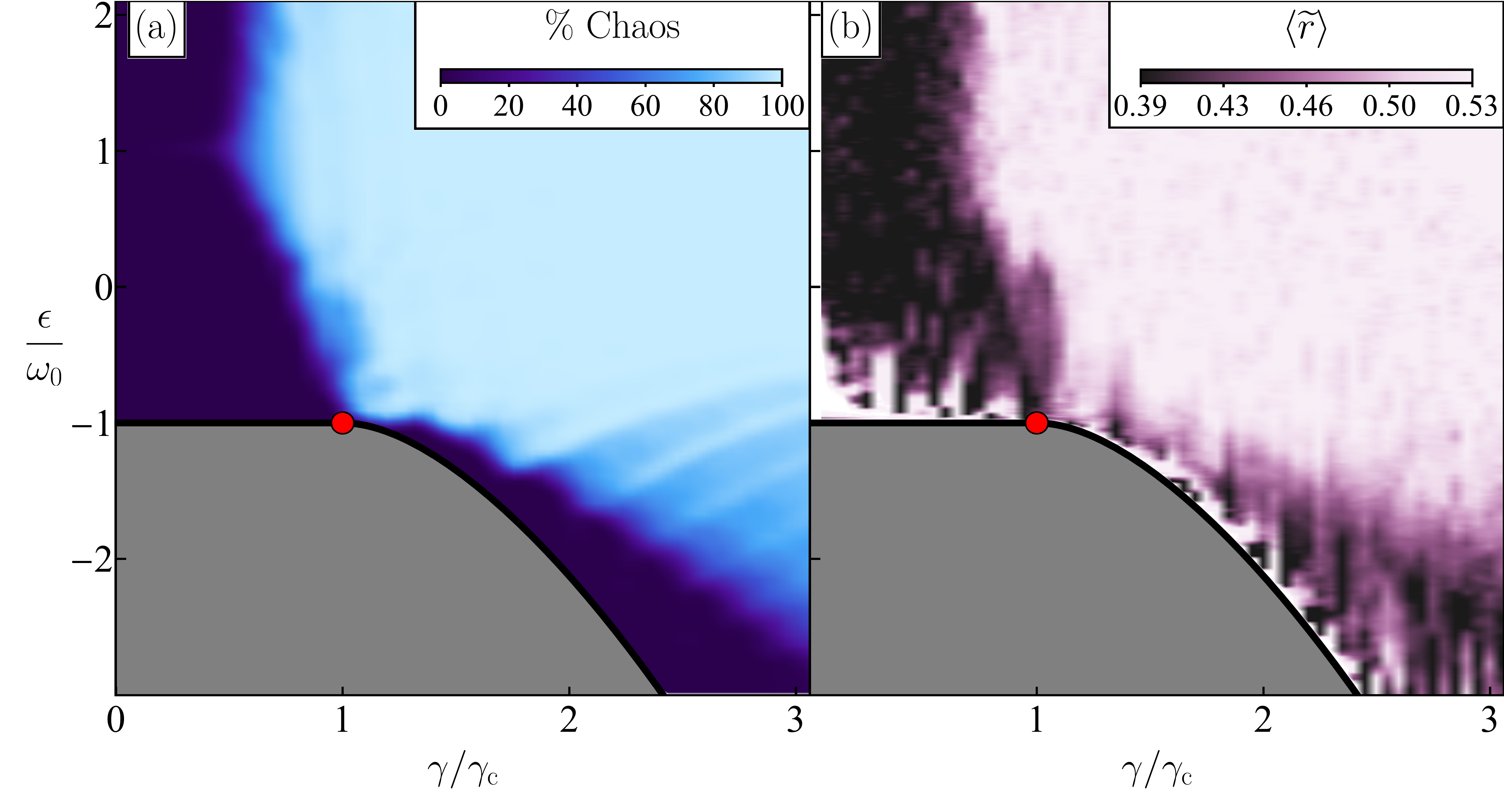}
    \caption{\textbf{Panel (a):} Map of the percentage of chaos for the classical trajectories as a function of the excitation energy $\epsilon$ and the coupling strength $\gamma$. \textbf{Panel (b):} Map of the average values of the ratio of consecutive energy levels $\langle\widetilde{r}\rangle$ [see Eq.~\eqref{eqn:spectral_factor}] obtained for moving windows of eigenenergies, which contain approximately 1000 energy levels. The low energy regions contain few energy levels and were averaged using windows of eigenenergies with the available energy levels. Because of this, the fluctuations of $\langle\widetilde{r}\rangle$ are more pronounced in these regions. In both panels (a) and (b), the red dot represents the critical point for the transition from the normal to the superradiant phase for the ground state. The system size in panel (b) is $j=100$. 
    \label{fig1}}
\end{figure}

The classical Hamiltonian of the Dicke model is presented in App.~\ref{app:Classical_Dicke_Model}. The generation of 
Poincar\'e sections and the computation of Lyapunov exponents for trajectories in phase space~\cite{Bastarrachea2014b,Chavez2016} reveal the onset of chaos for high energies  and strong couplings. In Fig.~\ref{fig1}~(a), we show a classical map of the percentage of chaos as a function of excitation energies and coupling strengths. The percentage of chaos is defined as the ratio of the number of chaotic initial conditions over the total number of initial conditions used in the sample and is illustrated with a color gradient, where dark indicates that the majority of the initial conditions are regular and light indicates that most initial conditions are chaotic.

Signatures of the onset of chaos in  classical systems are found in the quantum domain, as suggested by Casati {\em et al}'s  1980 paper~\cite{Casati1980} and  Bohigas {\em et al}'s 1984 work~\cite{Bohigas1984}, and became known as ``quantum chaos''. The eigenvalues of a quantum system become correlated when its classical counterpart is chaotic. The degree of correlations between neighboring levels can be detected with the level spacing distribution~\cite{MehtaBook} or the ratio of consecutive energy levels~\cite{Oganesyan2007,Atas2013},
\begin{align}
    \label{eqn:spectral_factor}
    \widetilde{r}_{k} & = \min(r_{k},r_{k}^{-1})
    = \frac{\min(s_{k},s_{k-1})}{\max(s_{k},s_{k-1})},
\end{align}
where $s_{k}=E_{k+1}-E_{k}$ is the nearest-neighbor spacing between energy levels and $r_{k}=s_{k}/s_{k-1}$; while both short- and long-range correlations are measured with quantities such as the level number variance~\cite{Guhr1998} or the spectral form factor~\cite{MehtaBook}. In the case of the Dicke model, level spacing distributions~\cite{Bastarrachea2014b,Bastarrachea2015,Wang2018} and spectral form factors~\cite{Lerma2019,Villasenor2020} in agreement with random matrix theory (RMT) were verified for the energies and parameters associated with the onset of chaos in the classical limit.

Other tests of quantum chaos include Peres lattices~\cite{Peres1984PRL} and the evolution of OTOCs~\cite{Larkin1969,Maldacena2016PRD,Maldacena2016JHEP}. The Peres lattice is a plot of the eigenstate expectation values of an observable as a function of the eigenvalues. It has been widely employed (not always under this name) in studies of ETH and  provides visual evidence of the loss of integrability~\cite{Bastarrachea2014b,Bastarrachea2014c}. As for the OTOCs, their exponential growth rates are seen as quantum analogs of the classical Lyapunov exponents. This was indeed confirmed for the Dicke model in the chaotic region~\cite{Chavez2019}, although its exponential growth  happens also in regular regions due to instability~\cite{Pilatowsky2020}.

In Fig.~\ref{fig1}, we compare the onset of classical chaos, investigated in Fig.~\ref{fig1}~(a), with the degree of correlations between neighboring energy levels, as captured by the ratio of consecutive energies in Fig.~\ref{fig1}~(b). To smooth spectral fluctuations, an average is performed and denoted by $\langle\widetilde{r}\rangle$. In the regular regime, where the eigenvalues are uncorrelated and the level spacing distribution is Poissonian, $\langle\widetilde{r}\rangle_{\text{P}}\approx0.39$, while in the chaotic region, where the eigenvalues are correlated as in RMT and the level spacing distribution follows the Wigner-Dyson distribution,  $\langle\widetilde{r}\rangle_{\text{WD}}\approx0.53$. Our results in Fig.~\ref{fig1} exhibit an evident classical-quantum correspondence. There is a visible relationship between the classical map of Fig.~\ref{fig1}~(a) and the quantum map of Fig.~\ref{fig1}~(b), when the percentage of chaos is large (small) in Fig.~\ref{fig1}~(a), $\langle\widetilde{r}\rangle$ approaches 0.53 (0.39) in Fig.~\ref{fig1}~(b).

\section{Thermalization in the Dicke Model}
\label{sec:ETH}

For an isolated quantum system initially in a pure state, the evolution is governed by the unitary time operator as
\begin{equation}
    |\Psi(t)\rangle  = e^{-i\hat{H} t}|\Psi(0)\rangle = \sum_{k}c_{k}e^{-iE_{k}t}|E_{k}\rangle,   
\end{equation}
where $\hat{H}|E_k  \rangle = E_k |E_k  \rangle$ and  $c_k = \langle E_k|\Psi(0)\rangle $.  Thus, the expectation value of a given operator $\hat{O}$ under states evolved in time can be calculated as follows
\begin{equation}
    O(t)  = \langle\Psi(t)|\hat{O}|\Psi(t)\rangle  = \sum_{k \neq k'}c_{k}^{\ast}c_{k'}e^{i(E_{k}-E_{k'})t}O_{k,k'} + \sum_{k} |c_{k}|^2 O_{k,k}\ \ \ ,
    \label{Eq:Otime}
\end{equation}
where $O_{k,k'}=\langle E_{k}|\hat{O}|E_{k'}\rangle$ are the matrix elements of the operator expressed in the energy eigenbasis.

The onset of thermalization according to the ETH happens for local observables if  two assumptions are satisfied: 

(i) The infinite-time average, 
\begin{equation}
    \label{eqn:infinite_time_average}
    \overline{O} = \lim_{t\to+\infty}\frac{1}{t}\int_{0}^{t}dt'O(t')  = \sum_{k}|c_{k}|^{2}O_{k,k},
\end{equation}
coincides with the microcanonical average,
\begin{equation}
    O_{\text{mic}}  = \frac{1}{W_{E,\Delta E}}\sum_{k}O_{k,k} \approx  \overline{O},
    \label{eq:W}
\end{equation}
or more precisely, $\overline{O}$ approaches the thermodynamic average as the system size grows. This is sometimes referred to as ``diagonal ETH''. In the equation above, $W_{E,\Delta E}$ is the number of eigenstates $|E_{k}\rangle$ contained in the energy window $E_{k}\in[E-\Delta E,E+\Delta E]$ with $|E-E_{k}|<\Delta E$. 

As explained above, the ETH should be valid when the eigenstates are chaotic (ergodic), filling the energy shell. In this case, one gets Gaussian distributions for few-body observables in many-body quantum systems, which can be understood as follows. Assume that the few-body observable $\hat{O}$ has $N$ nonzero eigenvalues $O_n$ in $\hat{O}|O_n\rangle = O_n|O_n \rangle$, where $N$ is large. We can project the energy eigenstates in the basis $|O_n \rangle$ as $|E_{k}\rangle=\sum_{n}C_{O_{n}}^{k}|O_{n}\rangle$, and write
\begin{equation}
O_{k k}=\langle E_k | \hat{O} |E_{k} \rangle = \sum_{n=1}^N  |C_{O_n}^{k}|^2 \langle O_n| \hat{O} |O_n\rangle  
= 
O_{1,1} |C_{O_1}^{k}|^2
+ \ldots + 
O_{N,N} |C_{O_N}^{k}|^2.
\label{Eq:diagDist}
\end{equation}
If the eigenstates are fully chaotic (ergodic), then $C_{O_n}^{k}$'s are independent Gaussian random numbers. According to the central limit theorem, a large sum of independent random numbers $O_{n,n}|C_{O_n}^{k}|^2$ will also follow a Gaussian distribution. Therefore, in the region where the eigenstates are chaotic, the distribution of $O_{k,k}$ should be Gaussian.

(ii) The fluctuations around $\overline{O}$, which are determined by the phases $e^{i(E_{k}-E_{k'})t}$, the coefficients $c_{k}$, and the off-diagonal elements $O_{k,k'}$ in Eq.~\eqref{Eq:Otime}, decrease with system size and cancel out on average. This is sometimes referred to as ``off-diagonal ETH''. A Gaussian distribution of $O_{k,k'}$ indicates that the eigenstates are strongly chaotic (ergodic), so condition (ii) should be satisfied if the energy of the initial state (of a system perturbed far from equilibrium) is in the chaotic region. 

The relationship between the Gaussian distribution of the off-diagonal elements of a few-body observable and ergodicity can be understood as above, using $\hat{O}|O_n\rangle = O_n|O_n \rangle$ and  $|E_{k}\rangle=\sum_{n}C_{O_{n}}^{k}|O_{n}\rangle$ in
\begin{equation}
O_{k k'}=\langle E_k | \hat{O} |E_{k'} \rangle = \sum_{n=1}^N  (C_{O_n}^{k})^{\ast} C_{O_n}^{k'}  \langle O_n| \hat{O} |O_n\rangle  
= 
O_{1,1} (C_{O_1}^{k})^{\ast} C_{O_1}^{k'}
+ \ldots + 
O_{N,N} (C_{O_N}^{k})^{\ast} C_{O_N}^{k'}.
\label{Eq:off}
\end{equation}
For chaotic eigenstates,  $C_{O_n}^{k}$'s are independent Gaussian random numbers and the product of two independent Gaussian random numbers, $(C_{O_n}^{k})^{\ast} C_{O_n}^{k'}$, which appears in each term of  Eq.~\eqref{Eq:off}, is also an independent random number.  According to the central limit theorem, a large sum of independent random numbers follows a Gaussian distribution, so in the region where the eigenstates are chaotic, the distribution of $O_{k,k'}$ should be Gaussian. This has been confirmed for different chaotic systems with many-degrees of freedom~\cite{Beugeling2015,LeBlond2019,Santos2020}, but not in chaotic systems with one~\cite{Lydzba2021} or few particles~\cite{Zisling2021}, few degrees of freedom~\cite{Wittmann2022} or in many-body systems when $\hat{O}$ is not few-body~\cite{Khaymovich2019}.

\begin{figure}[H]
    \centering
    \includegraphics[width=\textwidth]{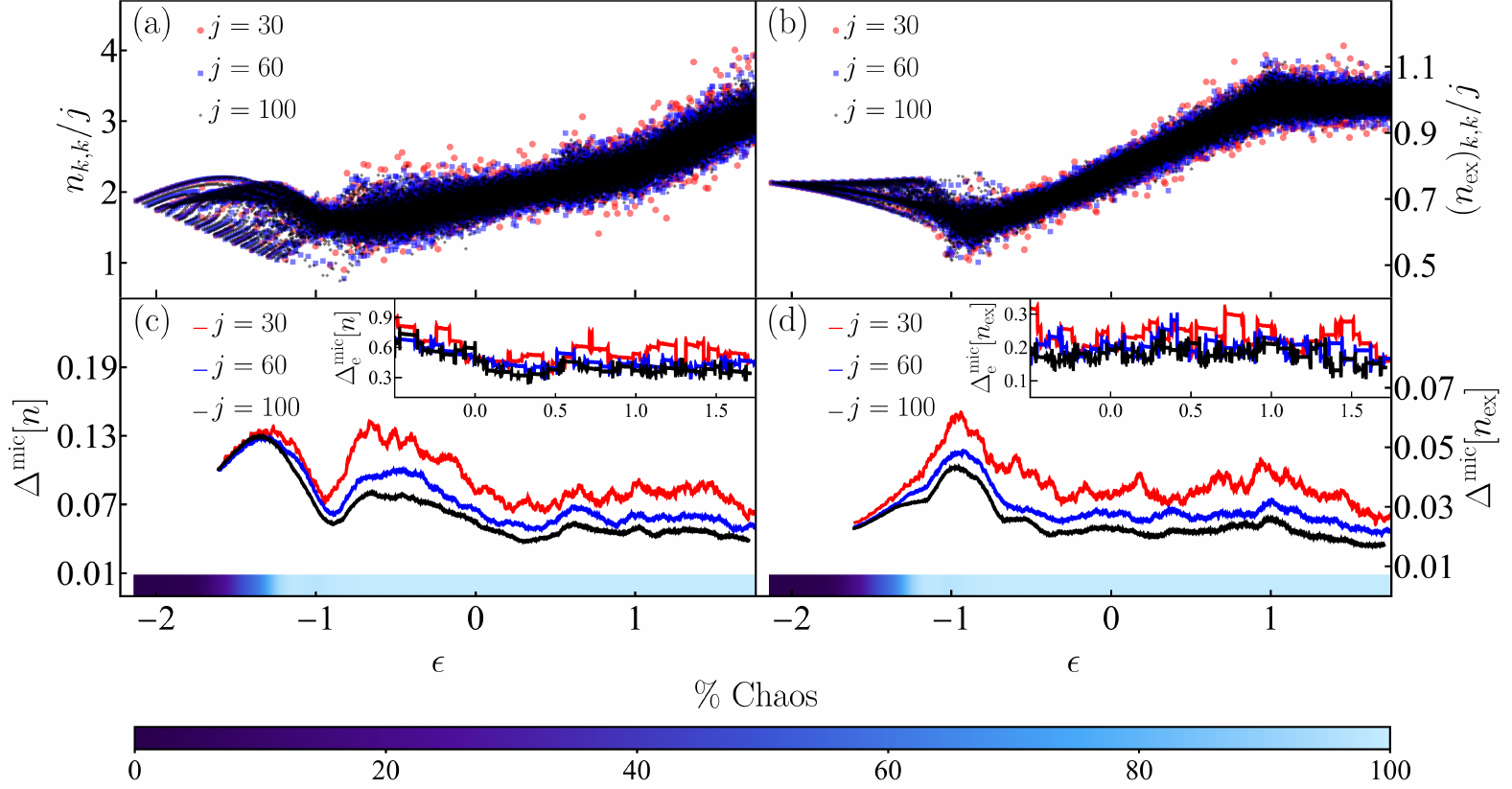}
    \caption{\textbf{Panels (a) and (b):} Peres lattice of the eigenstate expectation values of the number of photons $n_{k,k}$ (a) and the number of excited atoms $(n_{\text{ex}})_{k,k}$ (b)  scaled to the system size $j$ for eigenstates $|E_{k}\rangle$ with positive parity of the Dicke model. \textbf{Panels (c) and (d):} Deviations of the expectation values of the same observables $n_{k,k}$ (c) and $(n_{\text{ex}})_{k,k}$ (d) with respect to their microcanonical value [see Eq.~\eqref{eqn:delta_MIC}]. The insets in (c) and (d) show the extremal deviations of the respective quantities in the chaotic energy regime [see Eq.~\eqref{eqn:delta_MICe}]. The color scales contained within the panels (c) and (d) correspond to the values of the percentage of classical chaos shown at the bottom in the numbered color scale. The system size in all panels (a)-(d) is indicated with a given color for three values $j=30,60,100$.
    \label{fig2}}
\end{figure}

\subsection{Diagonal ETH}

To test the validity of the diagonal ETH, we compute the deviation of the eigenstate expectation values with respect to the microcanonical value~\cite{RigolSantos2010,Santos2010PREb}
\begin{equation}
    \label{eqn:delta_MIC}
    \Delta^{\text{mic}}[O] = \frac{\sum_{k}|O_{k,k}-O_{\text{mic}}|}{\sum_{k}O_{k,k}},
\end{equation}
and use also a stronger test that takes into account the normalized extremal fluctuations and is given by~\cite{Santos2010PREb}
\begin{equation}
    \label{eqn:delta_MICe}
    \Delta^{\text{mic}}_{e}[O] = \left|\frac{\max(O) - \min(O)}{O_{\text{mic}}}\right|,
\end{equation}
where $\max(O)$ and $\min(O)$ are taken from the same energy window $E_{k}\in[E-\Delta E,E+\Delta E]$ used in Eq.~\eqref{eq:W}. We consider as observables, the number of excited atoms, $\hat{n}_{\text{ex}}=\hat{J}_{z}+j\hat{1}$, and the number of photons, $\hat{n}=\hat{a}^{\dagger}\hat{a}$, of the Dicke model.

In Fig.~\ref{fig2}, we show the Peres lattices for the expectation values of the number of photons, $n_{k,k}=\langle E_k|\hat{n}|E_k \rangle$ [Fig.~\ref{fig2}~(a)], and of the number of excited atoms, $(n_{\text{ex}})_{k,k}=\langle E_k|\hat{n}_{\text{ex}}|E_k \rangle$ [Fig.~\ref{fig2}~(b)], for all eigenstates of the Dicke model with positive parity that range from the ground state energy, $\epsilon_{\text{GS}}=-2.125$, up to a maximal converged eigenstate with eigenenergy $\epsilon_{\text{T}}=1.755$. In the regular region of low energies, the Peres lattices present a clear pattern related with the quasi-conserved quantities. Above $\epsilon\approx-0.8$, where the system becomes chaotic, the lattices become smoother in energy. It is visible that the spread of the expectation values in the chaotic region decreases as the system size increases, that is, for high excitation energies, the fluctuations are larger  for $j=30$ than for $j=100$. It is also noticeable that the number of excited atoms fluctuate less than the number of  photons.

The reduction of the fluctuations with the increase of system size, which is a necessary condition for the validity of the ETH, is better quantified in  Fig.~\ref{fig2}~(c) and Fig.~\ref{fig2}~(d), where we show
$\Delta^{\text{mic}}[n]$ and $\Delta^{\text{mic}}[n_\text{ex}]$, respectively. To compute these quantities, we used moving windows of eigenenergies for each system size $j=30,60,100$, which contain $100,350,900$ energy levels. In the chaotic region, both deviations clearly decrease as $j$ increases, while at low energies ($\epsilon<-0.8$), where chaos and regularity coexist, the fluctuations decrease very slowly or  do not decrease at all.

\begin{figure}[H]
    \centering
    \includegraphics[width=\textwidth]{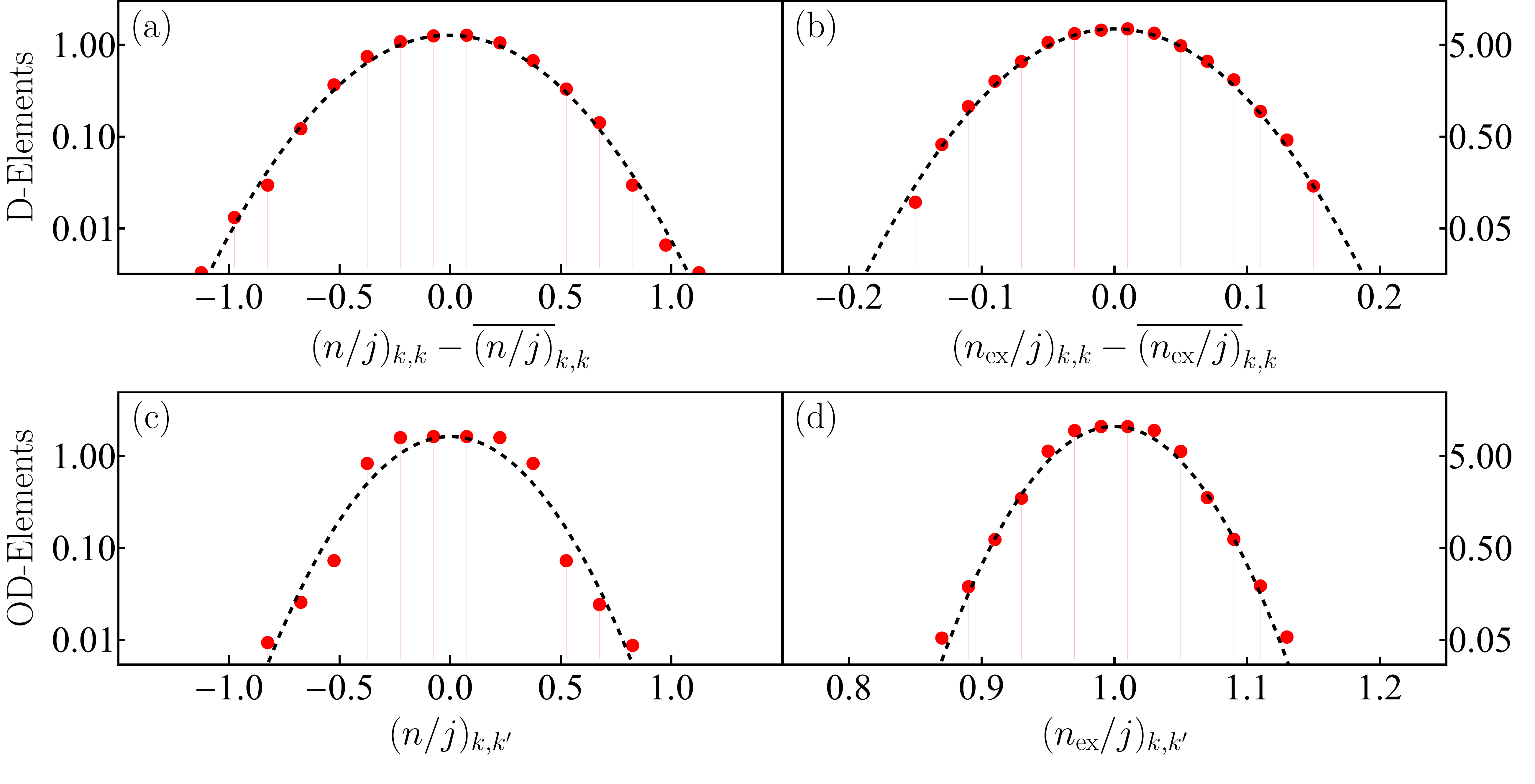}
    \caption{\textbf{Panels (a) and (b):} Statistical distribution (red solid dots) in semi-logarithmic scale of the diagonal matrix elements of number of photons $n_{k,k}$ (a) and excited atoms $(n_{\text{ex}})_{k,k}$ (b) scaled to the system size $j$ for eigenstates $|E_{k}\rangle$ with positive parity of the Dicke model contained in the chaotic energy interval $\epsilon_{k}\in(0.5,1.0)$. The term $\overline{O}_{k,k}$ represents the average of the expectation value $O_{k,k}$ in the same energy region for each operator $\hat{n}$ and $\hat{n}_{\text{ex}}$. The black dashed line depicts a Gaussian fit. \textbf{Panels (c) and (d):} The same as panels (a) and (b) for the off-diagonal matrix elements of the same observables $n_{k,k'}$ (c) and $(n_{\text{ex}})_{k,k'}$ (d). The system size in all panels (a)-(d) is $j=30$.
    \label{fig3}}
\end{figure}

The insets of Fig.~\ref{fig2}~(c) and Fig.~\ref{fig2}~(d) contain the results for the normalized extremal fluctuation in Eq.~\eqref{eqn:delta_MICe}. Only the chaotic energy interval defined by $\epsilon\in[-0.5,1.755]$ is shown. The reduction of the extremal fluctuations as $j$ increases is also perceptible in these insets, thus confirming the validity of the diagonal ETH in the chaotic region of the Dicke model.

In Fig.~\ref{fig3}, we consider the energy interval $\epsilon\in[0.5,1]$, where hard chaos manifests itself classically, and use only the positive parity sector of eigenstates. In Fig.~\ref{fig3}~(a)  and Fig.~\ref{fig3}~(b), we show respectively the distribution of the diagonal elements of the number of photons and the number of excited atom, which are presented in a semi-logarithmic scale. The figures confirm that the distribution shape is Gaussian, as explained in Eq.~(\ref{Eq:diagDist}).

\subsection{Off-Diagonal ETH}

We now analyze the off-diagonal elements of the number of photons in Fig.~\ref{fig3}~(c) and the number of excited atoms in Fig.~\ref{fig3}~(d). As discussed in Eq.~\eqref{Eq:off}, a Gaussian distribution validates the off-diagonal ETH. 
This is indeed the case of Fig.~\ref{fig3}~(c) and  Fig.~\ref{fig3}~(d).

\section{Entropies of the Eigenstates of the Dicke Model}
\label{sec:Entropies}

In the last section, we verified numerically the validity of the ETH in the chaotic region of the Dicke model using the matrix elements of the number of photons and the number of excited atoms. In this section, we use the von Neumann entanglement entropy and the Shannon entropy to analyze the structure of the eigenstates of the model. 
The von Neumann entanglement entropy  can be regarded as the limit of higher-order observables in a replicated Hilbert space, and one may generalize the ETH to include higher-order statistical moments~\cite{Kaneko2020} arising from these replicated spaces, which allows to explain the so-called Page correction~\cite{Page1993} to the volume law. The  von Neumann entanglement entropy has been linked to the onset of chaos in quantum systems~\cite{Miller1999,Lakshminarayan2001,Bandyopadhyay2002,Bandyopadhyay2004,Wang2004}. 
The Shannon entropy is a basis-dependent quantity. We study this entropy with respect to  both the Fock and an  efficient basis.

The Dicke model is a bipartite system, whose Hilbert space is a tensor product of the atomic $\mathcal{H}_{\text{A}}$ and bosonic $\mathcal{H}_{\text{B}}$ sectors, $\mathcal{H}_{\text{D}}=\mathcal{H}_{\text{A}}\otimes\mathcal{H}_{\text{B}}$. The bosonic sector is infinite-dimensional, while the atomic one has dimension $d_{\text{A}}=2j+1$.
For a pure state expanded in the Fock basis $|n;j,m_z\rangle$,
\begin{equation}
    \label{eqn:pure_state}
    |\Psi\rangle = \sum_{n=0}^{\infty}\sum_{m_z=-j}^{j}c_{n,m_z}|n;j,m_z\rangle,
\end{equation}
the density matrix is the projector operator $\hat{\rho}=|\Psi\rangle\langle\Psi|$, and the reduced density matrix in the atomic sector is calculated as,
\begin{align}
    \hat{\rho}_{\text{A}}  = \text{Tr}_{\text{B}}[\hat{\rho}]  = \sum_{m_z=-j}^{j}\sum_{m'_z=-j}^{j}\left( \sum_{n=0}^{\infty}c_{n,m_z}c_{n,m'_z}^{\ast} \right)|j,m_z\rangle\langle j,m'_z |.
\end{align}

The von Neumann entanglement entropy is  given by
\begin{align}
    \label{eqn:entanglement_entropy}
    S_{\text{En}} = & -\text{Tr}[\hat{\rho}_{\text{A}}\ln(\hat{\rho}_{\text{A}})].
  \end{align}
For numerical convenience, we trace out the infinite bosonic sector first, but from the Schmidt decomposition~\cite{Horodecki2009}, the result for $S_{\text{En}}$ is the same if we would instead trace out the atomic sector first.  (See App.~\ref{app:Quantum_Entanglement} for a generalization of quantum entanglement to multipartite systems.)

The standard basis to compute the entanglement entropy is the Fock basis, which is defined as a decoupled basis between photons and atoms $|n;j,m_z\rangle=|n\rangle\otimes|j,m_z\rangle$ (see App.~\ref{app:Fock_Basis}). Nevertheless, alternative bases can be used to calculate this entropy by selecting the correct partition of the system. In this work, in addition to the the Fock basis, we use an efficient basis (see App.~\ref{app:ECB} for a full description), that allows us to reach larger system sizes ($j>30$) than we can reach with the Fock basis~\cite{Bastarrachea2014PSa,Bastarrachea2014PSb}. The efficient basis is defined as the tensor product $|N;j,m_x\rangle=|N\rangle_{m_x}\otimes|j,m_x\rangle$, where $|j,m_x\rangle$ are the atomic pseudo-spin states rotated by an angle $-\pi/2$ around the $y$ axis and $|N\rangle_{m_x}$ are displaced Fock states, whose displacing  depends explicitly on the atomic pseudo-spin eigenvalue $m_{x}$. In the following general notation for a pure state,
\begin{equation}
    \label{eqn:pure_stateAny}
    |\Psi\rangle = \sum_{x=0}^{\infty}\sum_{y=-j}^{j}c_{x,y}|x;j,y\rangle,
\end{equation}
$(x,y)=(n,m_{z})$ stands for the Fock basis and $(x,y)=(N,m_{x})$ for the efficient basis.

To select the correct atomic sector starting with the efficient basis, we need to perform an adequate trace over the modified bosonic sector related with this basis, such that, it is equivalent to the trace over the bosonic sector of the Fock basis. This can be accomplished by mapping one basis into the other (see App.~\ref{app:ECB_FockBasis}). As the Hilbert space associated with the atomic pseudo-spin states $|j,m_{z}\rangle$ is the same as the Hilbert space of the rotated ones $|j,m_{x}\rangle$, the atomic sector for both subspaces is the same and the entanglement entropy can be computed properly. 
The Shannon entropy of the pure state in Eq.~\eqref{eqn:pure_stateAny} is given by
\begin{equation}
    \label{eqn:shannon_entropy}
    S_{\text{Sh}} = -\sum_{x=0}^{\infty}\sum_{y=-j}^{j}|c_{x,y}|^{2}\ln(|c_{x,y}|^{2}).
\end{equation}

\subsection{Results for the Entanglement Entropy}

We start by comparing the von Neumann entanglement entropy for the Dicke model, which is nonintegrable, with the results for one of the integrable limits of the model, namely the Tavis-Cummings model (see App.~\ref{app:Tavis_Cummings_Model} for the derivation of this model). This comparison has also been done in Ref.~\cite{Kloc2017}.

\begin{figure}[H]
    \centering
    \includegraphics[width=\textwidth]{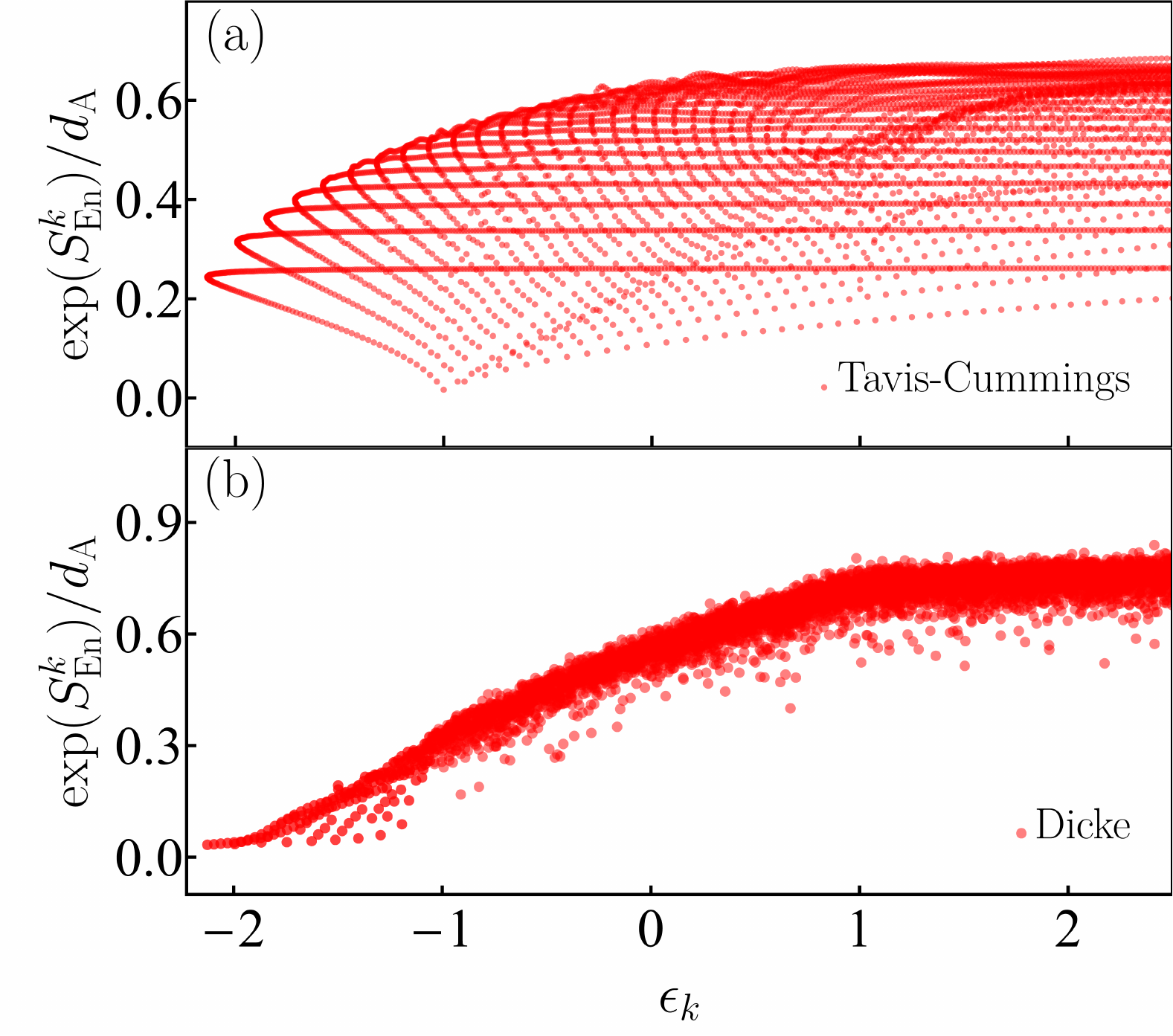}
    \caption{\textbf{Panels (a) and (b):} Peres lattice of the exponential of the von Neumann entanglement entropy $S_{\text{En}}^{k}$ [see Eq.~\eqref{eqn:entanglement_entropy}] scaled to the atomic Hilbert-space dimension $d_{\text{A}}=2j+1$ for eigenstates $|E_{k}\rangle$ from both parity sectors of the integrable Tavis-Cummings model (a) [see Eq.~\eqref{eqn:tavis_cummings_hamiltonian}] and the nonintegrable Dicke model (b) [see Eq.~\eqref{eqn:dicke_hamiltonian}]. The system size in both panels (a) and (b) is $j=30$.
    \label{fig4}}
\end{figure}

In Fig.~\ref{fig4}, we plot the Peres lattice of the exponential of the von Neumann entanglement entropy in Eq.~\eqref{eqn:entanglement_entropy} for the integrable Tavis-Cummings model [Fig.~\ref{fig4}~(a)] and the nonintegrable Dicke model [Fig.~\ref{fig4}~(b)]. Results for all states from both parity sectors, from the ground-state of the Tavis-Cummings model up to a maximal converged eigenstate with eigenenergies $\epsilon_{k}\in[\epsilon_{\text{GS}}=-2.136,\epsilon_{\text{T}}=2.5]$, are shown. As evident in Fig.~\ref{fig4}~(a), the values of the entropy for the integrable case show large fluctuations and patterns associated with regularity are visible. The fluctuations indicate that even states very close in energy may have very different structures, so ETH should not be satisfied. In contrast, $S_{\text{En}}$ becomes a smoother function of energy in the chaotic region of the Dicke model ($\epsilon>-0.8$), as seen in Fig.~\ref{fig4}~(b), which indicates the states close in energy are very similar. In the case of the Dicke model, regular patterns are restricted to the low energies, where the model is not chaotic.

We stress, however, that some isolated eigenstates located in the chaotic regime of the Dicke model present low entanglement. They should be related with strongly scarred states, which are known to exist in this model~\cite{Pilatowsky2021NatCommun,Pilatowsky2021,Pilatowsky2022}.

\begin{figure}[H]
    \centering
    \includegraphics[width=\textwidth]{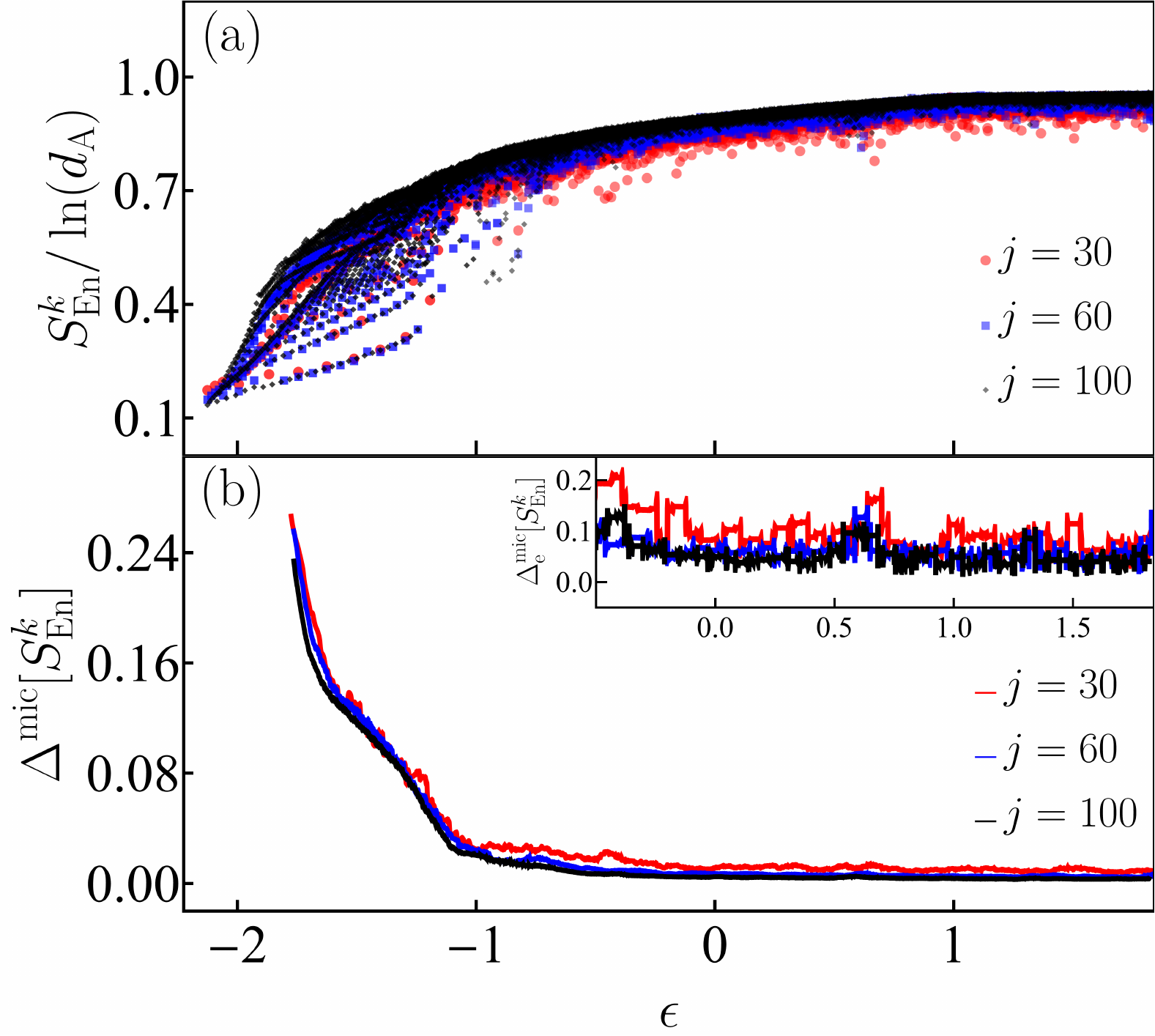}
    \caption{\textbf{Panel (a):} Peres lattice of the von Neumann entanglement entropy $S_{\text{En}}^{k}$ scaled to the atomic Hilbert-space dimension $d_{\text{A}}=2j+1$ for eigenstates $|E_{k}\rangle$ from both parity sectors of the Dicke model. \textbf{Panel (b):} Deviations of the von Neumann entanglement entropy with respect to its microcanonical value, computed with Eq.~\eqref{eqn:delta_MIC}. The inset shows the extremal deviations in the chaotic energy regime computed with Eq.~\eqref{eqn:delta_MICe}. The system size in both panels (a) and (b) is indicated with a given color for three values $j=30,60,100$.
    \label{fig5}}
\end{figure}

In Fig.~\ref{fig5}~(a), we analyze the von Neumann entanglement entropy for the eigenstates of the Dicke model, ranging from the ground state until a maximal converged eigenstate with eigenenergies $\epsilon_{k}\in[\epsilon_{\text{GS}}=-2.125,\epsilon_{\text{T}}=1.841]$, for three values of the system size $j=30$, $60$, $100$. As discussed in Sec.~\ref{sec:ETH} and in Fig.~\ref{fig2}, thermalization requires the convergence of the infinite-time average of a few-body observable towards the microcanonical average in the thermodynamic limit, so scaling analysis needs to be performed. As seen in Fig.~\ref{fig5}~(a), the patterns at the low energies of the regular region of the Dicke model do not disappear as the system size increases, but in the chaotic region, the fluctuations clearly shrink as the system size increases. This behavior is quantified in Fig.~\ref{fig5}~(b), where we present the deviation of the entanglement entropy from the microcanonical average, as computed in Eq.~\eqref{eqn:delta_MIC} for the three  system sizes $j=30,60,100$. The fluctuations decay to values close to zero for energies above $\epsilon\approx-1.2$, indicating the transition to a region where the majority of the eigenstates are chaotic. In the inset of Fig.~\ref{fig5}~(b), we present the extremal fluctuations calculated with Eq.~\eqref{eqn:delta_MICe} for the chaotic energy interval only, $\epsilon\in[-0.5,1.841]$. We avoid the regular region, where the changes in the extremal values are abrupt. The inset confirms the decrease of fluctuations with the increasing of system size.

By comparing Fig.~\ref{fig5}~(b) with Figs.~\ref{fig3}~(c) and (d), we observe that the deviation from the microcanonical average of the von Neumann entanglement entropy decreases to zero more abruptly as compared to the the number of photons and excited atoms. This different behavior between entropies and expectation values of observables is not well understood yet and motivates future work.

\begin{figure}[H]
    \centering
    \includegraphics[width=\textwidth]{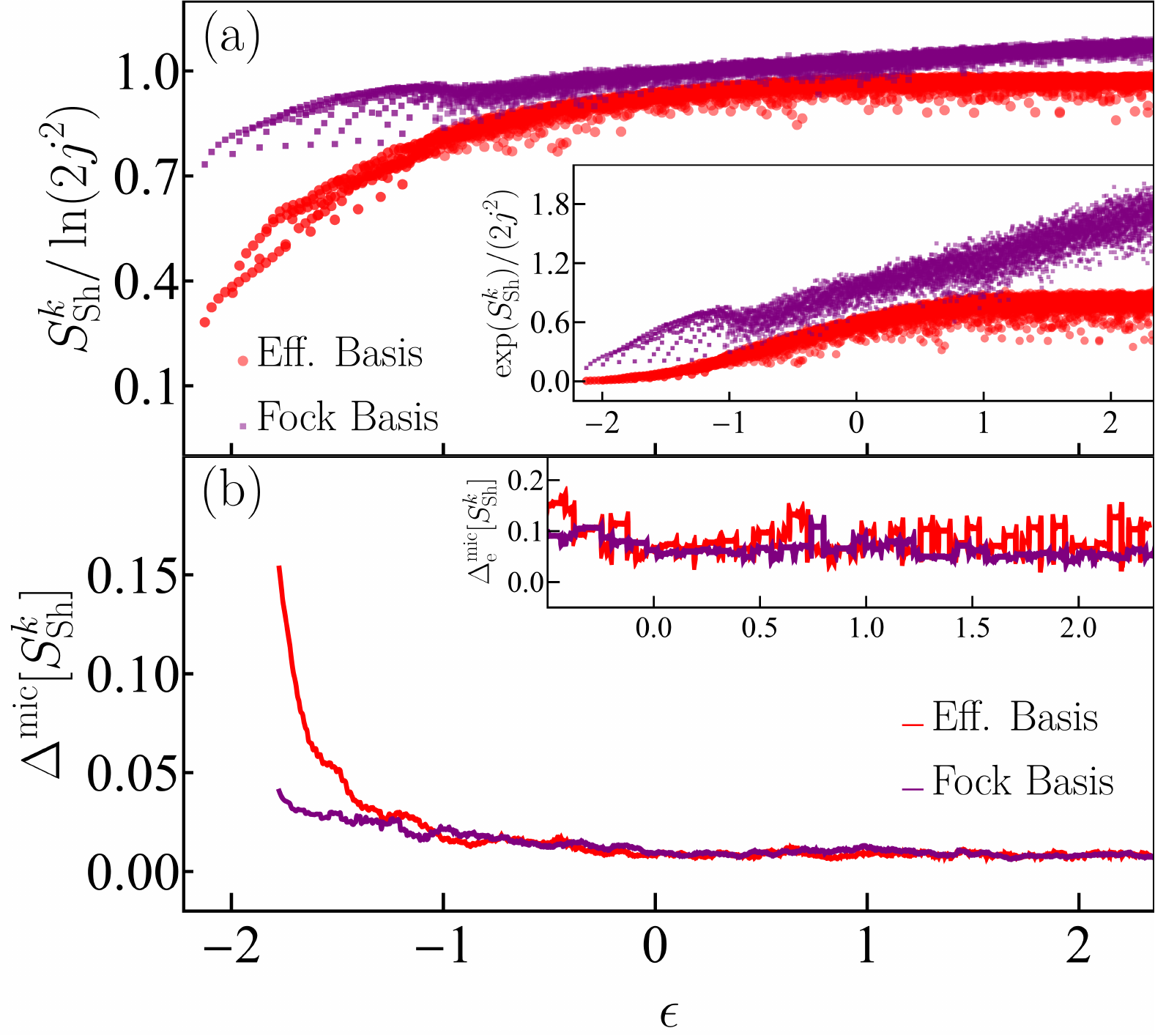}
    \caption{\textbf{Panel (a):} Peres lattices of the Shannon entropy $S_{\text{Sn}}^{k}$ [see Eq.~\eqref{eqn:shannon_entropy}] scaled to the system-size dependence of the density of states, $\nu(\epsilon)\propto 2j^{2}$ (see App.~\ref{app:Classical_Dicke_Model}), for eigenstates $|E_{k}\rangle$ from both parity sectors of the Dicke model written in the efficient basis (red dots) and the Fock basis (purple dots). The inset shows the exponential values of the Shannon entropy for both bases. \textbf{Panel (b):} Deviation of Shannon entropy respect to its microcanonical value [see Eq.~\eqref{eqn:delta_MIC}]. The inset shows the extremal deviation of the same quantity in the chaotic energy regime [see Eq.~\eqref{eqn:delta_MICe}]. The system size in both panels (a) and (b) is $j=30$.
    \label{fig6}}
\end{figure}

\subsection{Results for the Shannon Entropy}

We proceed with the analysis of the structure of the eigenstates of the Dicke model making use now of the basis-dependent Shannon entropy. 
In Fig.~\ref{fig6}~(a), we plot the Peres lattice of the Shannon entropy for the eigenstates of the Dicke model in the Fock and efficient bases. The eigenstates range from the ground state until a maximal converged eigenstate with eigenenergies $\epsilon_{k}\in[\epsilon_{\text{GS}}=-2.125,\epsilon_{\text{T}}=2.356]$. The inset in Fig.~\ref{fig6}~(a) contains the same data, but shows the exponential of the entropy. The values of the Shannon entropy computed in the Fock basis are larger than in the efficient basis. For the Fock basis, the entropy grows unboundedly with energy, while for the efficient basis, $S_{\text{Sh}}$ saturates at high energies. These features make  evident the advantages of using the efficient basis, since it requires fewer  basis states to build a given eigenstate than what is needed by the Fock basis.   While in Fig.~\ref{fig6}, our system size was restricted to $j=30$, when we employ the efficient basis, the same computational resources allow us to go up to $j=100$, as used in Fig.~\ref{fig2}, Fig.~\ref{fig5}, and Fig.~\ref{fig7} below.

\begin{figure}[H]
    \centering
    \includegraphics[width=\textwidth]{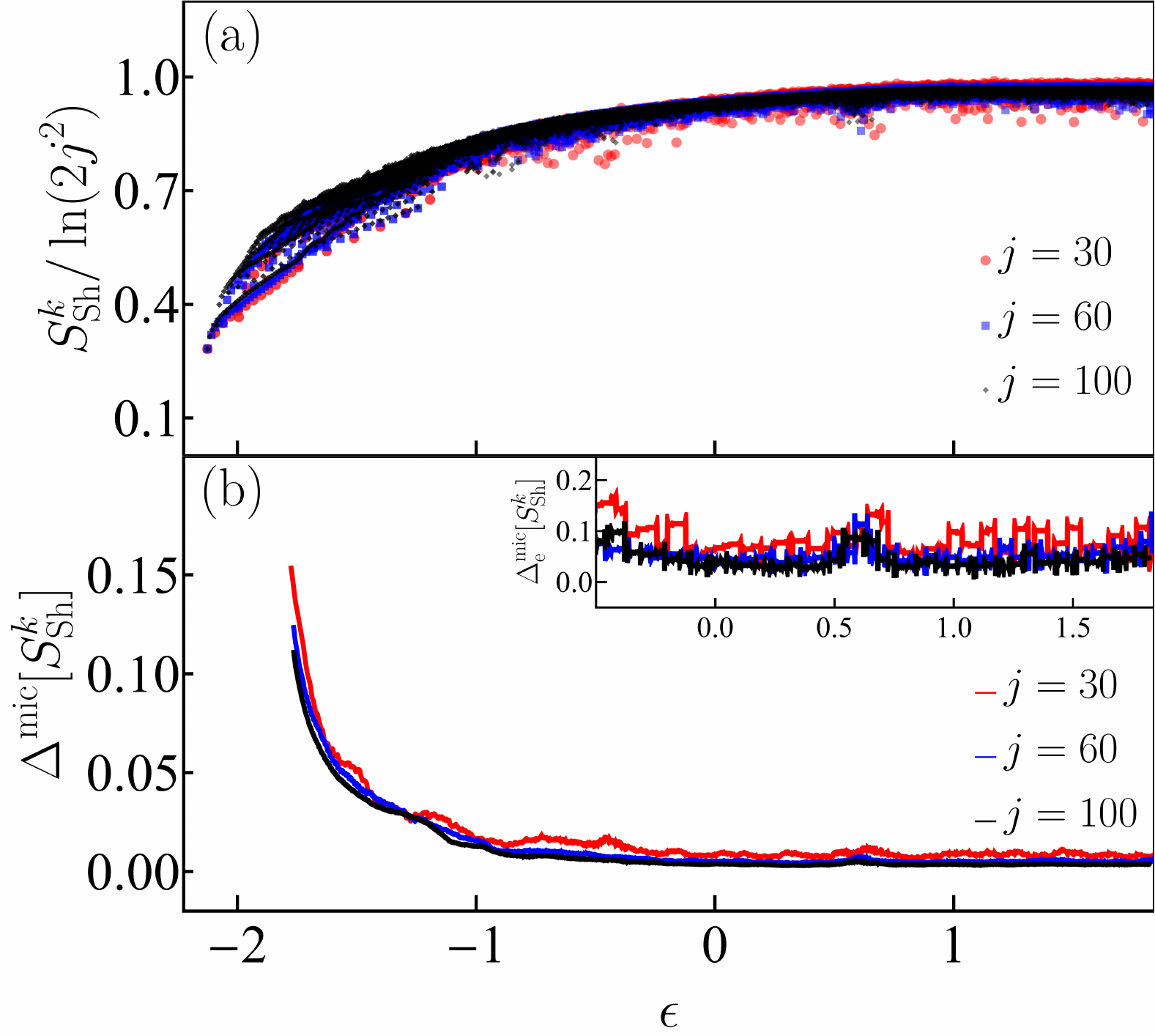}
    \caption{\textbf{Panel (a):} Peres lattice of Shannon entropy $S_{\text{Sh}}^{k}$ scaled to the system-size dependence of the density of states $\nu(\epsilon)\propto 2j^{2}$ for eigenstates $|E_{k}\rangle$ from both parity sectors of the Dicke model. The Shannon entropy for these eigenstates was built in the efficient basis (see App.~\ref{app:ECB}). \textbf{Panel (b):} Deviations of the Shannon entropy with respect to its microcanonical value, computed with Eq.~\eqref{eqn:delta_MIC}. The inset shows the extremal deviations in the chaotic energy regime computed with Eq.~\eqref{eqn:delta_MICe}. The system size in both panels (a) and (b) is indicated with a given color for three values $j=30,60,100$.
    \label{fig7}}
\end{figure}

The analysis of the fluctuations of the Shannon entropy in Fig.~\ref{fig6}~(b) and its inset shows that, similarly to what we observed for the entanglement entropy in Fig.~\ref{fig5}, they approach zero in the chaotic region. The results are comparable for both the Fock and efficient bases. In what follows, we examine the fluctuations of the Shannon entropy for different system sizes obtained for eigenstates written in the efficient basis only. 

In Fig.~\ref{fig7}~(a), we plot the Peres lattice of the Shannon entropy for the eigenstates of the Dicke model written in the efficient basis. These eigenstates are contained in the energy interval $\epsilon_{k}\in[\epsilon_{\text{GS}}=-2.125,\epsilon_{\text{T}}=1.841]$ for the same values of the system size which were considered previously, $j=30,60,100$. Similarly to what we observe for the entanglement entropy in Fig.~\ref{fig5}~(a), the regular patterns and large fluctuations are visible at low energies, where we do not expect the validity of ETH. The fluctuations decrease as we move to high energies and as the system size is increased. This behavior is quantified with the deviations from the microcanonical average shown in Fig.~\ref{fig7}~(b) and its inset. The decay of $\Delta^{\text{mic}}[S_{\text{En}}]$ to values close to zero for the entanglement in Fig.~\ref{fig5}~(b) is more abrupt than what we see for $\Delta^{\text{mic}}[S_{\text{Sh}}]$ in Fig.~\ref{fig7}~(b), but both cases signal the transition to chaos. 

We close this section mentioning that the regular patterns seen at low energies in Fig.~\ref{fig2} and Figs.~\ref{fig5}-\ref{fig7} reflect the existence of families of periodic orbits, which were studied in previous works~\cite{Pilatowsky2021NatCommun,Pilatowsky2021,Pilatowsky2022}. We also note that for high energies, the behavior of the Shannon entropy computed in the efficient basis and of the von Neumann entanglement entropy are similar in the sense that both measures nearly saturate. In contrast, the Shannon entropy in the Fock basis grows rapidly with energy, because of the infinite bosonic Hilbert subspace of the Dicke model. The modified bosonic  sector of the efficient basis is responsible for the smaller values of the Shannon entropy when compared to the values for the Fock basis.

\section{Conclusions}
\label{sec:Conclusions}

In this work, we confirmed the validity of the ETH in the chaotic region of the Dicke model. This was done by analyzing the diagonal and off-diagonal elements of the number of photons and the number of excited atoms for different system sizes. We corroborated that the validity of the ETH stems from the presence of chaotic eigenstates, which we showed by analyzing their components and measures of entanglement and delocalization. 

The Shannon entropy, used to quantify the level of delocalization of the eigenstates in a given basis, made evident the advantages of using the efficient basis over the Fock basis. For high energies, the first leads to a slower growth of the entropy than the Fock basis, allowing us to reach converged states for larger system sizes than accessible with the Fock basis.

\vspace{12pt} 


\textbf{Author Contributions:} All authors have contributed equally to the conceptualization, development, and writing of the work. All authors have read and agreed to the published version of the manuscript.
\vspace{12pt}

\textbf{Funding:} This research was funded by the DGAPA- UNAM project number IN104020, the Mexican CONACyT project number CB2015-01/255702, and the United States NSF, Grant No. DMR-1936006. L.F.S. had support from the MPS Simons Foundation Award ID: 678586.
\vspace{12pt}

\textbf{Institutional Review Board Statement:} Not applicable.
\vspace{12pt}

\textbf{Data Availability:} All the data that support the results and plots showed within this work are available from the corresponding authors upon request. 
\vspace{12pt}

\textbf{Acknowledgments:} We acknowledge the support of the Computation Center - ICN, in particular to Enrique Palacios, Luciano D\'iaz, and Eduardo Murrieta.
\vspace{12pt}

\textbf{Conflicts of Interest:} The authors declare no conflict of interest.
\vspace{12pt}

\textbf{Abbreviations}

The following abbreviations are used in this manuscript:

ETH \hspace{0.6cm} eigenstate thermalization hypothesis

RMT \hspace{0.5cm} random matrix theory

OTOC \hspace{0.25cm} out-of-time-ordered correlator




\appendix

\section[\appendixname~\thesection]{Diagonalization Bases}
\label{app:Diagonalization_Bases}

\subsection[\appendixname~\thesubsection]{Fock Basis}
\label{app:Fock_Basis}

The Fock basis is the natural choice to diagonalize the Dicke Hamiltonian, since it is composed by the tensor product between bosonic Fock states $|n\rangle$ and atomic pseudo-spin states $|j,m_{z}\rangle$, $|n;j,m_{z}\rangle=|n\rangle\otimes|j,m_{z}\rangle$. On the one hand, the eigenvalues $n=0,1,2,\ldots$ of the number operator $\hat{n}=\hat{a}^{\dagger}\hat{a}$ (with eigenvalue equation $\hat{n}|n\rangle=n|n\rangle$) provide an infinite bosonic Hilbert subspace. On the other hand, the eigenvalues $m_{z}=-j,-j+1,\ldots,j-1,j$ of the pseudo-spin operator $\hat{J}_{z}$ (with eigenvalue equation $\hat{J}_{z}|j,m_{z}\rangle=m_{z}|j,m_{z}\rangle$) provide a finite atomic Hilbert subspace with dimension $d_{\text{A}}^{\text{FB}}=2j+1$. In order to diagonalize the Dicke Hamiltonian a truncation value $n_{\max}$ of the bosonic Hilbert subspace has to be chosen, which allows to define a finite bosonic dimension $d_{\text{B}}^{\text{FB}}=n_{\max}+1$. In this way, the global Hilbert-space dimension of the Dicke model is given by the product $d_{\text{D}}^{\text{FB}}=d_{\text{A}}^{\text{FB}}\times d_{\text{B}}^{\text{FB}}$.

\subsection[\appendixname~\thesubsection]{Efficient Basis}
\label{app:ECB}

The truncation required by the Fock basis for the convergence of the high-energy eigenstates rapidly increases with the system size $j$. This can be circumvented by using the so-called efficient basis, which was originally obtained by studying the integrable limit $\omega_{0}\rightarrow0$~\cite{Chen2008,Bastarrachea2011}. This basis can be written in terms of a displaced bosonic annihilation operator $\hat{A}=\hat{a}+G\hat{J}_{x}$ with $G=2\gamma/(\omega\sqrt{\mathcal{N}})$ and the eigenstates $|j,{m_{x}}\rangle$ of $\hat{J}_x$ ($\hat{J}_x|j,{m_{x}}\rangle=m_{x}|j,{m_{x}}\rangle$). The efficient basis states are defined by
\begin{align}
    |N;j,m_{x}\rangle & = \frac{(\hat{A}^{\dagger})^{N}}{\sqrt{N!}}|\alpha_{m_{x}}\rangle\otimes|j,m_{x}\rangle,
\end{align}
where $|\alpha_{m_{x}}\rangle=D(\alpha_{m_{x}})|0\rangle$ is a coherent state centered at  $\alpha_{m_{x}}=\alpha_{m_{x}}^{\ast}=-Gm_{x}$, and $\hat{D}(\alpha_{m_{x}})=\text{exp}(\alpha_{m_{x}}\hat{a}^{\dagger}-\alpha_{m_{x}}^{\ast}\hat{a})$ is the displacement operator. By commuting $\hat{A}^\dagger$ with the displacement operator and using that $G\hat{J}_{x}|j,m_{x}\rangle =-\alpha_{m_x}|j,m_{x}\rangle$, one may further write
\begin{align}
|N;j,m_{x}\rangle  
    & = \frac{(\hat{a}^{\dagger}-\alpha_{m_{x}})^{N}}{\sqrt{N!}}|\alpha_{m_{x}}\rangle\otimes|j,m_{x}\rangle \nonumber \\
    &= \hat{D}(\alpha_{m_{x}})|N\rangle\otimes|j,m_{x}\rangle \nonumber \\
    & =   |N\rangle_{m_{x}}\otimes|j,m_{x}\rangle \nonumber \\
\end{align}
where $|N\rangle_{m_{x}}=\hat{D}(\alpha_{m_{x}})|N\rangle$ are displaced Fock states, also called generalized coherent states. 

The eigenvalues $N=0,1,2,\ldots$ of the displaced number operator $\hat{N}=\hat{A}^{\dagger}\hat{A}$ (with eigenvalue equation $\hat{N}|N\rangle_{m_{x}}=N|N\rangle_{m_{x}}$) label a modified bosonic Hilbert subspace. The eigenvalues $m_{x}=-j,-j+1,\ldots,j-1,j$ of the pseudo-spin operator $\hat{J}_{x}$ label the same finite atomic Hilbert sector with dimension $d_{\text{A}}^{\text{EB}}=2j+1$. As for the Fock basis, the modified bosonic sector must be truncated to some $N_{\max}$ for diagonalization. This yields a finite modified bosonic dimension $d_{\text{B}}^{\text{EB}}=N_{\max}+1$ and a global Hilbert-space dimension $d_{\text{D}}^{\text{EB}}=d_{\text{A}}^{\text{EB}}\times d_{\text{B}}^{\text{EB}}$.

In contrast to the Fock basis, where convergence of high-energy eigenstates is infeasible for large system sizes (usually $j>30$), the efficient basis allows to get thousands of converged eigenstates in high-energy regimes, even beyond $j=100$~\cite{Bastarrachea2014PSa,Bastarrachea2014PSb}.

\subsection[\appendixname~\thesubsection]{Mapping from Efficient Basis to Fock Basis}
\label{app:ECB_FockBasis}

To compute the entanglement entropies shown in Figs.~\ref{fig6} and~\ref{fig7}, we perform a partial trace over the usual atomic sector of the Fock basis, but we diagonalize in the efficient basis. Thus, we need to map the wave function of the eigenstates from the efficient basis to the Fock basis. This is done as follows.

A general pure quantum state $|\Psi\rangle$  can be expanded in the efficient basis and the rotated Fock basis ($|n;j,m_{x}\rangle$), respectively
\begin{equation}
    |\Psi\rangle = \sum_{x=0}^{x_{\max}}\sum_{m_{x}=-j}^{j}C_{x,m_{x}}|x;j,m_{x}\rangle,
\end{equation}
where $x=N$ defines the efficient basis and $x=n$ the Fock basis. Moreover, $C_{x,m_{x}}=\langle x;j,m_{x}|\Psi\rangle$ are the coefficients of the arbitrary state in each basis, which must satisfy the normalization condition $\sum_{x,m_{x}}|C_{x,m_{x}}|^{2}=\hat{1}$.

Note that
\begin{align}
\label{eq:coefeqcnmx}
    C_{n,m_{x}} & = \langle n;j,m_{x}|\Psi\rangle \nonumber \\
    & = \sum_{N=0}^{N_{\max}}\sum_{m'_{x}=-j}^{j}C_{N,m'_{x}}\langle n;j,m_{x}|N;j,m'_{x}\rangle \nonumber \\
    & = \sum_{N=0}^{N_{\max}}\sum_{m'_{x}=-j}^{j}C_{N,m'_{x}}\langle n|N\rangle_{m'_{x}}\otimes\langle j,m_{x}|j,m'_{x}\rangle \nonumber \\
    & = \sum_{N=0}^{N_{\max}}C_{N,m_{x}}\langle n|\hat{D}(\alpha_{m_{x}})|N\rangle,
\end{align}
using that $\langle j,m_{x}|j,m'_{x}\rangle=\delta_{m_{x},m'_{x}}$. The term $\langle n|\hat{D}(\alpha_{m_{x}})|N\rangle$ for $n>N$ is given by~\cite{Cahill1969a,Cahill1969b,Oliveira1990}
\begin{align}
    \langle n|\hat{D}(\alpha_{m_{x}})|N\rangle = \sqrt{\frac{N!}{n!}}\alpha_{m_{x}}^{n-N}e^{-|\alpha_{m_{x}}|^{2}/2}\mathcal{L}_{N}^{n-N}(|\alpha_{m_{x}}|^{2}),
\end{align}
where $\mathcal{L}_{n_{0}}^{n_{1}}(x)$ is an associated Laguerre polynomial given by the Rodrigues formula
\begin{equation}
    \mathcal{L}_{n_{0}}^{n_{1}}(x)=\frac{x^{-n_{1}}e^{x}}{n_{0}!}\frac{d^{n_{0}}}{dx^{n_{0}}}(e^{-x}x^{n_{0}+n_{1}}).
\end{equation}

We numerically found that, in order to ensure a correct convergence of the coefficients $C_{n,m_{x}}$ given by Eq. \eqref{eq:coefeqcnmx}, the truncation value must be chosen as $n_{\max}\approx3N_{\max}$. The associated Laguerre polynomials can be efficiently calculated to arbitrary precision with a package included in the Wolfram Mathematica software~\cite{Mathematica}.

\section[\appendixname~\thesection]{Classical Limit of the Dicke Model}
\label{app:Classical_Dicke_Model}

The classical limit of the Dicke model can be obtained taking the expectation value of the quantum Hamiltonian $\hat{H}_{\text{D}}$ under the tensor product of Glauber and Bloch coherent states $|\mathbf{x}\rangle=|q,p\rangle\otimes|Q,P\rangle$, and dividing it by the system size $j$ ~\cite{Deaguiar1991,Deaguiar1992,Bastarrachea2014a,Bastarrachea2014b,Bastarrachea2015,Chavez2016,Villasenor2020}
\begin{align}
    \label{eq:classical_dicke_hamiltonian}
    h_\text{D}(\mathbf{x}) & = \frac{\langle\mathbf{x}|\hat{H}_{\text{D}}|\mathbf{x}\rangle}{j} = h_{\text{F}}(\mathbf{x}) + h_{\text{A}}(\mathbf{x}) + h_{\text{I}}(\mathbf{x}), \\
    h_{\text{F}}(\mathbf{x}) & = \frac{\omega}{2}\big(q^{2}+p^{2}\big), \\
    h_{\text{A}}(\mathbf{x}) & = \frac{\omega_{0}}{2}\big(Q^2+P^2\big)-\omega_{0}, \\
    h_{\text{I}}(\mathbf{x}) & =2\gamma qQ\sqrt{1-\frac{Q^2+P^2}{4}},
\end{align}
where $h_{\text{F}}(\mathbf{x})$ and $h_{\text{A}}(\mathbf{x})$ represent the Hamiltonians of two classical harmonic oscillators, and $h_{\text{I}}(\mathbf{x})$ the coupling between them. The bosonic Glauber and the atomic Bloch coherent states, represented by the canonical variables $(q,p)$ and $(Q,P)$ respectively, are given explicitly by 
\begin{align}
    \label{eq:coherent_states}
    |q,p\rangle & = e^{-(j/4)\left(q^{2}+p^{2}\right)}e^{\left[\sqrt{j/2}\left(q+ip\right)\right]\hat{a}^{\dagger}}|0\rangle, \\
    |Q,P\rangle & = \Big(1-\frac{Q^2+P^2}{4}\Big)^{j}e^{\left[\left(Q+iP\right)/\sqrt{4-Q^2-P^2}\right]\hat{J}_{+}}|j,-j\rangle,
\end{align}
where $|0\rangle$ is the photon vacuum and $|j,-j\rangle$ is the state with all the atoms in the ground state.

The classical Dicke Hamiltonian $h_\text{D}(\mathbf{x})$, obtained with the latter method, has an infinite four-dimensional phase space $\mathcal{M}$ in the canonical variables $\mathbf{x}=(q,p;Q,P)$, where the atomic variables are bounded ($Q^{2} + P^{2} \leq 4$). A useful property of this phase space is that it can be partitioned into a family of classical energy shells with finite volume $\mathcal{V}(\epsilon)<\infty$, given by
\begin{equation}
    \mathcal{M}(\epsilon) = \{\mathbf{x}\in \mathcal{M} \,\mid\, h_\text{D}(\mathbf{x})=\epsilon\}, 
\end{equation}
where $\epsilon=E/j$ is the classical energy of the shell scaled to the system size $j$, which defines an effective Planck constant $\hbar_{\text{eff}}=1/j$~\cite{Ribeiro2006}. The finite volume $\mathcal{V}(\epsilon)$ of the classical energy shells $\mathcal{M}(\epsilon)$ is obtained with a semiclassical approximation to the quantum density of states $\nu(\epsilon)$, using the Gutzwiller trace formula~\cite{Gutzwiller1971,Gutzwiller1990book,Bastarrachea2014a}. The explicit expression is given by 
\begin{equation}
    \label{eq:volume_classical_energy_shell}
    \mathcal{V}(\epsilon) = \int_{\mathcal{M}}d \mathbf{x}\,  \delta(h_\text{D}(\mathbf{x})-\epsilon) =(2\pi\hbar_{\text{eff}})^2\nu(\epsilon),
\end{equation}
where the density of states is proportional to the system size $\nu(\epsilon)\propto2j^{2}$, and can be derived explicitly following Ref.~\cite{Bastarrachea2014a}.

\section[\appendixname~\thesection]{Quantum Entanglement}
\label{app:Quantum_Entanglement}

An arbitrary pure state $|\Psi\rangle$ of a multipartite system composed of $\mathcal{S}$ subsystems, whose Hilbert space is given by the tensor product $\mathcal{H}=\mathcal{H}_{1}\otimes\mathcal{H}_{2}\otimes\ldots\otimes\mathcal{H}_{\mathcal{S}}$, can be expanded due to the superposition principle in a tensor-product basis $\{|\psi_{k_{1}}^{1}\rangle\otimes|\psi_{k_{2}}^{2}\rangle\otimes\ldots\otimes|\psi_{k_{\mathcal{S}}}^{\mathcal{S}}\rangle\}$ as
\begin{equation}
    |\Psi\rangle = \sum_{k_{1},k_{2},\ldots,k_{\mathcal{S}}}^{d_{1},d_{2},\ldots,d_{\mathcal{S}}}c_{k_{1},k_{2},\ldots,k_{\mathcal{S}}}|\psi_{k_{1}}^{1}\rangle\otimes|\psi_{k_{2}}^{2}\rangle\otimes\ldots\otimes|\psi_{k_{\mathcal{S}}}^{\mathcal{S}}\rangle,
\end{equation}
where $d_{1},d_{2},\ldots,d_{\mathcal{S}}$ are the dimensions of each subspace and the coefficients $c_{k_{1},k_{2},\ldots,k_{\mathcal{S}}}$ satisfy the normalization condition
\begin{equation}
    \sum_{k_{1},k_{2},\ldots,k_{\mathcal{S}}}^{^{d_{1},d_{2},\ldots,d_{\mathcal{S}}}}|c_{k_{1},k_{2},\ldots,k_{\mathcal{S}}}|^{2} = \hat{1}.
\end{equation}

Thus, the state $|\Psi\rangle$ is called separable (entangled) if it can (cannot) be written as a tensor product of states corresponding to each subspace
\begin{equation}
    |\Psi\rangle = |\Psi_{1}\rangle\otimes|\Psi_{2}\rangle\otimes\ldots\otimes|\Psi_{\mathcal{S}}\rangle.
\end{equation}

For a mixed state $\hat{\rho}$ the definition of entanglement is no longer equivalent to that of a pure state. In this way, the mixed state $\hat{\rho}$ is called separable (entangled) if it can (cannot) be written as a convex combination of tensor-product states~\cite{Werner1989,Horodecki2009}, corresponding to each subspace
\begin{equation}
    \hat{\rho} = \sum_{k_{1},k_{2},\ldots,k_{\mathcal{S}}}^{d_{1},d_{2},\ldots,d_{\mathcal{S}}}p_{k_{1},k_{2},\ldots,k_{\mathcal{S}}}\hat{\rho}_{k_{1}}^{1}\otimes\hat{\rho}_{k_{2}}^{2}\otimes\ldots\otimes\hat{\rho}_{k_{\mathcal{S}}}^{\mathcal{S}},
\end{equation}
where $d_{1},d_{2},\ldots,d_{\mathcal{S}}$ are the dimensions of each subspace and the probabilities $p_{k_{1},k_{2},\ldots,k_{\mathcal{S}}}$ satisfy the normalization condition
\begin{equation}
    \sum_{k_{1},k_{2},\ldots,k_{\mathcal{S}}}^{d_{1},d_{2},\ldots,d_{\mathcal{S}}}p_{k_{1},k_{2},\ldots,k_{\mathcal{S}}} = \hat{1}.
\end{equation}

When multipartite systems are studied, a useful tool to work with is the reduced density matrix corresponding to a given subspace, which is obtained by taking the partial trace of the whole density matrix $\hat{\rho}$. For example, the reduced density matrix of the $i$-th subspace is given by
\begin{equation}
    \hat{\rho}_{i} = \text{Tr}_{(1,2,\ldots,\mathcal{S}) \neq i}[\hat{\rho}] = \sum_{k_{i}=1}^{d_{i}}P_{k_{i}}\hat{\rho}_{k_{i}}^{i},
\end{equation}
where $\sum_{k_{i}=1}^{d_{i}}P_{k_{i}} = \hat{1}$ and
\begin{equation}
    P_{k_{i}} = \sum_{(k_{1},k_{2},\ldots,k_{\mathcal{S}}) \neq k_{i} }^{(d_{1},d_{2},\ldots,d_{\mathcal{S}}) \neq d_{i}}p_{k_{1},k_{2},\ldots,k_{\mathcal{S}}}.
\end{equation}

\section[\appendixname~\thesection]{Integrable Dicke Model: Tavis-Cummings Model}
\label{app:Tavis_Cummings_Model}

The integrable limit of the Dicke model is known as the Tavis-Cummings model~\cite{Tavis1968}, and can be obtained by applying the rotating wave approximation (RWA) to the Dicke Hamiltonian~\eqref{eqn:dicke_hamiltonian}. The RWA consists of ignoring the interacting terms in $\hat{H}_{\text{I}}$ (see Eq.~\eqref{eqn:dicke_hamiltonian_hi}) that oscillates very fast, that is, $\hat{a}^{\dagger}\hat{J}_{+}$ and $\hat{a}\hat{J}_{-}$. The last results in a modified interacting Hamiltonian given by
\begin{equation}
    \hat{H}_{\text{I}}^{\text{RWA}} = \frac{\gamma}{\sqrt{\mathcal{N}}}(\hat{a}^{\dagger}\hat{J}_{-}+\hat{a}\hat{J}_{+}),
\end{equation}
such that, the complete Tavis-Cummings Hamiltonian is given by
\begin{align}
    \label{eqn:tavis_cummings_hamiltonian}
    \hat{H}_{\text{TC}} & = \hat{H}_{\text{F}} + \hat{H}_{\text{A}} + \hat{H}_{\text{I}}^{\text{RWA}},
\end{align}
where the Hamiltonians $\hat{H}_{\text{F}}$ and $\hat{H}_{\text{A}}$ are the same terms given in Eqs.~\eqref{eqn:dicke_hamiltonian_hf} and~\eqref{eqn:dicke_hamiltonian_ha}. The particularity of the Tavis-Cummings Hamiltonian is that it commutes with the operator $\hat{\Lambda}$ [see Eq.~\eqref{eqn:lambda_operator}], which is a conserved quantity that defines the number of excitations. The last feature allows to diagonalize the Tavis-Cummings Hamiltonian in finite subspaces of such operator~\cite{Tavis1968}.

\bibliography{main}

\begin{thebibliography}{119}%
\makeatletter
\providecommand \@ifxundefined [1]{%
 \@ifx{#1\undefined}
}%
\providecommand \@ifnum [1]{%
 \ifnum #1\expandafter \@firstoftwo
 \else \expandafter \@secondoftwo
 \fi
}%
\providecommand \@ifx [1]{%
 \ifx #1\expandafter \@firstoftwo
 \else \expandafter \@secondoftwo
 \fi
}%
\providecommand \natexlab [1]{#1}%
\providecommand \enquote  [1]{``#1''}%
\providecommand \bibnamefont  [1]{#1}%
\providecommand \bibfnamefont [1]{#1}%
\providecommand \citenamefont [1]{#1}%
\providecommand \href@noop [0]{\@secondoftwo}%
\providecommand \href [0]{\begingroup \@sanitize@url \@href}%
\providecommand \@href[1]{\@@startlink{#1}\@@href}%
\providecommand \@@href[1]{\endgroup#1\@@endlink}%
\providecommand \@sanitize@url [0]{\catcode `\\12\catcode `\$12\catcode
  `\&12\catcode `\#12\catcode `\^12\catcode `\_12\catcode `\%12\relax}%
\providecommand \@@startlink[1]{}%
\providecommand \@@endlink[0]{}%
\providecommand \url  [0]{\begingroup\@sanitize@url \@url }%
\providecommand \@url [1]{\endgroup\@href {#1}{\urlprefix }}%
\providecommand \urlprefix  [0]{URL }%
\providecommand \Eprint [0]{\href }%
\providecommand \doibase [0]{http://dx.doi.org/}%
\providecommand \selectlanguage [0]{\@gobble}%
\providecommand \bibinfo  [0]{\@secondoftwo}%
\providecommand \bibfield  [0]{\@secondoftwo}%
\providecommand \translation [1]{[#1]}%
\providecommand \BibitemOpen [0]{}%
\providecommand \bibitemStop [0]{}%
\providecommand \bibitemNoStop [0]{.\EOS\space}%
\providecommand \EOS [0]{\spacefactor3000\relax}%
\providecommand \BibitemShut  [1]{\csname bibitem#1\endcsname}%
\let\auto@bib@innerbib\@empty
\bibitem [{\citenamefont {von Neumann}(1929)}]{vonNeumann1929}%
  \BibitemOpen
  \bibfield  {author} {\bibinfo {author} {\bibfnamefont {J.}~\bibnamefont {von
  Neumann}},\ }\bibfield  {title} {\enquote {\bibinfo {title} {Beweis des
  ergodensatzes und des h-theorems in der neuen mechanik},}\ }\href@noop {}
  {\bibfield  {journal} {\bibinfo  {journal} {Zeitschrift f\"ur Physik}\
  }\textbf {\bibinfo {volume} {57}},\ \bibinfo {pages} {30--70} (\bibinfo
  {year} {1929})}\BibitemShut {NoStop}%
\bibitem [{\citenamefont {von Neumann}(2010)}]{vonNeumann2010}%
  \BibitemOpen
  \bibfield  {author} {\bibinfo {author} {\bibfnamefont {J.}~\bibnamefont {von
  Neumann}},\ }\bibfield  {title} {\enquote {\bibinfo {title} {Proof of the
  ergodic theorem and the h-theorem in quantum mechanics},}\ }\href {\doibase
  10.1140/epjh/e2010-00008-5} {\bibfield  {journal} {\bibinfo  {journal} {Eur.
  Phys. J. H}\ }\textbf {\bibinfo {volume} {35}},\ \bibinfo {pages} {201--237}
  (\bibinfo {year} {2010})}\BibitemShut {NoStop}%
\bibitem [{\citenamefont {Goldstein}\ \emph
  {et~al.}(2010{\natexlab{a}})\citenamefont {Goldstein}, \citenamefont
  {Lebowitz}, \citenamefont {Tumulka},\ and\ \citenamefont
  {Zangh\`\i}}]{Goldstein2010}%
  \BibitemOpen
  \bibfield  {author} {\bibinfo {author} {\bibfnamefont {S.}~\bibnamefont
  {Goldstein}}, \bibinfo {author} {\bibfnamefont {J.~L.}\ \bibnamefont
  {Lebowitz}}, \bibinfo {author} {\bibfnamefont {R.}~\bibnamefont {Tumulka}}, \
  and\ \bibinfo {author} {\bibfnamefont {N.}~\bibnamefont {Zangh\`\i}},\
  }\bibfield  {title} {\enquote {\bibinfo {title} {Long-time behavior of
  macroscopic quantum systems},}\ }\href {\doibase 10.1140/epjh/e2010-00007-7}
  {\bibfield  {journal} {\bibinfo  {journal} {Eur. Phys. J. H}\ }\textbf
  {\bibinfo {volume} {35}},\ \bibinfo {pages} {173--200} (\bibinfo {year}
  {2010}{\natexlab{a}})}\BibitemShut {NoStop}%
\bibitem [{\citenamefont {Goldstein}\ \emph
  {et~al.}(2010{\natexlab{b}})\citenamefont {Goldstein}, \citenamefont
  {Lebowitz}, \citenamefont {Mastrodonato}, \citenamefont {Tumulka},\ and\
  \citenamefont {Zangh\`\i}}]{Goldstein2010b}%
  \BibitemOpen
  \bibfield  {author} {\bibinfo {author} {\bibfnamefont {Sheldon}\ \bibnamefont
  {Goldstein}}, \bibinfo {author} {\bibfnamefont {Joel~L.}\ \bibnamefont
  {Lebowitz}}, \bibinfo {author} {\bibfnamefont {Christian}\ \bibnamefont
  {Mastrodonato}}, \bibinfo {author} {\bibfnamefont {Roderich}\ \bibnamefont
  {Tumulka}}, \ and\ \bibinfo {author} {\bibfnamefont {Nino}\ \bibnamefont
  {Zangh\`\i}},\ }\bibfield  {title} {\enquote {\bibinfo {title} {Normal
  typicality and von neumann's quantum ergodic theorem},}\ }\href {\doibase
  10.1098/rspa.2009.0635} {\bibfield  {journal} {\bibinfo  {journal} {Proc. R.
  Soc. A: Math. Phys. Eng. Sci.}\ }\textbf {\bibinfo {volume} {466}},\ \bibinfo
  {pages} {3203--3224} (\bibinfo {year} {2010}{\natexlab{b}})}\BibitemShut
  {NoStop}%
\bibitem [{\citenamefont {Pechukas}(1984{\natexlab{a}})}]{Pechukas1984}%
  \BibitemOpen
  \bibfield  {author} {\bibinfo {author} {\bibfnamefont {P.}~\bibnamefont
  {Pechukas}},\ }\bibfield  {title} {\enquote {\bibinfo {title} {Remarks on
  ``quantum chaos''},}\ }\href@noop {} {\bibfield  {journal} {\bibinfo
  {journal} {J. Phys. Chem.}\ }\textbf {\bibinfo {volume} {88}},\ \bibinfo
  {pages} {4823} (\bibinfo {year} {1984}{\natexlab{a}})}\BibitemShut {NoStop}%
\bibitem [{\citenamefont {Bocchieri}\ and\ \citenamefont
  {Loinger}(1958)}]{Bocchieri1958}%
  \BibitemOpen
  \bibfield  {author} {\bibinfo {author} {\bibfnamefont {P.}~\bibnamefont
  {Bocchieri}}\ and\ \bibinfo {author} {\bibfnamefont {A.}~\bibnamefont
  {Loinger}},\ }\bibfield  {title} {\enquote {\bibinfo {title} {Ergodic theorem
  in quantum mechanics},}\ }\href {\doibase 10.1103/PhysRev.111.668} {\bibfield
   {journal} {\bibinfo  {journal} {Phys. Rev.}\ }\textbf {\bibinfo {volume}
  {111}},\ \bibinfo {pages} {668--670} (\bibinfo {year} {1958})}\BibitemShut
  {NoStop}%
\bibitem [{\citenamefont {Pechukas}(1984{\natexlab{b}})}]{Pechukas1984b}%
  \BibitemOpen
  \bibfield  {author} {\bibinfo {author} {\bibfnamefont {Philip}\ \bibnamefont
  {Pechukas}},\ }\bibfield  {title} {\enquote {\bibinfo {title} {Sharpening an
  inequality in quantum ergodic theory},}\ }\href {\doibase 10.1063/1.526202}
  {\bibfield  {journal} {\bibinfo  {journal} {J. Math. Phys.}\ }\textbf
  {\bibinfo {volume} {25}},\ \bibinfo {pages} {532--534} (\bibinfo {year}
  {1984}{\natexlab{b}})},\ \Eprint
  {http://arxiv.org/abs/https://doi.org/10.1063/1.526202}
  {https://doi.org/10.1063/1.526202} \BibitemShut {NoStop}%
\bibitem [{\citenamefont {Jensen}\ and\ \citenamefont
  {Shankar}(1985)}]{Jensen1985}%
  \BibitemOpen
  \bibfield  {author} {\bibinfo {author} {\bibfnamefont {R.~V.}\ \bibnamefont
  {Jensen}}\ and\ \bibinfo {author} {\bibfnamefont {R.}~\bibnamefont
  {Shankar}},\ }\bibfield  {title} {\enquote {\bibinfo {title} {Statistical
  behavior in deterministic quantum systems with few degrees of freedom},}\
  }\href@noop {} {\bibfield  {journal} {\bibinfo  {journal} {Phys. Rev. Lett.}\
  }\textbf {\bibinfo {volume} {54}},\ \bibinfo {pages} {1879--1882} (\bibinfo
  {year} {1985})}\BibitemShut {NoStop}%
\bibitem [{\citenamefont {D'Alessio}\ \emph {et~al.}(2016)\citenamefont
  {D'Alessio}, \citenamefont {Kafri}, \citenamefont {Polkovnikov},\ and\
  \citenamefont {Rigol}}]{Dalessio2016}%
  \BibitemOpen
  \bibfield  {author} {\bibinfo {author} {\bibfnamefont {Luca}\ \bibnamefont
  {D'Alessio}}, \bibinfo {author} {\bibfnamefont {Yariv}\ \bibnamefont
  {Kafri}}, \bibinfo {author} {\bibfnamefont {Anatoli}\ \bibnamefont
  {Polkovnikov}}, \ and\ \bibinfo {author} {\bibfnamefont {Marcos}\
  \bibnamefont {Rigol}},\ }\bibfield  {title} {\enquote {\bibinfo {title} {From
  quantum chaos and eigenstate thermalization to statistical mechanics and
  thermodynamics},}\ }\href {\doibase 10.1080/00018732.2016.1198134} {\bibfield
   {journal} {\bibinfo  {journal} {Adv. Phys.}\ }\textbf {\bibinfo {volume}
  {65}},\ \bibinfo {pages} {239--362} (\bibinfo {year} {2016})}\BibitemShut
  {NoStop}%
\bibitem [{\citenamefont {Deutsch}(2018)}]{Deutsch2018}%
  \BibitemOpen
  \bibfield  {author} {\bibinfo {author} {\bibfnamefont {Joshua~M}\
  \bibnamefont {Deutsch}},\ }\bibfield  {title} {\enquote {\bibinfo {title}
  {Eigenstate thermalization hypothesis},}\ }\href {\doibase
  10.1088/1361-6633/aac9f1} {\bibfield  {journal} {\bibinfo  {journal} {Rep.
  Prog. Phys.}\ }\textbf {\bibinfo {volume} {81}},\ \bibinfo {pages} {082001}
  (\bibinfo {year} {2018})}\BibitemShut {NoStop}%
\bibitem [{\citenamefont {Rigol}\ \emph {et~al.}(2008)\citenamefont {Rigol},
  \citenamefont {Dunjko},\ and\ \citenamefont {Olshanii}}]{Rigol2008}%
  \BibitemOpen
  \bibfield  {author} {\bibinfo {author} {\bibfnamefont {M.}~\bibnamefont
  {Rigol}}, \bibinfo {author} {\bibfnamefont {V.}~\bibnamefont {Dunjko}}, \
  and\ \bibinfo {author} {\bibfnamefont {M.}~\bibnamefont {Olshanii}},\
  }\bibfield  {title} {\enquote {\bibinfo {title} {Thermalization and its
  mechanism for generic isolated quantum systems},}\ }\href {\doibase
  10.1038/nature06838} {\bibfield  {journal} {\bibinfo  {journal} {Nature}\
  }\textbf {\bibinfo {volume} {452}},\ \bibinfo {pages} {854} (\bibinfo {year}
  {2008})}\BibitemShut {NoStop}%
\bibitem [{\citenamefont {Srednicki}(1994)}]{Srednicki1994}%
  \BibitemOpen
  \bibfield  {author} {\bibinfo {author} {\bibfnamefont {Mark}\ \bibnamefont
  {Srednicki}},\ }\bibfield  {title} {\enquote {\bibinfo {title} {Chaos and
  quantum thermalization},}\ }\href {\doibase 10.1103/PhysRevE.50.888}
  {\bibfield  {journal} {\bibinfo  {journal} {Phys. Rev. E}\ }\textbf {\bibinfo
  {volume} {50}},\ \bibinfo {pages} {888--901} (\bibinfo {year}
  {1994})}\BibitemShut {NoStop}%
\bibitem [{\citenamefont {Deutsch}(1991)}]{Deutsch1991}%
  \BibitemOpen
  \bibfield  {author} {\bibinfo {author} {\bibfnamefont {J.~M.}\ \bibnamefont
  {Deutsch}},\ }\bibfield  {title} {\enquote {\bibinfo {title} {Quantum
  statistical mechanics in a closed system},}\ }\href {\doibase
  10.1103/PhysRevA.43.2046} {\bibfield  {journal} {\bibinfo  {journal} {Phys.
  Rev. A}\ }\textbf {\bibinfo {volume} {43}},\ \bibinfo {pages} {2046--2049}
  (\bibinfo {year} {1991})}\BibitemShut {NoStop}%
\bibitem [{\citenamefont {Beugeling}\ \emph {et~al.}(2014)\citenamefont
  {Beugeling}, \citenamefont {Moessner},\ and\ \citenamefont
  {Haque}}]{Beugeling2014}%
  \BibitemOpen
  \bibfield  {author} {\bibinfo {author} {\bibfnamefont {W.}~\bibnamefont
  {Beugeling}}, \bibinfo {author} {\bibfnamefont {R.}~\bibnamefont {Moessner}},
  \ and\ \bibinfo {author} {\bibfnamefont {Masudul}\ \bibnamefont {Haque}},\
  }\bibfield  {title} {\enquote {\bibinfo {title} {Finite-size scaling of
  eigenstate thermalization},}\ }\href {\doibase 10.1103/PhysRevE.89.042112}
  {\bibfield  {journal} {\bibinfo  {journal} {Phys. Rev. E}\ }\textbf {\bibinfo
  {volume} {89}},\ \bibinfo {pages} {042112} (\bibinfo {year}
  {2014})}\BibitemShut {NoStop}%
\bibitem [{\citenamefont {LeBlond}\ \emph {et~al.}(2019)\citenamefont
  {LeBlond}, \citenamefont {Mallayya}, \citenamefont {Vidmar},\ and\
  \citenamefont {Rigol}}]{LeBlond2019}%
  \BibitemOpen
  \bibfield  {author} {\bibinfo {author} {\bibfnamefont {Tyler}\ \bibnamefont
  {LeBlond}}, \bibinfo {author} {\bibfnamefont {Krishnanand}\ \bibnamefont
  {Mallayya}}, \bibinfo {author} {\bibfnamefont {Lev}\ \bibnamefont {Vidmar}},
  \ and\ \bibinfo {author} {\bibfnamefont {Marcos}\ \bibnamefont {Rigol}},\
  }\bibfield  {title} {\enquote {\bibinfo {title} {Entanglement and matrix
  elements of observables in interacting integrable systems},}\ }\href
  {\doibase 10.1103/PhysRevE.100.062134} {\bibfield  {journal} {\bibinfo
  {journal} {Phys. Rev. E}\ }\textbf {\bibinfo {volume} {100}},\ \bibinfo
  {pages} {062134} (\bibinfo {year} {2019})}\BibitemShut {NoStop}%
\bibitem [{\citenamefont {Torres-Herrera}\ and\ \citenamefont
  {Santos}(2013)}]{Torres2013}%
  \BibitemOpen
  \bibfield  {author} {\bibinfo {author} {\bibfnamefont {E.~J.}\ \bibnamefont
  {Torres-Herrera}}\ and\ \bibinfo {author} {\bibfnamefont {Lea~F.}\
  \bibnamefont {Santos}},\ }\bibfield  {title} {\enquote {\bibinfo {title}
  {Effects of the interplay between initial state and {H}amiltonian on the
  thermalization of isolated quantum many-body systems},}\ }\href {\doibase
  10.1103/PhysRevE.88.042121} {\bibfield  {journal} {\bibinfo  {journal} {Phys.
  Rev. E}\ }\textbf {\bibinfo {volume} {88}},\ \bibinfo {pages} {042121}
  (\bibinfo {year} {2013})}\BibitemShut {NoStop}%
\bibitem [{\citenamefont {He}\ and\ \citenamefont {Rigol}(2013)}]{He2013}%
  \BibitemOpen
  \bibfield  {author} {\bibinfo {author} {\bibfnamefont {Kai}\ \bibnamefont
  {He}}\ and\ \bibinfo {author} {\bibfnamefont {Marcos}\ \bibnamefont
  {Rigol}},\ }\bibfield  {title} {\enquote {\bibinfo {title} {Initial-state
  dependence of the quench dynamics in integrable quantum systems. iii. chaotic
  states},}\ }\href@noop {} {\bibfield  {journal} {\bibinfo  {journal} {Phys.
  Rev. A}\ }\textbf {\bibinfo {volume} {87}},\ \bibinfo {pages} {043615}
  (\bibinfo {year} {2013})}\BibitemShut {NoStop}%
\bibitem [{\citenamefont {Santos}\ and\ \citenamefont
  {Rigol}(2010{\natexlab{a}})}]{Santos2010PREa}%
  \BibitemOpen
  \bibfield  {author} {\bibinfo {author} {\bibfnamefont {Lea~F.}\ \bibnamefont
  {Santos}}\ and\ \bibinfo {author} {\bibfnamefont {Marcos}\ \bibnamefont
  {Rigol}},\ }\bibfield  {title} {\enquote {\bibinfo {title} {Onset of quantum
  chaos in one-dimensional bosonic and fermionic systems and its relation to
  thermalization},}\ }\href {\doibase 10.1103/PhysRevE.81.036206} {\bibfield
  {journal} {\bibinfo  {journal} {Phys. Rev. E}\ }\textbf {\bibinfo {volume}
  {81}},\ \bibinfo {pages} {036206} (\bibinfo {year}
  {2010}{\natexlab{a}})}\BibitemShut {NoStop}%
\bibitem [{\citenamefont {Santos}\ and\ \citenamefont
  {Rigol}(2010{\natexlab{b}})}]{Santos2010PREb}%
  \BibitemOpen
  \bibfield  {author} {\bibinfo {author} {\bibfnamefont {Lea~F.}\ \bibnamefont
  {Santos}}\ and\ \bibinfo {author} {\bibfnamefont {Marcos}\ \bibnamefont
  {Rigol}},\ }\bibfield  {title} {\enquote {\bibinfo {title} {Localization and
  the effects of symmetries in the thermalization properties of one-dimensional
  quantum systems},}\ }\href {\doibase 10.1103/PhysRevE.82.031130} {\bibfield
  {journal} {\bibinfo  {journal} {Phys. Rev. E}\ }\textbf {\bibinfo {volume}
  {82}},\ \bibinfo {pages} {031130} (\bibinfo {year}
  {2010}{\natexlab{b}})}\BibitemShut {NoStop}%
\bibitem [{\citenamefont {Rigol}\ and\ \citenamefont
  {Santos}(2010)}]{RigolSantos2010}%
  \BibitemOpen
  \bibfield  {author} {\bibinfo {author} {\bibfnamefont {M.}~\bibnamefont
  {Rigol}}\ and\ \bibinfo {author} {\bibfnamefont {L.~F.}\ \bibnamefont
  {Santos}},\ }\bibfield  {title} {\enquote {\bibinfo {title} {Quantum chaos
  and thermalization in gapped systems},}\ }\href@noop {} {\bibfield  {journal}
  {\bibinfo  {journal} {Phys. Rev. A}\ }\textbf {\bibinfo {volume} {82}},\
  \bibinfo {pages} {011604(R)} (\bibinfo {year} {2010})}\BibitemShut {NoStop}%
\bibitem [{\citenamefont {Benenti}\ \emph {et~al.}(2001)\citenamefont
  {Benenti}, \citenamefont {Casati},\ and\ \citenamefont
  {Shepelyansky}}]{Benenti2001}%
  \BibitemOpen
  \bibfield  {author} {\bibinfo {author} {\bibfnamefont {G.}~\bibnamefont
  {Benenti}}, \bibinfo {author} {\bibfnamefont {G.}~\bibnamefont {Casati}}, \
  and\ \bibinfo {author} {\bibfnamefont {D.L.}\ \bibnamefont {Shepelyansky}},\
  }\bibfield  {title} {\enquote {\bibinfo {title} {Emergence of fermi-dirac
  thermalization in the quantum computer core},}\ }\href {\doibase
  10.1007/s100530170031} {\bibfield  {journal} {\bibinfo  {journal} {Eur. Phys.
  J. D}\ }\textbf {\bibinfo {volume} {17}},\ \bibinfo {pages} {265} (\bibinfo
  {year} {2001})}\BibitemShut {NoStop}%
\bibitem [{\citenamefont {Flambaum}\ \emph {et~al.}(1994)\citenamefont
  {Flambaum}, \citenamefont {Gribakina}, \citenamefont {Gribakin},\ and\
  \citenamefont {Kozlov}}]{Flambaum1994}%
  \BibitemOpen
  \bibfield  {author} {\bibinfo {author} {\bibfnamefont {V.~V.}\ \bibnamefont
  {Flambaum}}, \bibinfo {author} {\bibfnamefont {A.~A.}\ \bibnamefont
  {Gribakina}}, \bibinfo {author} {\bibfnamefont {G.~F.}\ \bibnamefont
  {Gribakin}}, \ and\ \bibinfo {author} {\bibfnamefont {M.~G.}\ \bibnamefont
  {Kozlov}},\ }\bibfield  {title} {\enquote {\bibinfo {title} {Structure of
  compound states in the chaotic spectrum of the ce atom: Localization
  properties, matrix elements, and enhancement of weak perturbations},}\
  }\href@noop {} {\bibfield  {journal} {\bibinfo  {journal} {Phys. Rev. A}\
  }\textbf {\bibinfo {volume} {50}},\ \bibinfo {pages} {267--296} (\bibinfo
  {year} {1994})}\BibitemShut {NoStop}%
\bibitem [{\citenamefont {Flambaum}\ \emph {et~al.}(1996)\citenamefont
  {Flambaum}, \citenamefont {Izrailev},\ and\ \citenamefont
  {Casati}}]{Flambaum1996b}%
  \BibitemOpen
  \bibfield  {author} {\bibinfo {author} {\bibfnamefont {V.~V.}\ \bibnamefont
  {Flambaum}}, \bibinfo {author} {\bibfnamefont {F.~M.}\ \bibnamefont
  {Izrailev}}, \ and\ \bibinfo {author} {\bibfnamefont {G.}~\bibnamefont
  {Casati}},\ }\bibfield  {title} {\enquote {\bibinfo {title} {Towards a
  statistical theory of finite fermi systems and compound states: Random
  two-body interaction approach},}\ }\href {\doibase 10.1103/PhysRevE.54.2136}
  {\bibfield  {journal} {\bibinfo  {journal} {Phys. Rev. E}\ }\textbf {\bibinfo
  {volume} {54}},\ \bibinfo {pages} {2136--2139} (\bibinfo {year}
  {1996})}\BibitemShut {NoStop}%
\bibitem [{\citenamefont {Flambaum}\ and\ \citenamefont
  {Izrailev}(1997)}]{Flambaum1997}%
  \BibitemOpen
  \bibfield  {author} {\bibinfo {author} {\bibfnamefont {V.~V.}\ \bibnamefont
  {Flambaum}}\ and\ \bibinfo {author} {\bibfnamefont {F.~M.}\ \bibnamefont
  {Izrailev}},\ }\bibfield  {title} {\enquote {\bibinfo {title} {Statistical
  theory of finite fermi systems based on the structure of chaotic
  eigenstates},}\ }\href {\doibase 10.1103/PhysRevE.56.5144} {\bibfield
  {journal} {\bibinfo  {journal} {Phys. Rev. E}\ }\textbf {\bibinfo {volume}
  {56}},\ \bibinfo {pages} {5144--5159} (\bibinfo {year} {1997})}\BibitemShut
  {NoStop}%
\bibitem [{\citenamefont {Borgonovi}\ \emph {et~al.}(1998)\citenamefont
  {Borgonovi}, \citenamefont {Guarneri}, \citenamefont {Izrailev},\ and\
  \citenamefont {Casati}}]{Borgonovi1998}%
  \BibitemOpen
  \bibfield  {author} {\bibinfo {author} {\bibfnamefont {F}~\bibnamefont
  {Borgonovi}}, \bibinfo {author} {\bibfnamefont {I}~\bibnamefont {Guarneri}},
  \bibinfo {author} {\bibfnamefont {F.M}\ \bibnamefont {Izrailev}}, \ and\
  \bibinfo {author} {\bibfnamefont {G}~\bibnamefont {Casati}},\ }\bibfield
  {title} {\enquote {\bibinfo {title} {Chaos and thermalization in a dynamical
  model of two interacting particles},}\ }\href {\doibase
  10.1016/s0375-9601(98)00545-3} {\bibfield  {journal} {\bibinfo  {journal}
  {Phys. Lett. A}\ }\textbf {\bibinfo {volume} {247}},\ \bibinfo {pages}
  {140--144} (\bibinfo {year} {1998})}\BibitemShut {NoStop}%
\bibitem [{\citenamefont {Jacquod}\ and\ \citenamefont
  {Shepelyansky}(1997)}]{Jacquod1997}%
  \BibitemOpen
  \bibfield  {author} {\bibinfo {author} {\bibfnamefont {P.}~\bibnamefont
  {Jacquod}}\ and\ \bibinfo {author} {\bibfnamefont {D.~L.}\ \bibnamefont
  {Shepelyansky}},\ }\bibfield  {title} {\enquote {\bibinfo {title} {Emergence
  of quantum chaos in finite interacting fermi systems},}\ }\href@noop {}
  {\bibfield  {journal} {\bibinfo  {journal} {Phys. Rev. Lett.}\ }\textbf
  {\bibinfo {volume} {79}},\ \bibinfo {pages} {1837} (\bibinfo {year}
  {1997})}\BibitemShut {NoStop}%
\bibitem [{\citenamefont {Horoi}\ \emph {et~al.}(1995)\citenamefont {Horoi},
  \citenamefont {Zelevinsky},\ and\ \citenamefont {Brown}}]{Horoi1995}%
  \BibitemOpen
  \bibfield  {author} {\bibinfo {author} {\bibfnamefont {Mihai}\ \bibnamefont
  {Horoi}}, \bibinfo {author} {\bibfnamefont {Vladimir}\ \bibnamefont
  {Zelevinsky}}, \ and\ \bibinfo {author} {\bibfnamefont {B.~Alex}\
  \bibnamefont {Brown}},\ }\bibfield  {title} {\enquote {\bibinfo {title}
  {Chaos vs thermalization in the nuclear shell model},}\ }\href {\doibase
  10.1103/PhysRevLett.74.5194} {\bibfield  {journal} {\bibinfo  {journal}
  {Phys. Rev. Lett.}\ }\textbf {\bibinfo {volume} {74}},\ \bibinfo {pages}
  {5194--5197} (\bibinfo {year} {1995})}\BibitemShut {NoStop}%
\bibitem [{\citenamefont {Zelevinsky}\ \emph {et~al.}(1996)\citenamefont
  {Zelevinsky}, \citenamefont {Brown}, \citenamefont {Frazier},\ and\
  \citenamefont {Horoi}}]{Zelevinsky1996b}%
  \BibitemOpen
  \bibfield  {author} {\bibinfo {author} {\bibfnamefont {V.}~\bibnamefont
  {Zelevinsky}}, \bibinfo {author} {\bibfnamefont {B.~A.}\ \bibnamefont
  {Brown}}, \bibinfo {author} {\bibfnamefont {N.}~\bibnamefont {Frazier}}, \
  and\ \bibinfo {author} {\bibfnamefont {M.}~\bibnamefont {Horoi}},\ }\bibfield
   {title} {\enquote {\bibinfo {title} {The nuclear shell model as a testing
  ground for many-body quantum chaos},}\ }\href {\doibase
  10.1016/S0370-1573(96)00007-5} {\bibfield  {journal} {\bibinfo  {journal}
  {Phys. Rep.}\ }\textbf {\bibinfo {volume} {276}},\ \bibinfo {pages} {85--176}
  (\bibinfo {year} {1996})}\BibitemShut {NoStop}%
\bibitem [{\citenamefont {Borgonovi}\ \emph {et~al.}(2016)\citenamefont
  {Borgonovi}, \citenamefont {Izrailev}, \citenamefont {Santos},\ and\
  \citenamefont {Zelevinsky}}]{Borgonovi2016}%
  \BibitemOpen
  \bibfield  {author} {\bibinfo {author} {\bibfnamefont {F.}~\bibnamefont
  {Borgonovi}}, \bibinfo {author} {\bibfnamefont {F.~M.}\ \bibnamefont
  {Izrailev}}, \bibinfo {author} {\bibfnamefont {L.~F.}\ \bibnamefont
  {Santos}}, \ and\ \bibinfo {author} {\bibfnamefont {V.~G.}\ \bibnamefont
  {Zelevinsky}},\ }\bibfield  {title} {\enquote {\bibinfo {title} {Quantum
  chaos and thermalization in isolated systems of interacting particles},}\
  }\href {\doibase 10.1016/j.physrep.2016.02.005} {\bibfield  {journal}
  {\bibinfo  {journal} {Phys. Rep.}\ }\textbf {\bibinfo {volume} {626}},\
  \bibinfo {pages} {1} (\bibinfo {year} {2016})}\BibitemShut {NoStop}%
\bibitem [{\citenamefont {Berry}(1977)}]{Berry1977}%
  \BibitemOpen
  \bibfield  {author} {\bibinfo {author} {\bibfnamefont {M~V}\ \bibnamefont
  {Berry}},\ }\bibfield  {title} {\enquote {\bibinfo {title} {Regular and
  irregular semiclassical wavefunctions},}\ }\href {\doibase
  10.1088/0305-4470/10/12/016} {\bibfield  {journal} {\bibinfo  {journal} {J.
  Phys. A: Math. Gen.}\ }\textbf {\bibinfo {volume} {10}},\ \bibinfo {pages}
  {2083--2091} (\bibinfo {year} {1977})}\BibitemShut {NoStop}%
\bibitem [{\citenamefont {Borgonovi}\ \emph {et~al.}(2017)\citenamefont
  {Borgonovi}, \citenamefont {Mattiotti},\ and\ \citenamefont
  {Izrailev}}]{Borgonovi2017}%
  \BibitemOpen
  \bibfield  {author} {\bibinfo {author} {\bibfnamefont {Fausto}\ \bibnamefont
  {Borgonovi}}, \bibinfo {author} {\bibfnamefont {Francesco}\ \bibnamefont
  {Mattiotti}}, \ and\ \bibinfo {author} {\bibfnamefont {Felix~M.}\
  \bibnamefont {Izrailev}},\ }\bibfield  {title} {\enquote {\bibinfo {title}
  {Temperature of a single chaotic eigenstate},}\ }\href {\doibase
  10.1103/PhysRevE.95.042135} {\bibfield  {journal} {\bibinfo  {journal} {Phys.
  Rev. E}\ }\textbf {\bibinfo {volume} {95}},\ \bibinfo {pages} {042135}
  (\bibinfo {year} {2017})}\BibitemShut {NoStop}%
\bibitem [{\citenamefont {Wang}\ \emph {et~al.}(2020)\citenamefont {Wang},
  \citenamefont {Benenti}, \citenamefont {Casati},\ and\ \citenamefont
  {Wang}}]{Wang2020b}%
  \BibitemOpen
  \bibfield  {author} {\bibinfo {author} {\bibfnamefont {Jiaozi}\ \bibnamefont
  {Wang}}, \bibinfo {author} {\bibfnamefont {Giuliano}\ \bibnamefont
  {Benenti}}, \bibinfo {author} {\bibfnamefont {Giulio}\ \bibnamefont
  {Casati}}, \ and\ \bibinfo {author} {\bibfnamefont {Wen-ge}\ \bibnamefont
  {Wang}},\ }\bibfield  {title} {\enquote {\bibinfo {title} {Complexity of
  quantum motion and quantum-classical correspondence: A phase-space
  approach},}\ }\href {\doibase 10.1103/PhysRevResearch.2.043178} {\bibfield
  {journal} {\bibinfo  {journal} {Phys. Rev. Research}\ }\textbf {\bibinfo
  {volume} {2}},\ \bibinfo {pages} {043178} (\bibinfo {year}
  {2020})}\BibitemShut {NoStop}%
\bibitem [{\citenamefont {Balachandran}\ \emph {et~al.}(2021)\citenamefont
  {Balachandran}, \citenamefont {Benenti}, \citenamefont {Casati},\ and\
  \citenamefont {Poletti}}]{Balachandran2021}%
  \BibitemOpen
  \bibfield  {author} {\bibinfo {author} {\bibfnamefont {Vinitha}\ \bibnamefont
  {Balachandran}}, \bibinfo {author} {\bibfnamefont {Giuliano}\ \bibnamefont
  {Benenti}}, \bibinfo {author} {\bibfnamefont {Giulio}\ \bibnamefont
  {Casati}}, \ and\ \bibinfo {author} {\bibfnamefont {Dario}\ \bibnamefont
  {Poletti}},\ }\bibfield  {title} {\enquote {\bibinfo {title} {From the
  eigenstate thermalization hypothesis to algebraic relaxation of otocs in
  systems with conserved quantities},}\ }\href {\doibase
  10.1103/PhysRevB.104.104306} {\bibfield  {journal} {\bibinfo  {journal}
  {Phys. Rev. B}\ }\textbf {\bibinfo {volume} {104}},\ \bibinfo {pages}
  {104306} (\bibinfo {year} {2021})}\BibitemShut {NoStop}%
\bibitem [{\citenamefont {Wang}\ \emph
  {et~al.}(2022{\natexlab{a}})\citenamefont {Wang}, \citenamefont {Casati},\
  and\ \citenamefont {Benenti}}]{Wang2022a}%
  \BibitemOpen
  \bibfield  {author} {\bibinfo {author} {\bibfnamefont {Jiao}\ \bibnamefont
  {Wang}}, \bibinfo {author} {\bibfnamefont {Giulio}\ \bibnamefont {Casati}}, \
  and\ \bibinfo {author} {\bibfnamefont {Giuliano}\ \bibnamefont {Benenti}},\
  }\bibfield  {title} {\enquote {\bibinfo {title} {Classical physics and
  blackbody radiation},}\ }\href {\doibase 10.1103/PhysRevLett.128.134101}
  {\bibfield  {journal} {\bibinfo  {journal} {Phys. Rev. Lett.}\ }\textbf
  {\bibinfo {volume} {128}},\ \bibinfo {pages} {134101} (\bibinfo {year}
  {2022}{\natexlab{a}})}\BibitemShut {NoStop}%
\bibitem [{\citenamefont {Wang}\ \emph
  {et~al.}(2022{\natexlab{b}})\citenamefont {Wang}, \citenamefont {Benenti},
  \citenamefont {Casati},\ and\ \citenamefont {ge~Wang}}]{Wang2022b}%
  \BibitemOpen
  \bibfield  {author} {\bibinfo {author} {\bibfnamefont {Jiaozi}\ \bibnamefont
  {Wang}}, \bibinfo {author} {\bibfnamefont {Giuliano}\ \bibnamefont
  {Benenti}}, \bibinfo {author} {\bibfnamefont {Giulio}\ \bibnamefont
  {Casati}}, \ and\ \bibinfo {author} {\bibfnamefont {Wen}\ \bibnamefont
  {ge~Wang}},\ }\bibfield  {title} {\enquote {\bibinfo {title} {Statistical and
  dynamical properties of the quantum triangle map},}\ }\href {\doibase
  10.1088/1751-8121/ac6a93} {\bibfield  {journal} {\bibinfo  {journal} {J.
  Phys. A}\ }\textbf {\bibinfo {volume} {55}},\ \bibinfo {pages} {234002}
  (\bibinfo {year} {2022}{\natexlab{b}})}\BibitemShut {NoStop}%
\bibitem [{\citenamefont {Santos}\ \emph {et~al.}(2012)\citenamefont {Santos},
  \citenamefont {Borgonovi},\ and\ \citenamefont {Izrailev}}]{Santos2012PRL}%
  \BibitemOpen
  \bibfield  {author} {\bibinfo {author} {\bibfnamefont {L.~F.}\ \bibnamefont
  {Santos}}, \bibinfo {author} {\bibfnamefont {F.}~\bibnamefont {Borgonovi}}, \
  and\ \bibinfo {author} {\bibfnamefont {F.~M.}\ \bibnamefont {Izrailev}},\
  }\bibfield  {title} {\enquote {\bibinfo {title} {Chaos and statistical
  relaxation in quantum systems of interacting particles},}\ }\href@noop {}
  {\bibfield  {journal} {\bibinfo  {journal} {Phys. Rev. Lett.}\ }\textbf
  {\bibinfo {volume} {108}},\ \bibinfo {pages} {094102} (\bibinfo {year}
  {2012})}\BibitemShut {NoStop}%
\bibitem [{\citenamefont {Borgonovi}\ \emph
  {et~al.}(2019{\natexlab{a}})\citenamefont {Borgonovi}, \citenamefont
  {Izrailev},\ and\ \citenamefont {Santos}}]{Borgonovi2019a}%
  \BibitemOpen
  \bibfield  {author} {\bibinfo {author} {\bibfnamefont {Fausto}\ \bibnamefont
  {Borgonovi}}, \bibinfo {author} {\bibfnamefont {Felix~M.}\ \bibnamefont
  {Izrailev}}, \ and\ \bibinfo {author} {\bibfnamefont {Lea~F.}\ \bibnamefont
  {Santos}},\ }\bibfield  {title} {\enquote {\bibinfo {title} {Exponentially
  fast dynamics of chaotic many-body systems},}\ }\href {\doibase
  10.1103/PhysRevE.99.010101} {\bibfield  {journal} {\bibinfo  {journal} {Phys.
  Rev. E}\ }\textbf {\bibinfo {volume} {99}},\ \bibinfo {pages} {010101}
  (\bibinfo {year} {2019}{\natexlab{a}})}\BibitemShut {NoStop}%
\bibitem [{\citenamefont {Borgonovi}\ \emph
  {et~al.}(2019{\natexlab{b}})\citenamefont {Borgonovi}, \citenamefont
  {Izrailev},\ and\ \citenamefont {Santos}}]{Borgonovi2019b}%
  \BibitemOpen
  \bibfield  {author} {\bibinfo {author} {\bibfnamefont {Fausto}\ \bibnamefont
  {Borgonovi}}, \bibinfo {author} {\bibfnamefont {Felix~M.}\ \bibnamefont
  {Izrailev}}, \ and\ \bibinfo {author} {\bibfnamefont {Lea~F.}\ \bibnamefont
  {Santos}},\ }\bibfield  {title} {\enquote {\bibinfo {title} {Timescales in
  the quench dynamics of many-body quantum systems: Participation ratio versus
  out-of-time ordered correlator},}\ }\href {\doibase
  10.1103/PhysRevE.99.052143} {\bibfield  {journal} {\bibinfo  {journal} {Phys.
  Rev. E}\ }\textbf {\bibinfo {volume} {99}},\ \bibinfo {pages} {052143}
  (\bibinfo {year} {2019}{\natexlab{b}})}\BibitemShut {NoStop}%
\bibitem [{\citenamefont {Dicke}(1954)}]{Dicke1954}%
  \BibitemOpen
  \bibfield  {author} {\bibinfo {author} {\bibfnamefont {R.~H.}\ \bibnamefont
  {Dicke}},\ }\bibfield  {title} {\enquote {\bibinfo {title} {Coherence in
  spontaneous radiation processes},}\ }\href {\doibase 10.1103/PhysRev.93.99}
  {\bibfield  {journal} {\bibinfo  {journal} {Phys. Rev.}\ }\textbf {\bibinfo
  {volume} {93}},\ \bibinfo {pages} {99} (\bibinfo {year} {1954})}\BibitemShut
  {NoStop}%
\bibitem [{\citenamefont {Garraway}(2011)}]{Garraway2011}%
  \BibitemOpen
  \bibfield  {author} {\bibinfo {author} {\bibfnamefont {Barry~M.}\
  \bibnamefont {Garraway}},\ }\bibfield  {title} {\enquote {\bibinfo {title}
  {The {D}icke model in quantum optics: {D}icke model revisited},}\ }\href
  {\doibase 10.1098/rsta.2010.0333} {\bibfield  {journal} {\bibinfo  {journal}
  {Philos. Trans. Royal Soc. A}\ }\textbf {\bibinfo {volume} {369}},\ \bibinfo
  {pages} {1137} (\bibinfo {year} {2011})}\BibitemShut {NoStop}%
\bibitem [{\citenamefont {Kirton}\ \emph {et~al.}(2019)\citenamefont {Kirton},
  \citenamefont {Roses}, \citenamefont {Keeling},\ and\ \citenamefont
  {Dalla~Torre}}]{Kirton2019}%
  \BibitemOpen
  \bibfield  {author} {\bibinfo {author} {\bibfnamefont {Peter}\ \bibnamefont
  {Kirton}}, \bibinfo {author} {\bibfnamefont {Mor~M.}\ \bibnamefont {Roses}},
  \bibinfo {author} {\bibfnamefont {Jonathan}\ \bibnamefont {Keeling}}, \ and\
  \bibinfo {author} {\bibfnamefont {Emanuele~G.}\ \bibnamefont {Dalla~Torre}},\
  }\bibfield  {title} {\enquote {\bibinfo {title} {Introduction to the {D}icke
  model: From equilibrium to nonequilibrium, and vice versa},}\ }\href
  {\doibase 10.1002/qute.201800043} {\bibfield  {journal} {\bibinfo  {journal}
  {Adv. Quantum Technol.}\ }\textbf {\bibinfo {volume} {2}},\ \bibinfo {pages}
  {1800043} (\bibinfo {year} {2019})}\BibitemShut {NoStop}%
\bibitem [{\citenamefont {Emary}\ and\ \citenamefont
  {Brandes}(2003{\natexlab{a}})}]{Emary2003}%
  \BibitemOpen
  \bibfield  {author} {\bibinfo {author} {\bibfnamefont {Clive}\ \bibnamefont
  {Emary}}\ and\ \bibinfo {author} {\bibfnamefont {Tobias}\ \bibnamefont
  {Brandes}},\ }\bibfield  {title} {\enquote {\bibinfo {title} {Chaos and the
  quantum phase transition in the {D}icke model},}\ }\href {\doibase
  10.1103/PhysRevE.67.066203} {\bibfield  {journal} {\bibinfo  {journal} {Phys.
  Rev. E}\ }\textbf {\bibinfo {volume} {67}},\ \bibinfo {pages} {066203}
  (\bibinfo {year} {2003}{\natexlab{a}})}\BibitemShut {NoStop}%
\bibitem [{\citenamefont {Emary}\ and\ \citenamefont
  {Brandes}(2003{\natexlab{b}})}]{Emary2003PRL}%
  \BibitemOpen
  \bibfield  {author} {\bibinfo {author} {\bibfnamefont {Clive}\ \bibnamefont
  {Emary}}\ and\ \bibinfo {author} {\bibfnamefont {Tobias}\ \bibnamefont
  {Brandes}},\ }\bibfield  {title} {\enquote {\bibinfo {title} {Quantum chaos
  triggered by precursors of a quantum phase transition: The {D}icke model},}\
  }\href {\doibase 10.1103/PhysRevLett.90.044101} {\bibfield  {journal}
  {\bibinfo  {journal} {Phys. Rev. Lett.}\ }\textbf {\bibinfo {volume} {90}},\
  \bibinfo {pages} {044101} (\bibinfo {year} {2003}{\natexlab{b}})}\BibitemShut
  {NoStop}%
\bibitem [{\citenamefont {Brandes}(2013)}]{Brandes2013}%
  \BibitemOpen
  \bibfield  {author} {\bibinfo {author} {\bibfnamefont {Tobias}\ \bibnamefont
  {Brandes}},\ }\bibfield  {title} {\enquote {\bibinfo {title} {Excited-state
  quantum phase transitions in {D}icke superradiance models},}\ }\href
  {\doibase 10.1103/PhysRevE.88.032133} {\bibfield  {journal} {\bibinfo
  {journal} {Phys. Rev. E}\ }\textbf {\bibinfo {volume} {88}},\ \bibinfo
  {pages} {032133} (\bibinfo {year} {2013})}\BibitemShut {NoStop}%
\bibitem [{\citenamefont {Furuya}\ \emph {et~al.}(1998)\citenamefont {Furuya},
  \citenamefont {Nemes},\ and\ \citenamefont {Pellegrino}}]{Furuya1998}%
  \BibitemOpen
  \bibfield  {author} {\bibinfo {author} {\bibfnamefont {K.}~\bibnamefont
  {Furuya}}, \bibinfo {author} {\bibfnamefont {M.~C.}\ \bibnamefont {Nemes}}, \
  and\ \bibinfo {author} {\bibfnamefont {G.~Q.}\ \bibnamefont {Pellegrino}},\
  }\bibfield  {title} {\enquote {\bibinfo {title} {Quantum dynamical
  manifestation of chaotic behavior in the process of entanglement},}\ }\href
  {\doibase 10.1103/PhysRevLett.80.5524} {\bibfield  {journal} {\bibinfo
  {journal} {Phys. Rev. Lett.}\ }\textbf {\bibinfo {volume} {80}},\ \bibinfo
  {pages} {5524--5527} (\bibinfo {year} {1998})}\BibitemShut {NoStop}%
\bibitem [{\citenamefont {L\'obez}\ and\ \citenamefont
  {Rela\~no}(2016)}]{Lobez2016}%
  \BibitemOpen
  \bibfield  {author} {\bibinfo {author} {\bibfnamefont {C.~M.}\ \bibnamefont
  {L\'obez}}\ and\ \bibinfo {author} {\bibfnamefont {A.}~\bibnamefont
  {Rela\~no}},\ }\bibfield  {title} {\enquote {\bibinfo {title} {Entropy,
  chaos, and excited-state quantum phase transitions in the dicke model},}\
  }\href {\doibase 10.1103/PhysRevE.94.012140} {\bibfield  {journal} {\bibinfo
  {journal} {Phys. Rev. E}\ }\textbf {\bibinfo {volume} {94}},\ \bibinfo
  {pages} {012140} (\bibinfo {year} {2016})}\BibitemShut {NoStop}%
\bibitem [{\citenamefont {Ch\'avez-Carlos}\ \emph {et~al.}(2016)\citenamefont
  {Ch\'avez-Carlos}, \citenamefont {Bastarrachea-Magnani}, \citenamefont
  {Lerma-Hern\'andez},\ and\ \citenamefont {Hirsch}}]{Chavez2016}%
  \BibitemOpen
  \bibfield  {author} {\bibinfo {author} {\bibfnamefont {J.}~\bibnamefont
  {Ch\'avez-Carlos}}, \bibinfo {author} {\bibfnamefont {M.~A.}\ \bibnamefont
  {Bastarrachea-Magnani}}, \bibinfo {author} {\bibfnamefont {S.}~\bibnamefont
  {Lerma-Hern\'andez}}, \ and\ \bibinfo {author} {\bibfnamefont {J.~G.}\
  \bibnamefont {Hirsch}},\ }\bibfield  {title} {\enquote {\bibinfo {title}
  {Classical chaos in atom-field systems},}\ }\href {\doibase
  10.1103/PhysRevE.94.022209} {\bibfield  {journal} {\bibinfo  {journal} {Phys.
  Rev. E}\ }\textbf {\bibinfo {volume} {94}},\ \bibinfo {pages} {022209}
  (\bibinfo {year} {2016})}\BibitemShut {NoStop}%
\bibitem [{\citenamefont {Sinha}\ and\ \citenamefont
  {Sinha}(2020)}]{Sinha2020}%
  \BibitemOpen
  \bibfield  {author} {\bibinfo {author} {\bibfnamefont {Sudip}\ \bibnamefont
  {Sinha}}\ and\ \bibinfo {author} {\bibfnamefont {S.}~\bibnamefont {Sinha}},\
  }\bibfield  {title} {\enquote {\bibinfo {title} {Chaos and quantum scars in
  bose-josephson junction coupled to a bosonic mode},}\ }\href {\doibase
  10.1103/PhysRevLett.125.134101} {\bibfield  {journal} {\bibinfo  {journal}
  {Phys. Rev. Lett.}\ }\textbf {\bibinfo {volume} {125}},\ \bibinfo {pages}
  {134101} (\bibinfo {year} {2020})}\BibitemShut {NoStop}%
\bibitem [{\citenamefont {Valencia-Tortora}\ \emph {et~al.}(2022)\citenamefont
  {Valencia-Tortora}, \citenamefont {Kelly}, \citenamefont {Donner},
  \citenamefont {Morigi}, \citenamefont {Fazio},\ and\ \citenamefont
  {Marino}}]{Valencia2022}%
  \BibitemOpen
  \bibfield  {author} {\bibinfo {author} {\bibfnamefont {Riccardo~J.}\
  \bibnamefont {Valencia-Tortora}}, \bibinfo {author} {\bibfnamefont
  {Shane~P.}\ \bibnamefont {Kelly}}, \bibinfo {author} {\bibfnamefont {Tobias}\
  \bibnamefont {Donner}}, \bibinfo {author} {\bibfnamefont {Giovanna}\
  \bibnamefont {Morigi}}, \bibinfo {author} {\bibfnamefont {Rosario}\
  \bibnamefont {Fazio}}, \ and\ \bibinfo {author} {\bibfnamefont {Jamir}\
  \bibnamefont {Marino}},\ }\href {\doibase 10.48550/ARXIV.2210.14224}
  {\enquote {\bibinfo {title} {Crafting the dynamical structure of
  synchronization by harnessing bosonic multi-level cavity qed},}\ } (\bibinfo
  {year} {2022})\BibitemShut {NoStop}%
\bibitem [{\citenamefont {Altland}\ and\ \citenamefont
  {Haake}(2012)}]{Altland2012NJP}%
  \BibitemOpen
  \bibfield  {author} {\bibinfo {author} {\bibfnamefont {Alexander}\
  \bibnamefont {Altland}}\ and\ \bibinfo {author} {\bibfnamefont {Fritz}\
  \bibnamefont {Haake}},\ }\bibfield  {title} {\enquote {\bibinfo {title}
  {Equilibration and macroscopic quantum fluctuations in the dicke model},}\
  }\href {\doibase 10.1088/1367-2630/14/7/073011} {\bibfield  {journal}
  {\bibinfo  {journal} {New J. Phys.}\ }\textbf {\bibinfo {volume} {14}},\
  \bibinfo {pages} {073011} (\bibinfo {year} {2012})}\BibitemShut {NoStop}%
\bibitem [{\citenamefont {Kloc}\ \emph {et~al.}(2018)\citenamefont {Kloc},
  \citenamefont {Str\'ansk\'y},\ and\ \citenamefont {Cejnar}}]{Kloc2018}%
  \BibitemOpen
  \bibfield  {author} {\bibinfo {author} {\bibfnamefont {Michal}\ \bibnamefont
  {Kloc}}, \bibinfo {author} {\bibfnamefont {Pavel}\ \bibnamefont
  {Str\'ansk\'y}}, \ and\ \bibinfo {author} {\bibfnamefont {Pavel}\
  \bibnamefont {Cejnar}},\ }\bibfield  {title} {\enquote {\bibinfo {title}
  {Quantum quench dynamics in {D}icke superradiance models},}\ }\href {\doibase
  10.1103/PhysRevA.98.013836} {\bibfield  {journal} {\bibinfo  {journal} {Phys.
  Rev. A}\ }\textbf {\bibinfo {volume} {98}},\ \bibinfo {pages} {013836}
  (\bibinfo {year} {2018})}\BibitemShut {NoStop}%
\bibitem [{\citenamefont {Lerma-Hern{\'{a}}ndez}\ \emph
  {et~al.}(2018)\citenamefont {Lerma-Hern{\'{a}}ndez}, \citenamefont
  {Ch{\'{a}}vez-Carlos}, \citenamefont {Bastarrachea-Magnani}, \citenamefont
  {Santos},\ and\ \citenamefont {Hirsch}}]{Lerma2018}%
  \BibitemOpen
  \bibfield  {author} {\bibinfo {author} {\bibfnamefont {Sergio}\ \bibnamefont
  {Lerma-Hern{\'{a}}ndez}}, \bibinfo {author} {\bibfnamefont {Jorge}\
  \bibnamefont {Ch{\'{a}}vez-Carlos}}, \bibinfo {author} {\bibfnamefont
  {Miguel~A}\ \bibnamefont {Bastarrachea-Magnani}}, \bibinfo {author}
  {\bibfnamefont {Lea~F}\ \bibnamefont {Santos}}, \ and\ \bibinfo {author}
  {\bibfnamefont {Jorge~G}\ \bibnamefont {Hirsch}},\ }\bibfield  {title}
  {\enquote {\bibinfo {title} {Analytical description of the survival
  probability of coherent states in regular regimes},}\ }\href {\doibase
  10.1088/1751-8121/aae2c3} {\bibfield  {journal} {\bibinfo  {journal} {J.
  Phys. A: Math. Theor.}\ }\textbf {\bibinfo {volume} {51}},\ \bibinfo {pages}
  {475302} (\bibinfo {year} {2018})}\BibitemShut {NoStop}%
\bibitem [{\citenamefont {Lerma-Hern\'andez}\ \emph {et~al.}(2019)\citenamefont
  {Lerma-Hern\'andez}, \citenamefont {Villase\~nor}, \citenamefont
  {Bastarrachea-Magnani}, \citenamefont {Torres-Herrera}, \citenamefont
  {Santos},\ and\ \citenamefont {Hirsch}}]{Lerma2019}%
  \BibitemOpen
  \bibfield  {author} {\bibinfo {author} {\bibfnamefont {S.}~\bibnamefont
  {Lerma-Hern\'andez}}, \bibinfo {author} {\bibfnamefont {D.}~\bibnamefont
  {Villase\~nor}}, \bibinfo {author} {\bibfnamefont {M.~A.}\ \bibnamefont
  {Bastarrachea-Magnani}}, \bibinfo {author} {\bibfnamefont {E.~J.}\
  \bibnamefont {Torres-Herrera}}, \bibinfo {author} {\bibfnamefont {L.~F.}\
  \bibnamefont {Santos}}, \ and\ \bibinfo {author} {\bibfnamefont {J.~G.}\
  \bibnamefont {Hirsch}},\ }\bibfield  {title} {\enquote {\bibinfo {title}
  {Dynamical signatures of quantum chaos and relaxation time scales in a
  spin-boson system},}\ }\href {\doibase 10.1103/PhysRevE.100.012218}
  {\bibfield  {journal} {\bibinfo  {journal} {Phys. Rev. E}\ }\textbf {\bibinfo
  {volume} {100}},\ \bibinfo {pages} {012218} (\bibinfo {year}
  {2019})}\BibitemShut {NoStop}%
\bibitem [{\citenamefont {Villase{\~{n}}or}\ \emph {et~al.}(2020)\citenamefont
  {Villase{\~{n}}or}, \citenamefont {Pilatowsky-Cameo}, \citenamefont
  {Bastarrachea-Magnani}, \citenamefont {Lerma-Hern{\'{a}}ndez}, \citenamefont
  {Santos},\ and\ \citenamefont {Hirsch}}]{Villasenor2020}%
  \BibitemOpen
  \bibfield  {author} {\bibinfo {author} {\bibfnamefont {D}~\bibnamefont
  {Villase{\~{n}}or}}, \bibinfo {author} {\bibfnamefont {S}~\bibnamefont
  {Pilatowsky-Cameo}}, \bibinfo {author} {\bibfnamefont {M~A}\ \bibnamefont
  {Bastarrachea-Magnani}}, \bibinfo {author} {\bibfnamefont {S}~\bibnamefont
  {Lerma-Hern{\'{a}}ndez}}, \bibinfo {author} {\bibfnamefont {L~F}\
  \bibnamefont {Santos}}, \ and\ \bibinfo {author} {\bibfnamefont {J~G}\
  \bibnamefont {Hirsch}},\ }\bibfield  {title} {\enquote {\bibinfo {title}
  {Quantum vs classical dynamics in a spin-boson system: manifestations of
  spectral correlations and scarring},}\ }\href {\doibase
  10.1088/1367-2630/ab8ef8} {\bibfield  {journal} {\bibinfo  {journal} {New J.
  Phys.}\ }\textbf {\bibinfo {volume} {22}},\ \bibinfo {pages} {063036}
  (\bibinfo {year} {2020})}\BibitemShut {NoStop}%
\bibitem [{\citenamefont {Ch\'avez-Carlos}\ \emph {et~al.}(2019)\citenamefont
  {Ch\'avez-Carlos}, \citenamefont {L\'opez-del Carpio}, \citenamefont
  {Bastarrachea-Magnani}, \citenamefont {Str\'ansk\'y}, \citenamefont
  {Lerma-Hern\'andez}, \citenamefont {Santos},\ and\ \citenamefont
  {Hirsch}}]{Chavez2019}%
  \BibitemOpen
  \bibfield  {author} {\bibinfo {author} {\bibfnamefont {Jorge}\ \bibnamefont
  {Ch\'avez-Carlos}}, \bibinfo {author} {\bibfnamefont {B.}~\bibnamefont
  {L\'opez-del Carpio}}, \bibinfo {author} {\bibfnamefont {Miguel~A.}\
  \bibnamefont {Bastarrachea-Magnani}}, \bibinfo {author} {\bibfnamefont
  {Pavel}\ \bibnamefont {Str\'ansk\'y}}, \bibinfo {author} {\bibfnamefont
  {Sergio}\ \bibnamefont {Lerma-Hern\'andez}}, \bibinfo {author} {\bibfnamefont
  {Lea~F.}\ \bibnamefont {Santos}}, \ and\ \bibinfo {author} {\bibfnamefont
  {Jorge~G.}\ \bibnamefont {Hirsch}},\ }\bibfield  {title} {\enquote {\bibinfo
  {title} {Quantum and classical {L}yapunov exponents in atom-field interaction
  systems},}\ }\href {\doibase 10.1103/PhysRevLett.122.024101} {\bibfield
  {journal} {\bibinfo  {journal} {Phys. Rev. Lett.}\ }\textbf {\bibinfo
  {volume} {122}},\ \bibinfo {pages} {024101} (\bibinfo {year}
  {2019})}\BibitemShut {NoStop}%
\bibitem [{\citenamefont {Lewis-Swan}\ \emph {et~al.}(2019)\citenamefont
  {Lewis-Swan}, \citenamefont {Safavi-Naini}, \citenamefont {Bollinger},\ and\
  \citenamefont {Rey}}]{Lewis-Swan2019}%
  \BibitemOpen
  \bibfield  {author} {\bibinfo {author} {\bibfnamefont {R.~J.}\ \bibnamefont
  {Lewis-Swan}}, \bibinfo {author} {\bibfnamefont {A.}~\bibnamefont
  {Safavi-Naini}}, \bibinfo {author} {\bibfnamefont {J.~J.}\ \bibnamefont
  {Bollinger}}, \ and\ \bibinfo {author} {\bibfnamefont {A.~M.}\ \bibnamefont
  {Rey}},\ }\bibfield  {title} {\enquote {\bibinfo {title} {Unifying
  thermalization and entanglement through measurement of fidelity
  out-of-time-order correlators in the {D}icke model},}\ }\href {\doibase
  10.1038/s41467-019-09436-y} {\bibfield  {journal} {\bibinfo  {journal} {Nat.
  Comm.}\ }\textbf {\bibinfo {volume} {10}},\ \bibinfo {pages} {1581} (\bibinfo
  {year} {2019})}\BibitemShut {NoStop}%
\bibitem [{\citenamefont {Pilatowsky-Cameo}\ \emph {et~al.}(2020)\citenamefont
  {Pilatowsky-Cameo}, \citenamefont {Ch\'avez-Carlos}, \citenamefont
  {Bastarrachea-Magnani}, \citenamefont {Str\'ansk\'y}, \citenamefont
  {Lerma-Hern\'andez}, \citenamefont {Santos},\ and\ \citenamefont
  {Hirsch}}]{Pilatowsky2020}%
  \BibitemOpen
  \bibfield  {author} {\bibinfo {author} {\bibfnamefont {Sa\'ul}\ \bibnamefont
  {Pilatowsky-Cameo}}, \bibinfo {author} {\bibfnamefont {Jorge}\ \bibnamefont
  {Ch\'avez-Carlos}}, \bibinfo {author} {\bibfnamefont {Miguel~A.}\
  \bibnamefont {Bastarrachea-Magnani}}, \bibinfo {author} {\bibfnamefont
  {Pavel}\ \bibnamefont {Str\'ansk\'y}}, \bibinfo {author} {\bibfnamefont
  {Sergio}\ \bibnamefont {Lerma-Hern\'andez}}, \bibinfo {author} {\bibfnamefont
  {Lea~F.}\ \bibnamefont {Santos}}, \ and\ \bibinfo {author} {\bibfnamefont
  {Jorge~G.}\ \bibnamefont {Hirsch}},\ }\bibfield  {title} {\enquote {\bibinfo
  {title} {Positive quantum {L}yapunov exponents in experimental systems with a
  regular classical limit},}\ }\href {\doibase 10.1103/PhysRevE.101.010202}
  {\bibfield  {journal} {\bibinfo  {journal} {Phys. Rev. E}\ }\textbf {\bibinfo
  {volume} {101}},\ \bibinfo {pages} {010202(R)} (\bibinfo {year}
  {2020})}\BibitemShut {NoStop}%
\bibitem [{\citenamefont {de~Aguiar}\ \emph {et~al.}(1992)\citenamefont
  {de~Aguiar}, \citenamefont {Furuya}, \citenamefont {Lewenkopf},\ and\
  \citenamefont {Nemes}}]{Deaguiar1992}%
  \BibitemOpen
  \bibfield  {author} {\bibinfo {author} {\bibfnamefont {M.A.M}\ \bibnamefont
  {de~Aguiar}}, \bibinfo {author} {\bibfnamefont {K}~\bibnamefont {Furuya}},
  \bibinfo {author} {\bibfnamefont {C.H}\ \bibnamefont {Lewenkopf}}, \ and\
  \bibinfo {author} {\bibfnamefont {M.C}\ \bibnamefont {Nemes}},\ }\bibfield
  {title} {\enquote {\bibinfo {title} {Chaos in a spin-boson system: Classical
  analysis},}\ }\href {\doibase https://doi.org/10.1016/0003-4916(92)90178-O}
  {\bibfield  {journal} {\bibinfo  {journal} {Ann. Phys.}\ }\textbf {\bibinfo
  {volume} {216}},\ \bibinfo {pages} {291 -- 312} (\bibinfo {year}
  {1992})}\BibitemShut {NoStop}%
\bibitem [{\citenamefont {Furuya}\ \emph {et~al.}(1992)\citenamefont {Furuya},
  \citenamefont {de~Aguiar}, \citenamefont {Lewenkopf},\ and\ \citenamefont
  {Nemes}}]{Furuya1992}%
  \BibitemOpen
  \bibfield  {author} {\bibinfo {author} {\bibfnamefont {K}~\bibnamefont
  {Furuya}}, \bibinfo {author} {\bibfnamefont {M.A.M}\ \bibnamefont
  {de~Aguiar}}, \bibinfo {author} {\bibfnamefont {C.H}\ \bibnamefont
  {Lewenkopf}}, \ and\ \bibinfo {author} {\bibfnamefont {M.C}\ \bibnamefont
  {Nemes}},\ }\bibfield  {title} {\enquote {\bibinfo {title} {{H}usimi
  distributions of a spin-boson system and the signatures of its classical
  dynamics},}\ }\href {\doibase 10.1016/0003-4916(92)90179-p} {\bibfield
  {journal} {\bibinfo  {journal} {Ann. of Phys.}\ }\textbf {\bibinfo {volume}
  {216}},\ \bibinfo {pages} {313--322} (\bibinfo {year} {1992})}\BibitemShut
  {NoStop}%
\bibitem [{\citenamefont {Bakemeier}\ \emph {et~al.}(2013)\citenamefont
  {Bakemeier}, \citenamefont {Alvermann},\ and\ \citenamefont
  {Fehske}}]{Bakemeier2013}%
  \BibitemOpen
  \bibfield  {author} {\bibinfo {author} {\bibfnamefont {L.}~\bibnamefont
  {Bakemeier}}, \bibinfo {author} {\bibfnamefont {A.}~\bibnamefont
  {Alvermann}}, \ and\ \bibinfo {author} {\bibfnamefont {H.}~\bibnamefont
  {Fehske}},\ }\bibfield  {title} {\enquote {\bibinfo {title} {Dynamics of the
  {D}icke model close to the classical limit},}\ }\href {\doibase
  10.1103/PhysRevA.88.043835} {\bibfield  {journal} {\bibinfo  {journal} {Phys.
  Rev. A}\ }\textbf {\bibinfo {volume} {88}},\ \bibinfo {pages} {043835}
  (\bibinfo {year} {2013})}\BibitemShut {NoStop}%
\bibitem [{\citenamefont {Pilatowsky-Cameo}\ \emph
  {et~al.}(2021{\natexlab{a}})\citenamefont {Pilatowsky-Cameo}, \citenamefont
  {Villase{\~{n}}or}, \citenamefont {Bastarrachea-Magnani}, \citenamefont
  {Lerma-Hern{\'{a}}ndez}, \citenamefont {Santos},\ and\ \citenamefont
  {Hirsch}}]{Pilatowsky2021NatCommun}%
  \BibitemOpen
  \bibfield  {author} {\bibinfo {author} {\bibfnamefont {Sa{\'{u}}l}\
  \bibnamefont {Pilatowsky-Cameo}}, \bibinfo {author} {\bibfnamefont {David}\
  \bibnamefont {Villase{\~{n}}or}}, \bibinfo {author} {\bibfnamefont
  {Miguel~A.}\ \bibnamefont {Bastarrachea-Magnani}}, \bibinfo {author}
  {\bibfnamefont {Sergio}\ \bibnamefont {Lerma-Hern{\'{a}}ndez}}, \bibinfo
  {author} {\bibfnamefont {Lea~F.}\ \bibnamefont {Santos}}, \ and\ \bibinfo
  {author} {\bibfnamefont {Jorge~G.}\ \bibnamefont {Hirsch}},\ }\bibfield
  {title} {\enquote {\bibinfo {title} {Ubiquitous quantum scarring does not
  prevent ergodicity},}\ }\href {\doibase 10.1038/s41467-021-21123-5}
  {\bibfield  {journal} {\bibinfo  {journal} {Nat. Commun.}\ }\textbf {\bibinfo
  {volume} {12}} (\bibinfo {year} {2021}{\natexlab{a}}),\
  10.1038/s41467-021-21123-5}\BibitemShut {NoStop}%
\bibitem [{\citenamefont {Pilatowsky-Cameo}\ \emph
  {et~al.}(2021{\natexlab{b}})\citenamefont {Pilatowsky-Cameo}, \citenamefont
  {Villase{\~{n}}or}, \citenamefont {Bastarrachea-Magnani}, \citenamefont
  {Lerma-Hern\'andez}, \citenamefont {Santos},\ and\ \citenamefont
  {Hirsch}}]{Pilatowsky2021}%
  \BibitemOpen
  \bibfield  {author} {\bibinfo {author} {\bibfnamefont {Sa\'ul}\ \bibnamefont
  {Pilatowsky-Cameo}}, \bibinfo {author} {\bibfnamefont {David}\ \bibnamefont
  {Villase{\~{n}}or}}, \bibinfo {author} {\bibfnamefont {Miguel~A.}\
  \bibnamefont {Bastarrachea-Magnani}}, \bibinfo {author} {\bibfnamefont
  {Sergio}\ \bibnamefont {Lerma-Hern\'andez}}, \bibinfo {author} {\bibfnamefont
  {Lea~F.}\ \bibnamefont {Santos}}, \ and\ \bibinfo {author} {\bibfnamefont
  {Jorge~G.}\ \bibnamefont {Hirsch}},\ }\bibfield  {title} {\enquote {\bibinfo
  {title} {Quantum scarring in a spin-boson system: fundamental families of
  periodic orbits},}\ }\href {\doibase 10.1088/1367-2630/abd2e6} {\bibfield
  {journal} {\bibinfo  {journal} {New J. Phys.}\ }\textbf {\bibinfo {volume}
  {23}},\ \bibinfo {pages} {033045} (\bibinfo {year}
  {2021}{\natexlab{b}})}\BibitemShut {NoStop}%
\bibitem [{\citenamefont {Wang}\ and\ \citenamefont {Robnik}(2020)}]{Wang2020}%
  \BibitemOpen
  \bibfield  {author} {\bibinfo {author} {\bibfnamefont {Qian}\ \bibnamefont
  {Wang}}\ and\ \bibinfo {author} {\bibfnamefont {Marko}\ \bibnamefont
  {Robnik}},\ }\bibfield  {title} {\enquote {\bibinfo {title} {Statistical
  properties of the localization measure of chaotic eigenstates in the {D}icke
  model},}\ }\href {\doibase 10.1103/PhysRevE.102.032212} {\bibfield  {journal}
  {\bibinfo  {journal} {Phys. Rev. E}\ }\textbf {\bibinfo {volume} {102}},\
  \bibinfo {pages} {032212} (\bibinfo {year} {2020})}\BibitemShut {NoStop}%
\bibitem [{\citenamefont {Jaako}\ \emph {et~al.}(2016)\citenamefont {Jaako},
  \citenamefont {Xiang}, \citenamefont {Garcia-Ripoll},\ and\ \citenamefont
  {Rabl}}]{Jaako2016}%
  \BibitemOpen
  \bibfield  {author} {\bibinfo {author} {\bibfnamefont {Tuomas}\ \bibnamefont
  {Jaako}}, \bibinfo {author} {\bibfnamefont {Ze-Liang}\ \bibnamefont {Xiang}},
  \bibinfo {author} {\bibfnamefont {Juan~Jos\'e}\ \bibnamefont
  {Garcia-Ripoll}}, \ and\ \bibinfo {author} {\bibfnamefont {Peter}\
  \bibnamefont {Rabl}},\ }\bibfield  {title} {\enquote {\bibinfo {title}
  {Ultrastrong-coupling phenomena beyond the {D}icke model},}\ }\href {\doibase
  10.1103/PhysRevA.94.033850} {\bibfield  {journal} {\bibinfo  {journal} {Phys.
  Rev. A}\ }\textbf {\bibinfo {volume} {94}},\ \bibinfo {pages} {033850}
  (\bibinfo {year} {2016})}\BibitemShut {NoStop}%
\bibitem [{\citenamefont {Baden}\ \emph {et~al.}(2014)\citenamefont {Baden},
  \citenamefont {Arnold}, \citenamefont {Grimsmo}, \citenamefont {Parkins},\
  and\ \citenamefont {Barrett}}]{Baden2014}%
  \BibitemOpen
  \bibfield  {author} {\bibinfo {author} {\bibfnamefont {Markus~P.}\
  \bibnamefont {Baden}}, \bibinfo {author} {\bibfnamefont {Kyle~J.}\
  \bibnamefont {Arnold}}, \bibinfo {author} {\bibfnamefont {Arne~L.}\
  \bibnamefont {Grimsmo}}, \bibinfo {author} {\bibfnamefont {Scott}\
  \bibnamefont {Parkins}}, \ and\ \bibinfo {author} {\bibfnamefont {Murray~D.}\
  \bibnamefont {Barrett}},\ }\bibfield  {title} {\enquote {\bibinfo {title}
  {Realization of the {D}icke model using cavity-assisted {R}aman
  transitions},}\ }\href {\doibase 10.1103/PhysRevLett.113.020408} {\bibfield
  {journal} {\bibinfo  {journal} {Phys. Rev. Lett.}\ }\textbf {\bibinfo
  {volume} {113}},\ \bibinfo {pages} {020408} (\bibinfo {year}
  {2014})}\BibitemShut {NoStop}%
\bibitem [{\citenamefont {Zhang}\ \emph {et~al.}(2018)\citenamefont {Zhang},
  \citenamefont {Lee}, \citenamefont {Kumar}, \citenamefont {Arnold},
  \citenamefont {Masson}, \citenamefont {Grimsmo}, \citenamefont {Parkins},\
  and\ \citenamefont {Barrett}}]{Zhang2018}%
  \BibitemOpen
  \bibfield  {author} {\bibinfo {author} {\bibfnamefont {Zhiqiang}\
  \bibnamefont {Zhang}}, \bibinfo {author} {\bibfnamefont {Chern~Hui}\
  \bibnamefont {Lee}}, \bibinfo {author} {\bibfnamefont {Ravi}\ \bibnamefont
  {Kumar}}, \bibinfo {author} {\bibfnamefont {K.~J.}\ \bibnamefont {Arnold}},
  \bibinfo {author} {\bibfnamefont {Stuart~J.}\ \bibnamefont {Masson}},
  \bibinfo {author} {\bibfnamefont {A.~L.}\ \bibnamefont {Grimsmo}}, \bibinfo
  {author} {\bibfnamefont {A.~S.}\ \bibnamefont {Parkins}}, \ and\ \bibinfo
  {author} {\bibfnamefont {M.~D.}\ \bibnamefont {Barrett}},\ }\bibfield
  {title} {\enquote {\bibinfo {title} {Dicke-model simulation via
  cavity-assisted {R}aman transitions},}\ }\href {\doibase
  10.1103/PhysRevA.97.043858} {\bibfield  {journal} {\bibinfo  {journal} {Phys.
  Rev. A}\ }\textbf {\bibinfo {volume} {97}},\ \bibinfo {pages} {043858}
  (\bibinfo {year} {2018})}\BibitemShut {NoStop}%
\bibitem [{\citenamefont {Cohn}\ \emph {et~al.}(2018)\citenamefont {Cohn},
  \citenamefont {Safavi-Naini}, \citenamefont {Lewis-Swan}, \citenamefont
  {Bohnet}, \citenamefont {Gärttner}, \citenamefont {Gilmore}, \citenamefont
  {Jordan}, \citenamefont {Rey}, \citenamefont {Bollinger},\ and\ \citenamefont
  {Freericks}}]{Cohn2018}%
  \BibitemOpen
  \bibfield  {author} {\bibinfo {author} {\bibfnamefont {J}~\bibnamefont
  {Cohn}}, \bibinfo {author} {\bibfnamefont {A}~\bibnamefont {Safavi-Naini}},
  \bibinfo {author} {\bibfnamefont {R~J}\ \bibnamefont {Lewis-Swan}}, \bibinfo
  {author} {\bibfnamefont {J~G}\ \bibnamefont {Bohnet}}, \bibinfo {author}
  {\bibfnamefont {M}~\bibnamefont {Gärttner}}, \bibinfo {author}
  {\bibfnamefont {K~A}\ \bibnamefont {Gilmore}}, \bibinfo {author}
  {\bibfnamefont {J~E}\ \bibnamefont {Jordan}}, \bibinfo {author}
  {\bibfnamefont {A~M}\ \bibnamefont {Rey}}, \bibinfo {author} {\bibfnamefont
  {J~J}\ \bibnamefont {Bollinger}}, \ and\ \bibinfo {author} {\bibfnamefont
  {J~K}\ \bibnamefont {Freericks}},\ }\bibfield  {title} {\enquote {\bibinfo
  {title} {Bang-bang shortcut to adiabaticity in the dicke model as realized in
  a penning trap experiment},}\ }\href {\doibase 10.1088/1367-2630/aac3fa}
  {\bibfield  {journal} {\bibinfo  {journal} {New J. Phys.}\ }\textbf {\bibinfo
  {volume} {20}},\ \bibinfo {pages} {055013} (\bibinfo {year}
  {2018})}\BibitemShut {NoStop}%
\bibitem [{\citenamefont {Safavi-Naini}\ \emph {et~al.}(2018)\citenamefont
  {Safavi-Naini}, \citenamefont {Lewis-Swan}, \citenamefont {Bohnet},
  \citenamefont {G\"arttner}, \citenamefont {Gilmore}, \citenamefont {Jordan},
  \citenamefont {Cohn}, \citenamefont {Freericks}, \citenamefont {Rey},\ and\
  \citenamefont {Bollinger}}]{Safavi2018}%
  \BibitemOpen
  \bibfield  {author} {\bibinfo {author} {\bibfnamefont {A.}~\bibnamefont
  {Safavi-Naini}}, \bibinfo {author} {\bibfnamefont {R.~J.}\ \bibnamefont
  {Lewis-Swan}}, \bibinfo {author} {\bibfnamefont {J.~G.}\ \bibnamefont
  {Bohnet}}, \bibinfo {author} {\bibfnamefont {M.}~\bibnamefont {G\"arttner}},
  \bibinfo {author} {\bibfnamefont {K.~A.}\ \bibnamefont {Gilmore}}, \bibinfo
  {author} {\bibfnamefont {J.~E.}\ \bibnamefont {Jordan}}, \bibinfo {author}
  {\bibfnamefont {J.}~\bibnamefont {Cohn}}, \bibinfo {author} {\bibfnamefont
  {J.~K.}\ \bibnamefont {Freericks}}, \bibinfo {author} {\bibfnamefont {A.~M.}\
  \bibnamefont {Rey}}, \ and\ \bibinfo {author} {\bibfnamefont {J.~J.}\
  \bibnamefont {Bollinger}},\ }\bibfield  {title} {\enquote {\bibinfo {title}
  {Verification of a many-ion simulator of the {D}icke model through slow
  quenches across a phase transition},}\ }\href {\doibase
  10.1103/PhysRevLett.121.040503} {\bibfield  {journal} {\bibinfo  {journal}
  {Phys. Rev. Lett.}\ }\textbf {\bibinfo {volume} {121}},\ \bibinfo {pages}
  {040503} (\bibinfo {year} {2018})}\BibitemShut {NoStop}%
\bibitem [{\citenamefont {Chelpanova}\ \emph {et~al.}(2021)\citenamefont
  {Chelpanova}, \citenamefont {Lerose}, \citenamefont {Zhang}, \citenamefont
  {Carusotto}, \citenamefont {Tserkovnyak},\ and\ \citenamefont
  {Marino}}]{Chelpanova2021}%
  \BibitemOpen
  \bibfield  {author} {\bibinfo {author} {\bibfnamefont {Oksana}\ \bibnamefont
  {Chelpanova}}, \bibinfo {author} {\bibfnamefont {Alessio}\ \bibnamefont
  {Lerose}}, \bibinfo {author} {\bibfnamefont {Shu}\ \bibnamefont {Zhang}},
  \bibinfo {author} {\bibfnamefont {Iacopo}\ \bibnamefont {Carusotto}},
  \bibinfo {author} {\bibfnamefont {Yaroslav}\ \bibnamefont {Tserkovnyak}}, \
  and\ \bibinfo {author} {\bibfnamefont {Jamir}\ \bibnamefont {Marino}},\
  }\href {\doibase 10.48550/ARXIV.2112.04509} {\enquote {\bibinfo {title}
  {Intertwining of lasing and superradiance under spintronic pumping},}\ }
  (\bibinfo {year} {2021})\BibitemShut {NoStop}%
\bibitem [{\citenamefont {Kirkova}\ and\ \citenamefont
  {Ivanov}(2022)}]{Kirkova2022}%
  \BibitemOpen
  \bibfield  {author} {\bibinfo {author} {\bibfnamefont {Aleksandrina~V.}\
  \bibnamefont {Kirkova}}\ and\ \bibinfo {author} {\bibfnamefont {Peter~A.}\
  \bibnamefont {Ivanov}},\ }\href {\doibase 10.48550/ARXIV.2207.03825}
  {\enquote {\bibinfo {title} {Quantum chaos and thermalization in the two-mode
  dicke model},}\ } (\bibinfo {year} {2022})\BibitemShut {NoStop}%
\bibitem [{\citenamefont {Ray}\ \emph {et~al.}(2016)\citenamefont {Ray},
  \citenamefont {Ghosh},\ and\ \citenamefont {Sinha}}]{Ray2016}%
  \BibitemOpen
  \bibfield  {author} {\bibinfo {author} {\bibfnamefont {S.}~\bibnamefont
  {Ray}}, \bibinfo {author} {\bibfnamefont {A.}~\bibnamefont {Ghosh}}, \ and\
  \bibinfo {author} {\bibfnamefont {S.}~\bibnamefont {Sinha}},\ }\bibfield
  {title} {\enquote {\bibinfo {title} {Quantum signature of chaos and
  thermalization in the kicked dicke model},}\ }\href {\doibase
  10.1103/PhysRevE.94.032103} {\bibfield  {journal} {\bibinfo  {journal} {Phys.
  Rev. E}\ }\textbf {\bibinfo {volume} {94}},\ \bibinfo {pages} {032103}
  (\bibinfo {year} {2016})}\BibitemShut {NoStop}%
\bibitem [{\citenamefont {Hepp}\ and\ \citenamefont
  {Lieb}(1973{\natexlab{a}})}]{Hepp1973a}%
  \BibitemOpen
  \bibfield  {author} {\bibinfo {author} {\bibfnamefont {Klaus}\ \bibnamefont
  {Hepp}}\ and\ \bibinfo {author} {\bibfnamefont {Elliott~H}\ \bibnamefont
  {Lieb}},\ }\bibfield  {title} {\enquote {\bibinfo {title} {On the
  superradiant phase transition for molecules in a quantized radiation field:
  the {D}icke maser model},}\ }\href {\doibase
  https://doi.org/10.1016/0003-4916(73)90039-0} {\bibfield  {journal} {\bibinfo
   {journal} {Ann. Phys. (N.Y.)}\ }\textbf {\bibinfo {volume} {76}},\ \bibinfo
  {pages} {360 -- 404} (\bibinfo {year} {1973}{\natexlab{a}})}\BibitemShut
  {NoStop}%
\bibitem [{\citenamefont {Hepp}\ and\ \citenamefont
  {Lieb}(1973{\natexlab{b}})}]{Hepp1973b}%
  \BibitemOpen
  \bibfield  {author} {\bibinfo {author} {\bibfnamefont {Klaus}\ \bibnamefont
  {Hepp}}\ and\ \bibinfo {author} {\bibfnamefont {Elliott~H.}\ \bibnamefont
  {Lieb}},\ }\bibfield  {title} {\enquote {\bibinfo {title} {Equilibrium
  statistical mechanics of matter interacting with the quantized radiation
  field},}\ }\href {\doibase 10.1103/PhysRevA.8.2517} {\bibfield  {journal}
  {\bibinfo  {journal} {Phys. Rev. A}\ }\textbf {\bibinfo {volume} {8}},\
  \bibinfo {pages} {2517--2525} (\bibinfo {year}
  {1973}{\natexlab{b}})}\BibitemShut {NoStop}%
\bibitem [{\citenamefont {Wang}\ and\ \citenamefont {Hioe}(1973)}]{Wang1973}%
  \BibitemOpen
  \bibfield  {author} {\bibinfo {author} {\bibfnamefont {Y.~K.}\ \bibnamefont
  {Wang}}\ and\ \bibinfo {author} {\bibfnamefont {F.~T.}\ \bibnamefont
  {Hioe}},\ }\bibfield  {title} {\enquote {\bibinfo {title} {Phase transition
  in the {D}icke model of superradiance},}\ }\href {\doibase
  10.1103/PhysRevA.7.831} {\bibfield  {journal} {\bibinfo  {journal} {Phys.
  Rev. A}\ }\textbf {\bibinfo {volume} {7}},\ \bibinfo {pages} {831--836}
  (\bibinfo {year} {1973})}\BibitemShut {NoStop}%
\bibitem [{\citenamefont {Bastarrachea-Magnani}\ \emph
  {et~al.}(2014{\natexlab{a}})\citenamefont {Bastarrachea-Magnani},
  \citenamefont {Lerma-Hern\'andez},\ and\ \citenamefont
  {Hirsch}}]{Bastarrachea2014b}%
  \BibitemOpen
  \bibfield  {author} {\bibinfo {author} {\bibfnamefont {M.~A.}\ \bibnamefont
  {Bastarrachea-Magnani}}, \bibinfo {author} {\bibfnamefont {S.}~\bibnamefont
  {Lerma-Hern\'andez}}, \ and\ \bibinfo {author} {\bibfnamefont {J.~G.}\
  \bibnamefont {Hirsch}},\ }\bibfield  {title} {\enquote {\bibinfo {title}
  {Comparative quantum and semiclassical analysis of atom-field systems. {II}.
  {C}haos and regularity},}\ }\href {\doibase 10.1103/PhysRevA.89.032102}
  {\bibfield  {journal} {\bibinfo  {journal} {Phys. Rev. A}\ }\textbf {\bibinfo
  {volume} {89}},\ \bibinfo {pages} {032102} (\bibinfo {year}
  {2014}{\natexlab{a}})}\BibitemShut {NoStop}%
\bibitem [{\citenamefont {Casati}\ \emph {et~al.}(1980)\citenamefont {Casati},
  \citenamefont {Valz-Gris},\ and\ \citenamefont {Guarnieri}}]{Casati1980}%
  \BibitemOpen
  \bibfield  {author} {\bibinfo {author} {\bibfnamefont {G.}~\bibnamefont
  {Casati}}, \bibinfo {author} {\bibfnamefont {F.}~\bibnamefont {Valz-Gris}}, \
  and\ \bibinfo {author} {\bibfnamefont {I.}~\bibnamefont {Guarnieri}},\
  }\bibfield  {title} {\enquote {\bibinfo {title} {On the connection between
  quantization of nonintegrable systems and statistical theory of spectra},}\
  }\href {\doibase 10.1007/BF02798790} {\bibfield  {journal} {\bibinfo
  {journal} {Lett. Nuov. Cim.}\ }\textbf {\bibinfo {volume} {28}},\ \bibinfo
  {pages} {279--282} (\bibinfo {year} {1980})}\BibitemShut {NoStop}%
\bibitem [{\citenamefont {Bohigas}\ \emph {et~al.}(1984)\citenamefont
  {Bohigas}, \citenamefont {Giannoni},\ and\ \citenamefont
  {Schmit}}]{Bohigas1984}%
  \BibitemOpen
  \bibfield  {author} {\bibinfo {author} {\bibfnamefont {O.}~\bibnamefont
  {Bohigas}}, \bibinfo {author} {\bibfnamefont {M.~J.}\ \bibnamefont
  {Giannoni}}, \ and\ \bibinfo {author} {\bibfnamefont {C.}~\bibnamefont
  {Schmit}},\ }\bibfield  {title} {\enquote {\bibinfo {title} {Characterization
  of chaotic quantum spectra and universality of level fluctuation laws},}\
  }\href {\doibase 10.1103/PhysRevLett.52.1} {\bibfield  {journal} {\bibinfo
  {journal} {Phys. Rev. Lett.}\ }\textbf {\bibinfo {volume} {52}},\ \bibinfo
  {pages} {1--4} (\bibinfo {year} {1984})}\BibitemShut {NoStop}%
\bibitem [{\citenamefont {Mehta}(1991)}]{MehtaBook}%
  \BibitemOpen
  \bibfield  {author} {\bibinfo {author} {\bibfnamefont {M.~L.}\ \bibnamefont
  {Mehta}},\ }\href@noop {} {\emph {\bibinfo {title} {Random Matrices}}}\
  (\bibinfo  {publisher} {Academic Press},\ \bibinfo {address} {Boston},\
  \bibinfo {year} {1991})\BibitemShut {NoStop}%
\bibitem [{\citenamefont {Oganesyan}\ and\ \citenamefont
  {Huse}(2007)}]{Oganesyan2007}%
  \BibitemOpen
  \bibfield  {author} {\bibinfo {author} {\bibfnamefont {Vadim}\ \bibnamefont
  {Oganesyan}}\ and\ \bibinfo {author} {\bibfnamefont {David~A.}\ \bibnamefont
  {Huse}},\ }\bibfield  {title} {\enquote {\bibinfo {title} {Localization of
  interacting fermions at high temperature},}\ }\href {\doibase
  10.1103/PhysRevB.75.155111} {\bibfield  {journal} {\bibinfo  {journal} {Phys.
  Rev. B}\ }\textbf {\bibinfo {volume} {75}},\ \bibinfo {pages} {155111}
  (\bibinfo {year} {2007})}\BibitemShut {NoStop}%
\bibitem [{\citenamefont {Atas}\ \emph {et~al.}(2013)\citenamefont {Atas},
  \citenamefont {Bogomolny}, \citenamefont {Giraud},\ and\ \citenamefont
  {Roux}}]{Atas2013}%
  \BibitemOpen
  \bibfield  {author} {\bibinfo {author} {\bibfnamefont {Y.~Y.}\ \bibnamefont
  {Atas}}, \bibinfo {author} {\bibfnamefont {E.}~\bibnamefont {Bogomolny}},
  \bibinfo {author} {\bibfnamefont {O.}~\bibnamefont {Giraud}}, \ and\ \bibinfo
  {author} {\bibfnamefont {G.}~\bibnamefont {Roux}},\ }\bibfield  {title}
  {\enquote {\bibinfo {title} {Distribution of the ratio of consecutive level
  spacings in random matrix ensembles},}\ }\href {\doibase
  10.1103/PhysRevLett.110.084101} {\bibfield  {journal} {\bibinfo  {journal}
  {Phys. Rev. Lett.}\ }\textbf {\bibinfo {volume} {110}},\ \bibinfo {pages}
  {084101} (\bibinfo {year} {2013})}\BibitemShut {NoStop}%
\bibitem [{\citenamefont {Guhr}\ \emph {et~al.}(1998)\citenamefont {Guhr},
  \citenamefont {M\"uller-Groeling},\ and\ \citenamefont
  {Weidenm\"uller}}]{Guhr1998}%
  \BibitemOpen
  \bibfield  {author} {\bibinfo {author} {\bibfnamefont {T.}~\bibnamefont
  {Guhr}}, \bibinfo {author} {\bibfnamefont {A.}~\bibnamefont
  {M\"uller-Groeling}}, \ and\ \bibinfo {author} {\bibfnamefont {H.~A.}\
  \bibnamefont {Weidenm\"uller}},\ }\bibfield  {title} {\enquote {\bibinfo
  {title} {Random matrix theories in quantum physics: Common concepts},}\
  }\href {\doibase https://doi.org/10.1016/S0370-1573(97)00088-4} {\bibfield
  {journal} {\bibinfo  {journal} {Phys. Rep.}\ }\textbf {\bibinfo {volume}
  {299}},\ \bibinfo {pages} {189} (\bibinfo {year} {1998})}\BibitemShut
  {NoStop}%
\bibitem [{\citenamefont {Bastarrachea-Magnani}\ \emph
  {et~al.}(2015)\citenamefont {Bastarrachea-Magnani}, \citenamefont {del
  Carpio}, \citenamefont {Lerma-Hern{\'{a}}ndez},\ and\ \citenamefont
  {Hirsch}}]{Bastarrachea2015}%
  \BibitemOpen
  \bibfield  {author} {\bibinfo {author} {\bibfnamefont {Miguel~Angel}\
  \bibnamefont {Bastarrachea-Magnani}}, \bibinfo {author} {\bibfnamefont
  {Baldemar~L{\'{o}}pez}\ \bibnamefont {del Carpio}}, \bibinfo {author}
  {\bibfnamefont {Sergio}\ \bibnamefont {Lerma-Hern{\'{a}}ndez}}, \ and\
  \bibinfo {author} {\bibfnamefont {Jorge~G}\ \bibnamefont {Hirsch}},\
  }\bibfield  {title} {\enquote {\bibinfo {title} {Chaos in the dicke model:
  quantum and semiclassical analysis},}\ }\href {\doibase
  10.1088/0031-8949/90/6/068015} {\bibfield  {journal} {\bibinfo  {journal}
  {Phys. Scr.}\ }\textbf {\bibinfo {volume} {90}},\ \bibinfo {pages} {068015}
  (\bibinfo {year} {2015})}\BibitemShut {NoStop}%
\bibitem [{\citenamefont {Wang}\ and\ \citenamefont {Wang}(2018)}]{Wang2018}%
  \BibitemOpen
  \bibfield  {author} {\bibinfo {author} {\bibfnamefont {Jiaozi}\ \bibnamefont
  {Wang}}\ and\ \bibinfo {author} {\bibfnamefont {Wen-ge}\ \bibnamefont
  {Wang}},\ }\bibfield  {title} {\enquote {\bibinfo {title} {Characterization
  of random features of chaotic eigenfunctions in unperturbed basis},}\ }\href
  {\doibase 10.1103/PhysRevE.97.062219} {\bibfield  {journal} {\bibinfo
  {journal} {Phys. Rev. E}\ }\textbf {\bibinfo {volume} {97}},\ \bibinfo
  {pages} {062219} (\bibinfo {year} {2018})}\BibitemShut {NoStop}%
\bibitem [{\citenamefont {Peres}(1984)}]{Peres1984PRL}%
  \BibitemOpen
  \bibfield  {author} {\bibinfo {author} {\bibfnamefont {Asher}\ \bibnamefont
  {Peres}},\ }\bibfield  {title} {\enquote {\bibinfo {title} {New conserved
  quantities and test for regular spectra},}\ }\href {\doibase
  10.1103/PhysRevLett.53.1711} {\bibfield  {journal} {\bibinfo  {journal}
  {Phys. Rev. Lett.}\ }\textbf {\bibinfo {volume} {53}},\ \bibinfo {pages}
  {1711--1713} (\bibinfo {year} {1984})}\BibitemShut {NoStop}%
\bibitem [{Lar()}]{Larkin1969}%
  \BibitemOpen
  \href@noop {} {}\bibinfo {note} {A. Larkin and Yu. N. Ovchinnikov, Zh. Eksp.
  Teor. Fiz. 55, 2262 (1969) [``Quasiclassical Method in the Theory of
  Superconductivity'', Sov. Phys. JETP {\bf 28}, 1200 (1969)].}\BibitemShut
  {Stop}%
\bibitem [{\citenamefont {Maldacena}\ and\ \citenamefont
  {Stanford}(2016)}]{Maldacena2016PRD}%
  \BibitemOpen
  \bibfield  {author} {\bibinfo {author} {\bibfnamefont {Juan}\ \bibnamefont
  {Maldacena}}\ and\ \bibinfo {author} {\bibfnamefont {Douglas}\ \bibnamefont
  {Stanford}},\ }\bibfield  {title} {\enquote {\bibinfo {title} {Remarks on the
  {S}achdev-{Y}e-{K}itaev model},}\ }\href {\doibase
  10.1103/PhysRevD.94.106002} {\bibfield  {journal} {\bibinfo  {journal} {Phys.
  Rev. D}\ }\textbf {\bibinfo {volume} {94}},\ \bibinfo {pages} {106002}
  (\bibinfo {year} {2016})}\BibitemShut {NoStop}%
\bibitem [{\citenamefont {Maldacena}\ \emph {et~al.}(2016)\citenamefont
  {Maldacena}, \citenamefont {Shenker},\ and\ \citenamefont
  {Stanford}}]{Maldacena2016JHEP}%
  \BibitemOpen
  \bibfield  {author} {\bibinfo {author} {\bibfnamefont {Juan}\ \bibnamefont
  {Maldacena}}, \bibinfo {author} {\bibfnamefont {Stephen~H.}\ \bibnamefont
  {Shenker}}, \ and\ \bibinfo {author} {\bibfnamefont {Douglas}\ \bibnamefont
  {Stanford}},\ }\bibfield  {title} {\enquote {\bibinfo {title} {A bound on
  chaos},}\ }\href {\doibase 10.1007/JHEP08(2016)106} {\bibfield  {journal}
  {\bibinfo  {journal} {J. High Energy Phys.}\ }\textbf {\bibinfo {volume}
  {2016}},\ \bibinfo {pages} {106} (\bibinfo {year} {2016})}\BibitemShut
  {NoStop}%
\bibitem [{\citenamefont {Bastarrachea-Magnani}\ and\ \citenamefont
  {Hirsch}(2014{\natexlab{a}})}]{Bastarrachea2014c}%
  \BibitemOpen
  \bibfield  {author} {\bibinfo {author} {\bibfnamefont {Miguel~Angel}\
  \bibnamefont {Bastarrachea-Magnani}}\ and\ \bibinfo {author} {\bibfnamefont
  {Jorge~G}\ \bibnamefont {Hirsch}},\ }\bibfield  {title} {\enquote {\bibinfo
  {title} {Peres lattices and chaos in the dicke model},}\ }\href {\doibase
  10.1088/1742-6596/512/1/012004} {\bibfield  {journal} {\bibinfo  {journal}
  {J. Phys.: Conf. Ser.}\ }\textbf {\bibinfo {volume} {512}},\ \bibinfo {pages}
  {012004} (\bibinfo {year} {2014}{\natexlab{a}})}\BibitemShut {NoStop}%
\bibitem [{\citenamefont {Beugeling}\ \emph {et~al.}(2015)\citenamefont
  {Beugeling}, \citenamefont {Moessner},\ and\ \citenamefont
  {Haque}}]{Beugeling2015}%
  \BibitemOpen
  \bibfield  {author} {\bibinfo {author} {\bibfnamefont {Wouter}\ \bibnamefont
  {Beugeling}}, \bibinfo {author} {\bibfnamefont {Roderich}\ \bibnamefont
  {Moessner}}, \ and\ \bibinfo {author} {\bibfnamefont {Masudul}\ \bibnamefont
  {Haque}},\ }\bibfield  {title} {\enquote {\bibinfo {title} {Off-diagonal
  matrix elements of local operators in many-body quantum systems},}\ }\href
  {\doibase 10.1103/PhysRevE.91.012144} {\bibfield  {journal} {\bibinfo
  {journal} {Phys. Rev. E}\ }\textbf {\bibinfo {volume} {91}},\ \bibinfo
  {pages} {012144} (\bibinfo {year} {2015})}\BibitemShut {NoStop}%
\bibitem [{\citenamefont {Santos}\ \emph {et~al.}(2020)\citenamefont {Santos},
  \citenamefont {P\'erez-Bernal},\ and\ \citenamefont
  {Torres-Herrera}}]{Santos2020}%
  \BibitemOpen
  \bibfield  {author} {\bibinfo {author} {\bibfnamefont {Lea~F.}\ \bibnamefont
  {Santos}}, \bibinfo {author} {\bibfnamefont {Francisco}\ \bibnamefont
  {P\'erez-Bernal}}, \ and\ \bibinfo {author} {\bibfnamefont {E.~Jonathan}\
  \bibnamefont {Torres-Herrera}},\ }\bibfield  {title} {\enquote {\bibinfo
  {title} {Speck of chaos},}\ }\href {\doibase
  10.1103/PhysRevResearch.2.043034} {\bibfield  {journal} {\bibinfo  {journal}
  {Phys. Rev. Research}\ }\textbf {\bibinfo {volume} {2}},\ \bibinfo {pages}
  {043034} (\bibinfo {year} {2020})}\BibitemShut {NoStop}%
\bibitem [{\citenamefont {\L{}yd\ifmmode~\dot{z}\else \.{z}\fi{}ba}\ \emph
  {et~al.}(2021)\citenamefont {\L{}yd\ifmmode~\dot{z}\else \.{z}\fi{}ba},
  \citenamefont {Zhang}, \citenamefont {Rigol},\ and\ \citenamefont
  {Vidmar}}]{Lydzba2021}%
  \BibitemOpen
  \bibfield  {author} {\bibinfo {author} {\bibfnamefont {Patrycja}\
  \bibnamefont {\L{}yd\ifmmode~\dot{z}\else \.{z}\fi{}ba}}, \bibinfo {author}
  {\bibfnamefont {Yicheng}\ \bibnamefont {Zhang}}, \bibinfo {author}
  {\bibfnamefont {Marcos}\ \bibnamefont {Rigol}}, \ and\ \bibinfo {author}
  {\bibfnamefont {Lev}\ \bibnamefont {Vidmar}},\ }\bibfield  {title} {\enquote
  {\bibinfo {title} {Single-particle eigenstate thermalization in
  quantum-chaotic quadratic hamiltonians},}\ }\href {\doibase
  10.1103/PhysRevB.104.214203} {\bibfield  {journal} {\bibinfo  {journal}
  {Phys. Rev. B}\ }\textbf {\bibinfo {volume} {104}},\ \bibinfo {pages}
  {214203} (\bibinfo {year} {2021})}\BibitemShut {NoStop}%
\bibitem [{\citenamefont {Zisling}\ \emph {et~al.}(2021)\citenamefont
  {Zisling}, \citenamefont {Santos},\ and\ \citenamefont {Lev}}]{Zisling2021}%
  \BibitemOpen
  \bibfield  {author} {\bibinfo {author} {\bibfnamefont {Guy}\ \bibnamefont
  {Zisling}}, \bibinfo {author} {\bibfnamefont {Lea~F.}\ \bibnamefont
  {Santos}}, \ and\ \bibinfo {author} {\bibfnamefont {Yevgeny~Bar}\
  \bibnamefont {Lev}},\ }\bibfield  {title} {\enquote {\bibinfo {title} {{How
  many particles make up a chaotic many-body quantum system?}}}\ }\href
  {\doibase 10.21468/SciPostPhys.10.4.088} {\bibfield  {journal} {\bibinfo
  {journal} {SciPost Phys.}\ }\textbf {\bibinfo {volume} {10}},\ \bibinfo
  {pages} {088} (\bibinfo {year} {2021})}\BibitemShut {NoStop}%
\bibitem [{\citenamefont {Wittmann~W.}\ \emph {et~al.}(2022)\citenamefont
  {Wittmann~W.}, \citenamefont {Castro}, \citenamefont {Foerster},\ and\
  \citenamefont {Santos}}]{Wittmann2022}%
  \BibitemOpen
  \bibfield  {author} {\bibinfo {author} {\bibfnamefont {Karin}\ \bibnamefont
  {Wittmann~W.}}, \bibinfo {author} {\bibfnamefont {E.~R.}\ \bibnamefont
  {Castro}}, \bibinfo {author} {\bibfnamefont {Angela}\ \bibnamefont
  {Foerster}}, \ and\ \bibinfo {author} {\bibfnamefont {Lea~F.}\ \bibnamefont
  {Santos}},\ }\bibfield  {title} {\enquote {\bibinfo {title} {Interacting
  bosons in a triple well: Preface of many-body quantum chaos},}\ }\href
  {\doibase 10.1103/PhysRevE.105.034204} {\bibfield  {journal} {\bibinfo
  {journal} {Phys. Rev. E}\ }\textbf {\bibinfo {volume} {105}},\ \bibinfo
  {pages} {034204} (\bibinfo {year} {2022})}\BibitemShut {NoStop}%
\bibitem [{\citenamefont {Khaymovich}\ \emph {et~al.}(2019)\citenamefont
  {Khaymovich}, \citenamefont {Haque},\ and\ \citenamefont
  {McClarty}}]{Khaymovich2019}%
  \BibitemOpen
  \bibfield  {author} {\bibinfo {author} {\bibfnamefont {Ivan~M.}\ \bibnamefont
  {Khaymovich}}, \bibinfo {author} {\bibfnamefont {Masudul}\ \bibnamefont
  {Haque}}, \ and\ \bibinfo {author} {\bibfnamefont {Paul~A.}\ \bibnamefont
  {McClarty}},\ }\bibfield  {title} {\enquote {\bibinfo {title} {Eigenstate
  thermalization, random matrix theory, and behemoths},}\ }\href {\doibase
  10.1103/PhysRevLett.122.070601} {\bibfield  {journal} {\bibinfo  {journal}
  {Phys. Rev. Lett.}\ }\textbf {\bibinfo {volume} {122}},\ \bibinfo {pages}
  {070601} (\bibinfo {year} {2019})}\BibitemShut {NoStop}%
\bibitem [{\citenamefont {Kaneko}\ \emph {et~al.}(2020)\citenamefont {Kaneko},
  \citenamefont {Iyoda},\ and\ \citenamefont {Sagawa}}]{Kaneko2020}%
  \BibitemOpen
  \bibfield  {author} {\bibinfo {author} {\bibfnamefont {Kazuya}\ \bibnamefont
  {Kaneko}}, \bibinfo {author} {\bibfnamefont {Eiki}\ \bibnamefont {Iyoda}}, \
  and\ \bibinfo {author} {\bibfnamefont {Takahiro}\ \bibnamefont {Sagawa}},\
  }\bibfield  {title} {\enquote {\bibinfo {title} {Characterizing complexity of
  many-body quantum dynamics by higher-order eigenstate thermalization},}\
  }\href {\doibase 10.1103/PhysRevA.101.042126} {\bibfield  {journal} {\bibinfo
   {journal} {Phys. Rev. A}\ }\textbf {\bibinfo {volume} {101}},\ \bibinfo
  {pages} {042126} (\bibinfo {year} {2020})}\BibitemShut {NoStop}%
\bibitem [{\citenamefont {Page}(1993)}]{Page1993}%
  \BibitemOpen
  \bibfield  {author} {\bibinfo {author} {\bibfnamefont {Don~N.}\ \bibnamefont
  {Page}},\ }\bibfield  {title} {\enquote {\bibinfo {title} {Average entropy of
  a subsystem},}\ }\href {\doibase 10.1103/PhysRevLett.71.1291} {\bibfield
  {journal} {\bibinfo  {journal} {Phys. Rev. Lett.}\ }\textbf {\bibinfo
  {volume} {71}},\ \bibinfo {pages} {1291--1294} (\bibinfo {year}
  {1993})}\BibitemShut {NoStop}%
\bibitem [{\citenamefont {Miller}\ and\ \citenamefont
  {Sarkar}(1999)}]{Miller1999}%
  \BibitemOpen
  \bibfield  {author} {\bibinfo {author} {\bibfnamefont {Paul~A.}\ \bibnamefont
  {Miller}}\ and\ \bibinfo {author} {\bibfnamefont {Sarben}\ \bibnamefont
  {Sarkar}},\ }\bibfield  {title} {\enquote {\bibinfo {title} {Signatures of
  chaos in the entanglement of two coupled quantum kicked tops},}\ }\href
  {\doibase 10.1103/PhysRevE.60.1542} {\bibfield  {journal} {\bibinfo
  {journal} {Phys. Rev. E}\ }\textbf {\bibinfo {volume} {60}},\ \bibinfo
  {pages} {1542--1550} (\bibinfo {year} {1999})}\BibitemShut {NoStop}%
\bibitem [{\citenamefont {Lakshminarayan}(2001)}]{Lakshminarayan2001}%
  \BibitemOpen
  \bibfield  {author} {\bibinfo {author} {\bibfnamefont {Arul}\ \bibnamefont
  {Lakshminarayan}},\ }\bibfield  {title} {\enquote {\bibinfo {title}
  {Entangling power of quantized chaotic systems},}\ }\href {\doibase
  10.1103/PhysRevE.64.036207} {\bibfield  {journal} {\bibinfo  {journal} {Phys.
  Rev. E}\ }\textbf {\bibinfo {volume} {64}},\ \bibinfo {pages} {036207}
  (\bibinfo {year} {2001})}\BibitemShut {NoStop}%
\bibitem [{\citenamefont {Bandyopadhyay}\ and\ \citenamefont
  {Lakshminarayan}(2002)}]{Bandyopadhyay2002}%
  \BibitemOpen
  \bibfield  {author} {\bibinfo {author} {\bibfnamefont {Jayendra~N.}\
  \bibnamefont {Bandyopadhyay}}\ and\ \bibinfo {author} {\bibfnamefont {Arul}\
  \bibnamefont {Lakshminarayan}},\ }\bibfield  {title} {\enquote {\bibinfo
  {title} {Testing statistical bounds on entanglement using quantum chaos},}\
  }\href {\doibase 10.1103/PhysRevLett.89.060402} {\bibfield  {journal}
  {\bibinfo  {journal} {Phys. Rev. Lett.}\ }\textbf {\bibinfo {volume} {89}},\
  \bibinfo {pages} {060402} (\bibinfo {year} {2002})}\BibitemShut {NoStop}%
\bibitem [{\citenamefont {Bandyopadhyay}\ and\ \citenamefont
  {Lakshminarayan}(2004)}]{Bandyopadhyay2004}%
  \BibitemOpen
  \bibfield  {author} {\bibinfo {author} {\bibfnamefont {Jayendra~N.}\
  \bibnamefont {Bandyopadhyay}}\ and\ \bibinfo {author} {\bibfnamefont {Arul}\
  \bibnamefont {Lakshminarayan}},\ }\bibfield  {title} {\enquote {\bibinfo
  {title} {Entanglement production in coupled chaotic systems: Case of the
  kicked tops},}\ }\href {\doibase 10.1103/PhysRevE.69.016201} {\bibfield
  {journal} {\bibinfo  {journal} {Phys. Rev. E}\ }\textbf {\bibinfo {volume}
  {69}},\ \bibinfo {pages} {016201} (\bibinfo {year} {2004})}\BibitemShut
  {NoStop}%
\bibitem [{\citenamefont {Wang}\ \emph {et~al.}(2004)\citenamefont {Wang},
  \citenamefont {Ghose}, \citenamefont {Sanders},\ and\ \citenamefont
  {Hu}}]{Wang2004}%
  \BibitemOpen
  \bibfield  {author} {\bibinfo {author} {\bibfnamefont {Xiaoguang}\
  \bibnamefont {Wang}}, \bibinfo {author} {\bibfnamefont {Shohini}\
  \bibnamefont {Ghose}}, \bibinfo {author} {\bibfnamefont {Barry~C.}\
  \bibnamefont {Sanders}}, \ and\ \bibinfo {author} {\bibfnamefont {Bambi}\
  \bibnamefont {Hu}},\ }\bibfield  {title} {\enquote {\bibinfo {title}
  {Entanglement as a signature of quantum chaos},}\ }\href {\doibase
  10.1103/PhysRevE.70.016217} {\bibfield  {journal} {\bibinfo  {journal} {Phys.
  Rev. E}\ }\textbf {\bibinfo {volume} {70}},\ \bibinfo {pages} {016217}
  (\bibinfo {year} {2004})}\BibitemShut {NoStop}%
\bibitem [{\citenamefont {Horodecki}\ \emph {et~al.}(2009)\citenamefont
  {Horodecki}, \citenamefont {Horodecki}, \citenamefont {Horodecki},\ and\
  \citenamefont {Horodecki}}]{Horodecki2009}%
  \BibitemOpen
  \bibfield  {author} {\bibinfo {author} {\bibfnamefont {Ryszard}\ \bibnamefont
  {Horodecki}}, \bibinfo {author} {\bibfnamefont {Pawe\l{}}\ \bibnamefont
  {Horodecki}}, \bibinfo {author} {\bibfnamefont {Micha\l{}}\ \bibnamefont
  {Horodecki}}, \ and\ \bibinfo {author} {\bibfnamefont {Karol}\ \bibnamefont
  {Horodecki}},\ }\bibfield  {title} {\enquote {\bibinfo {title} {Quantum
  entanglement},}\ }\href {\doibase 10.1103/RevModPhys.81.865} {\bibfield
  {journal} {\bibinfo  {journal} {Rev. Mod. Phys.}\ }\textbf {\bibinfo {volume}
  {81}},\ \bibinfo {pages} {865--942} (\bibinfo {year} {2009})}\BibitemShut
  {NoStop}%
\bibitem [{\citenamefont {Bastarrachea-Magnani}\ and\ \citenamefont
  {Hirsch}(2014{\natexlab{b}})}]{Bastarrachea2014PSa}%
  \BibitemOpen
  \bibfield  {author} {\bibinfo {author} {\bibfnamefont {Miguel~A}\
  \bibnamefont {Bastarrachea-Magnani}}\ and\ \bibinfo {author} {\bibfnamefont
  {Jorge~G}\ \bibnamefont {Hirsch}},\ }\bibfield  {title} {\enquote {\bibinfo
  {title} {Efficient basis for the dicke model: I. theory and convergence in
  energy},}\ }\href {\doibase 10.1088/0031-8949/2014/t160/014005} {\bibfield
  {journal} {\bibinfo  {journal} {Phys. Scr.}\ }\textbf {\bibinfo {volume}
  {T160}},\ \bibinfo {pages} {014005} (\bibinfo {year}
  {2014}{\natexlab{b}})}\BibitemShut {NoStop}%
\bibitem [{\citenamefont {Hirsch}\ and\ \citenamefont
  {Bastarrachea-Magnani}(2014)}]{Bastarrachea2014PSb}%
  \BibitemOpen
  \bibfield  {author} {\bibinfo {author} {\bibfnamefont {Jorge~G}\ \bibnamefont
  {Hirsch}}\ and\ \bibinfo {author} {\bibfnamefont {Miguel~A}\ \bibnamefont
  {Bastarrachea-Magnani}},\ }\bibfield  {title} {\enquote {\bibinfo {title}
  {Efficient basis for the dicke model: {II}. wave function convergence and
  excited states},}\ }\href {\doibase 10.1088/0031-8949/2014/t160/014018}
  {\bibfield  {journal} {\bibinfo  {journal} {Phys. Scr.}\ }\textbf {\bibinfo
  {volume} {T160}},\ \bibinfo {pages} {014018} (\bibinfo {year}
  {2014})}\BibitemShut {NoStop}%
\bibitem [{\citenamefont {Kloc}\ \emph {et~al.}(2017)\citenamefont {Kloc},
  \citenamefont {Str\'ansk\'y},\ and\ \citenamefont {Cejnar}}]{Kloc2017}%
  \BibitemOpen
  \bibfield  {author} {\bibinfo {author} {\bibfnamefont {Michal}\ \bibnamefont
  {Kloc}}, \bibinfo {author} {\bibfnamefont {Pavel}\ \bibnamefont
  {Str\'ansk\'y}}, \ and\ \bibinfo {author} {\bibfnamefont {Pavel}\
  \bibnamefont {Cejnar}},\ }\bibfield  {title} {\enquote {\bibinfo {title}
  {Quantum phases and entanglement properties of an extended {D}icke model},}\
  }\href {\doibase https://doi.org/10.1016/j.aop.2017.04.005} {\bibfield
  {journal} {\bibinfo  {journal} {Ann. Phys.}\ }\textbf {\bibinfo {volume}
  {382}},\ \bibinfo {pages} {85 -- 111} (\bibinfo {year} {2017})}\BibitemShut
  {NoStop}%
\bibitem [{\citenamefont {Pilatowsky-Cameo}\ \emph {et~al.}(2022)\citenamefont
  {Pilatowsky-Cameo}, \citenamefont {Villase{\~{n}}or}, \citenamefont
  {Bastarrachea-Magnani}, \citenamefont {Lerma-Hern{\'{a}}ndez}, \citenamefont
  {F.~Santos},\ and\ \citenamefont {Hirsch}}]{Pilatowsky2022}%
  \BibitemOpen
  \bibfield  {author} {\bibinfo {author} {\bibfnamefont {Sa{\'{u}}l}\
  \bibnamefont {Pilatowsky-Cameo}}, \bibinfo {author} {\bibfnamefont {David}\
  \bibnamefont {Villase{\~{n}}or}}, \bibinfo {author} {\bibfnamefont
  {Miguel~A.}\ \bibnamefont {Bastarrachea-Magnani}}, \bibinfo {author}
  {\bibfnamefont {Sergio}\ \bibnamefont {Lerma-Hern{\'{a}}ndez}}, \bibinfo
  {author} {\bibfnamefont {Lea}\ \bibnamefont {F.~Santos}}, \ and\ \bibinfo
  {author} {\bibfnamefont {Jorge~G.}\ \bibnamefont {Hirsch}},\ }\bibfield
  {title} {\enquote {\bibinfo {title} {Identification of quantum scars via
  phase-space localization measures},}\ }\href {\doibase
  10.22331/q-2022-02-08-644} {\bibfield  {journal} {\bibinfo  {journal}
  {{Quantum}}\ }\textbf {\bibinfo {volume} {6}},\ \bibinfo {pages} {644}
  (\bibinfo {year} {2022})}\BibitemShut {NoStop}%
\bibitem [{\citenamefont {Chen}\ \emph {et~al.}(2008)\citenamefont {Chen},
  \citenamefont {Zhang}, \citenamefont {Liu},\ and\ \citenamefont
  {Wang}}]{Chen2008}%
  \BibitemOpen
  \bibfield  {author} {\bibinfo {author} {\bibfnamefont {Qing-Hu}\ \bibnamefont
  {Chen}}, \bibinfo {author} {\bibfnamefont {Yu-Yu}\ \bibnamefont {Zhang}},
  \bibinfo {author} {\bibfnamefont {Tao}\ \bibnamefont {Liu}}, \ and\ \bibinfo
  {author} {\bibfnamefont {Ke-Lin}\ \bibnamefont {Wang}},\ }\bibfield  {title}
  {\enquote {\bibinfo {title} {Numerically exact solution to the finite-size
  {D}icke model},}\ }\href {\doibase 10.1103/PhysRevA.78.051801} {\bibfield
  {journal} {\bibinfo  {journal} {Phys. Rev. A}\ }\textbf {\bibinfo {volume}
  {78}},\ \bibinfo {pages} {051801} (\bibinfo {year} {2008})}\BibitemShut
  {NoStop}%
\bibitem [{\citenamefont {Bastarrachea-Magnani}\ and\ \citenamefont
  {Hirsch}(2011)}]{Bastarrachea2011}%
  \BibitemOpen
  \bibfield  {author} {\bibinfo {author} {\bibfnamefont {M.~A.}\ \bibnamefont
  {Bastarrachea-Magnani}}\ and\ \bibinfo {author} {\bibfnamefont {J.~G.}\
  \bibnamefont {Hirsch}},\ }\bibfield  {title} {\enquote {\bibinfo {title}
  {Numerical solutions of the {D}icke {H}amiltonian},}\ }\href@noop {}
  {\bibfield  {journal} {\bibinfo  {journal} {Rev. Mex. Fis. S}\ }\textbf
  {\bibinfo {volume} {57}},\ \bibinfo {pages} {69} (\bibinfo {year}
  {2011})}\BibitemShut {NoStop}%
\bibitem [{\citenamefont {Cahill}\ and\ \citenamefont
  {Glauber}(1969{\natexlab{a}})}]{Cahill1969a}%
  \BibitemOpen
  \bibfield  {author} {\bibinfo {author} {\bibfnamefont {K.~E.}\ \bibnamefont
  {Cahill}}\ and\ \bibinfo {author} {\bibfnamefont {R.~J.}\ \bibnamefont
  {Glauber}},\ }\bibfield  {title} {\enquote {\bibinfo {title} {Ordered
  expansions in boson amplitude operators},}\ }\href {\doibase
  10.1103/PhysRev.177.1857} {\bibfield  {journal} {\bibinfo  {journal} {Phys.
  Rev.}\ }\textbf {\bibinfo {volume} {177}},\ \bibinfo {pages} {1857--1881}
  (\bibinfo {year} {1969}{\natexlab{a}})}\BibitemShut {NoStop}%
\bibitem [{\citenamefont {Cahill}\ and\ \citenamefont
  {Glauber}(1969{\natexlab{b}})}]{Cahill1969b}%
  \BibitemOpen
  \bibfield  {author} {\bibinfo {author} {\bibfnamefont {K.~E.}\ \bibnamefont
  {Cahill}}\ and\ \bibinfo {author} {\bibfnamefont {R.~J.}\ \bibnamefont
  {Glauber}},\ }\bibfield  {title} {\enquote {\bibinfo {title} {Density
  operators and quasiprobability distributions},}\ }\href {\doibase
  10.1103/PhysRev.177.1882} {\bibfield  {journal} {\bibinfo  {journal} {Phys.
  Rev.}\ }\textbf {\bibinfo {volume} {177}},\ \bibinfo {pages} {1882--1902}
  (\bibinfo {year} {1969}{\natexlab{b}})}\BibitemShut {NoStop}%
\bibitem [{\citenamefont {de~Oliveira}\ \emph {et~al.}(1990)\citenamefont
  {de~Oliveira}, \citenamefont {Kim}, \citenamefont {Knight},\ and\
  \citenamefont {Buek}}]{Oliveira1990}%
  \BibitemOpen
  \bibfield  {author} {\bibinfo {author} {\bibfnamefont {F.~A.~M.}\
  \bibnamefont {de~Oliveira}}, \bibinfo {author} {\bibfnamefont {M.~S.}\
  \bibnamefont {Kim}}, \bibinfo {author} {\bibfnamefont {P.~L.}\ \bibnamefont
  {Knight}}, \ and\ \bibinfo {author} {\bibfnamefont {V.}~\bibnamefont
  {Buek}},\ }\bibfield  {title} {\enquote {\bibinfo {title} {Properties of
  displaced number states},}\ }\href {\doibase 10.1103/PhysRevA.41.2645}
  {\bibfield  {journal} {\bibinfo  {journal} {Phys. Rev. A}\ }\textbf {\bibinfo
  {volume} {41}},\ \bibinfo {pages} {2645--2652} (\bibinfo {year}
  {1990})}\BibitemShut {NoStop}%
\bibitem [{\citenamefont {Inc.}()}]{Mathematica}%
  \BibitemOpen
  \bibfield  {author} {\bibinfo {author} {\bibfnamefont {Wolfram~Research{,}}\
  \bibnamefont {Inc.}},\ }\href {https://www.wolfram.com/mathematica} {\enquote
  {\bibinfo {title} {Mathematica, {V}ersion 13.1},}\ }\bibinfo {note}
  {Champaign, IL, 2022}\BibitemShut {NoStop}%
\bibitem [{\citenamefont {de~Aguiar}\ \emph {et~al.}(1991)\citenamefont
  {de~Aguiar}, \citenamefont {Furuya}, \citenamefont {Lewenkopf},\ and\
  \citenamefont {Nemes}}]{Deaguiar1991}%
  \BibitemOpen
  \bibfield  {author} {\bibinfo {author} {\bibfnamefont {M.~A.~M.}\
  \bibnamefont {de~Aguiar}}, \bibinfo {author} {\bibfnamefont {K}~\bibnamefont
  {Furuya}}, \bibinfo {author} {\bibfnamefont {C.~H}\ \bibnamefont
  {Lewenkopf}}, \ and\ \bibinfo {author} {\bibfnamefont {M.~C}\ \bibnamefont
  {Nemes}},\ }\bibfield  {title} {\enquote {\bibinfo {title} {Particle-spin
  coupling in a chaotic system: Localization-delocalization in the husimi
  distributions},}\ }\href {\doibase 10.1209/0295-5075/15/2/003} {\bibfield
  {journal} {\bibinfo  {journal} {EPL}\ }\textbf {\bibinfo {volume} {15}},\
  \bibinfo {pages} {125--131} (\bibinfo {year} {1991})}\BibitemShut {NoStop}%
\bibitem [{\citenamefont {Bastarrachea-Magnani}\ \emph
  {et~al.}(2014{\natexlab{b}})\citenamefont {Bastarrachea-Magnani},
  \citenamefont {Lerma-Hern\'andez},\ and\ \citenamefont
  {Hirsch}}]{Bastarrachea2014a}%
  \BibitemOpen
  \bibfield  {author} {\bibinfo {author} {\bibfnamefont {M.~A.}\ \bibnamefont
  {Bastarrachea-Magnani}}, \bibinfo {author} {\bibfnamefont {S.}~\bibnamefont
  {Lerma-Hern\'andez}}, \ and\ \bibinfo {author} {\bibfnamefont {J.~G.}\
  \bibnamefont {Hirsch}},\ }\bibfield  {title} {\enquote {\bibinfo {title}
  {Comparative quantum and semiclassical analysis of atom-field systems. {I}.
  {D}ensity of states and excited-state quantum phase transitions},}\ }\href
  {\doibase 10.1103/PhysRevA.89.032101} {\bibfield  {journal} {\bibinfo
  {journal} {Phys. Rev. A}\ }\textbf {\bibinfo {volume} {89}},\ \bibinfo
  {pages} {032101} (\bibinfo {year} {2014}{\natexlab{b}})}\BibitemShut
  {NoStop}%
\bibitem [{\citenamefont {Ribeiro}\ \emph {et~al.}(2006)\citenamefont
  {Ribeiro}, \citenamefont {de~Aguiar},\ and\ \citenamefont
  {de~Toledo~Piza}}]{Ribeiro2006}%
  \BibitemOpen
  \bibfield  {author} {\bibinfo {author} {\bibfnamefont {A~D}\ \bibnamefont
  {Ribeiro}}, \bibinfo {author} {\bibfnamefont {M~A~M}\ \bibnamefont
  {de~Aguiar}}, \ and\ \bibinfo {author} {\bibfnamefont {A~F~R}\ \bibnamefont
  {de~Toledo~Piza}},\ }\bibfield  {title} {\enquote {\bibinfo {title} {The
  semiclassical coherent state propagator for systems with spin},}\ }\href
  {\doibase 10.1088/0305-4470/39/12/016} {\bibfield  {journal} {\bibinfo
  {journal} {J. Phys. A: Math. Gen.}\ }\textbf {\bibinfo {volume} {39}},\
  \bibinfo {pages} {3085--3097} (\bibinfo {year} {2006})}\BibitemShut {NoStop}%
\bibitem [{\citenamefont {Gutzwiller}(1971)}]{Gutzwiller1971}%
  \BibitemOpen
  \bibfield  {author} {\bibinfo {author} {\bibfnamefont {Martin~C.}\
  \bibnamefont {Gutzwiller}},\ }\bibfield  {title} {\enquote {\bibinfo {title}
  {Periodic orbits and classical quantization conditions},}\ }\href {\doibase
  10.1063/1.1665596} {\bibfield  {journal} {\bibinfo  {journal} {J. Math.
  Phys.}\ }\textbf {\bibinfo {volume} {12}},\ \bibinfo {pages} {343--358}
  (\bibinfo {year} {1971})}\BibitemShut {NoStop}%
\bibitem [{\citenamefont {Gutzwiller}(1990)}]{Gutzwiller1990book}%
  \BibitemOpen
  \bibfield  {author} {\bibinfo {author} {\bibfnamefont {M.~C.}\ \bibnamefont
  {Gutzwiller}},\ }\href@noop {} {\emph {\bibinfo {title} {Chaos in classical
  and quantum mechanics}}}\ (\bibinfo  {publisher} {Springer-Verlag},\ \bibinfo
  {address} {New York},\ \bibinfo {year} {1990})\BibitemShut {NoStop}%
\bibitem [{\citenamefont {Werner}(1989)}]{Werner1989}%
  \BibitemOpen
  \bibfield  {author} {\bibinfo {author} {\bibfnamefont {Reinhard~F.}\
  \bibnamefont {Werner}},\ }\bibfield  {title} {\enquote {\bibinfo {title}
  {Quantum states with einstein-podolsky-rosen correlations admitting a
  hidden-variable model},}\ }\href {\doibase 10.1103/PhysRevA.40.4277}
  {\bibfield  {journal} {\bibinfo  {journal} {Phys. Rev. A}\ }\textbf {\bibinfo
  {volume} {40}},\ \bibinfo {pages} {4277--4281} (\bibinfo {year}
  {1989})}\BibitemShut {NoStop}%
\bibitem [{\citenamefont {Tavis}\ and\ \citenamefont
  {Cummings}(1968)}]{Tavis1968}%
  \BibitemOpen
  \bibfield  {author} {\bibinfo {author} {\bibfnamefont {Michael}\ \bibnamefont
  {Tavis}}\ and\ \bibinfo {author} {\bibfnamefont {Frederick~W.}\ \bibnamefont
  {Cummings}},\ }\bibfield  {title} {\enquote {\bibinfo {title} {Exact solution
  for an $n$-molecule---radiation-field {H}amiltonian},}\ }\href {\doibase
  10.1103/PhysRev.170.379} {\bibfield  {journal} {\bibinfo  {journal} {Phys.
  Rev.}\ }\textbf {\bibinfo {volume} {170}},\ \bibinfo {pages} {379--384}
  (\bibinfo {year} {1968})}\BibitemShut {NoStop}%
\end{thebibliography}%

\end{document}